\documentclass[twocolumn]{aastex631}

\usepackage{amsmath}

\newcommand{\vsys}{v_{\text{sys}}}

\shorttitle{eDisk II: Dust Settling and Snow Surfaces of IRAS~04302+2247}
\shortauthors{Lin et al.}

\graphicspath{{./}{figures/}}
\begin{document}

%A First Dissection of the Butterfly Star IRAS~04302+2247 in Dazzling Detail --- 
%Dissecting the Butterfly Star IRAS~04302+2247 Featuring Non-Settled Dust Amid Complex Infall and Outflow Gas of IRAS~04302+2247
\title{
    Early Planet Formation in Embedded Disks (eDisk). II. \\
    Limited Dust Settling and Prominent Snow Surfaces in the Edge-on Class~I Disk IRAS~04302+2247
    }

\author[0000-0001-7233-4171]{Zhe-Yu Daniel Lin}
\affiliation{University of Virginia, 530 McCormick Rd., Charlottesville, Virginia 22904, USA}

\author[0000-0002-7402-6487]{Zhi-Yun Li}
\affiliation{University of Virginia, 530 McCormick Rd., Charlottesville, Virginia 22904, USA}

\author[0000-0002-6195-0152]{John J. Tobin}
\affiliation{National Radio Astronomy Observatory, 520 Edgemont Rd., Charlottesville, Virginia 22903, USA}

\author[0000-0003-0998-5064]{Nagayoshi Ohashi}
\affiliation{Academia Sinica Institute of Astronomy and Astrophysics, 11F of Astronomy-Mathematics Building, AS/NTU, No. 1, Sec. 4, Roosevelt Rd., Taipei 10617, Taiwan}

\author[0000-0001-9133-8047]{Jes Kristian Jørgensen}
\affiliation{Niels Bohr Institute, University of Copenhagen, {\O}ster Voldgade 5--7, 1350, Copenhagen K, Denmark}

%%%%%%%%%%%%%%%%%%%%%%%%%%%%%%%%%%%%%%%%%%%%%%
%%%%% AUTHOR LIST:
%%%%% Add your name + ORCID and affiliation here if you want to be part of and have contributed to the paper.
%%%%% The exact order of the author list will be determined later.

\author[0000-0002-4540-6587]{Leslie W. Looney}
\affiliation{Department of Astronomy, University of Illinois, 1002 West Green St., Urbana, Illinois 61801, USA}

\author[0000-0002-8238-7709]{Yusuke Aso}
\affiliation{Korea Astronomy and Space Science Institute, 776 Daedeok-daero, Yuseong-gu, Daejeon 34055, Republic of Korea}

\author[0000-0003-0845-128X]{Shigehisa Takakuwa}
\affiliation{Department of Physics and Astronomy, Graduate School of Science and Engineering, Kagoshima University, 1-21-35 Korimoto, Kagoshima, Kagoshima 890-0065, Japan}
\affiliation{Academia Sinica Institute of Astronomy and Astrophysics, 11F of Astronomy-Mathematics Building, AS/NTU, No. 1, Sec. 4, Roosevelt Rd., Taipei 10617, Taiwan}

\author[0000-0003-3283-6884]{Yuri Aikawa}
\affiliation{Department of Astronomy, Graduate School of Science, The University of Tokyo, 7-3-1 Hongo, Bunkyo-ku, Tokyo 113-0033, Japan}

\author[0000-0002-2555-9869]{Merel L.R. van 't Hoff}
\affiliation{Department of Astronomy, University of Michigan, 1085 S. University Ave., Ann Arbor, Michigan 48109-1107, USA}

\author[0000-0003-4518-407X]{Itziar de Gregorio-Monsalvo}
\affiliation{European Southern Observatory, European Southern Observatory, Alonso de Cordova 3107, Casilla 19, Vitacura, Santiago, Chile}

\author[0000-0002-3566-6270]{Frankie J. Encalada}
\affiliation{Department of Astronomy, University of Illinois, 1002 West Green St., Urbana, Illinois 61801, USA}

\author[0000-0002-8591-472X]{Christian Flores}
\affiliation{Academia Sinica Institute of Astronomy and Astrophysics, 11F of Astronomy-Mathematics Building, AS/NTU, No. 1, Sec. 4, Roosevelt Rd., Taipei 10617, Taiwan}

\author[0000-0001-5782-915X]{Sacha Gavino}
\affiliation{Niels Bohr Institute, University of Copenhagen, {\O}ster Voldgade 5--7, 1350, Copenhagen K, Denmark}

\author[0000-0002-9143-1433]{Ilseung Han}
\affiliation{Division of Astronomy and Space Science, University of Science and Technology, 217 Gajeong-ro, Yuseong-gu, Daejeon 34113, Republic of Korea}
\affiliation{Korea Astronomy and Space Science Institute, 776 Daedeok-daero, Yuseong-gu, Daejeon 34055, Republic of Korea}

\author[0000-0002-2902-4239]{Miyu Kido}
\affiliation{Department of Physics and Astronomy, Graduate School of Science and Engineering, Kagoshima University, 1-21-35 Korimoto, Kagoshima, Kagoshima 890-0065, Japan}

\author[0000-0003-2777-5861]{Patrick M. Koch}
\affil{Academia Sinica Institute of Astronomy and Astrophysics, 11F of Astronomy-Mathematics Building, AS/NTU, No. 1, Sec. 4, Roosevelt Rd., Taipei 10617, Taiwan}

\author[0000-0003-4022-4132]{Woojin Kwon}
\affil{Department of Earth Science Education, Seoul National University, 1 Gwanak-ro, Gwanak-gu, Seoul 08826, Republic of Korea}
\affil{SNU Astronomy Research Center, Seoul National University, 1 Gwanak-ro, Gwanak-gu, Seoul 08826, Republic of Korea}

\author[0000-0001-5522-486X]{Shih-Ping Lai}
\affiliation{Institute of Astronomy, National Tsing Hua University, No. 101, Section 2, Kuang-Fu Road, Hsinchu 30013, Taiwan}
\affiliation{Center for Informatics and Computation in Astronomy, National Tsing Hua University, No. 101, Section 2, Kuang-Fu Road, Hsinchu 30013, Taiwan}
\affiliation{Department of Physics, National Tsing Hua University, No. 101, Section 2, Kuang-Fu Road, Hsinchu 30013, Taiwan}
%\affiliation{Academia Sinica Institute of Astronomy and Astrophysics, P.O. Box 23-141, 10617 Taipei, Taiwan}
\affiliation{Academia Sinica Institute of Astronomy and Astrophysics, 11F of Astronomy-Mathematics Building, AS/NTU, No. 1, Sec. 4, Roosevelt Rd., Taipei 10617, Taiwan}

\author[0000-0002-3179-6334]{Chang Won Lee}
\affiliation{Division of Astronomy and Space Science, University of Science and Technology, 217 Gajeong-ro, Yuseong-gu, Daejeon 34113, Republic of Korea}
\affiliation{Korea Astronomy and Space Science Institute, 776 Daedeok-daero, Yuseong-gu, Daejeon 34055, Republic of Korea}

\author[0000-0003-3119-2087]{Jeong-Eun Lee}
\affiliation{Department of Physics and Astronomy, Seoul National University, 1 Gwanak-ro, Gwanak-gu, Seoul 08826, Korea}

\author[0000-0002-4372-5509]{Nguyen Thi Phuong}
\affiliation{Korea Astronomy and Space Science Institute, 776 Daedeokdae-ro, Yuseong-gu, Daejeon, Korea}
\affiliation{Department of Astrophysics, Vietnam National Space Center, Vietnam Academy of Science and Technology, 18 Hoang Quoc Viet, Cau Giay, Hanoi, Vietnam}

\author[0000-0003-4361-5577]{Jinshi Sai (Insa Choi)}
\affiliation{Academia Sinica Institute of Astronomy and Astrophysics, 11F of Astronomy-Mathematics Building, AS/NTU, No. 1, Sec. 4, Roosevelt Rd., Taipei 10617, Taiwan}

\author[0000-0002-0549-544X]{Rajeeb Sharma}
\affiliation{Niels Bohr Institute, University of Copenhagen, {\O}ster Voldgade 5--7, 1350, Copenhagen K, Denmark}

\author[0000-0002-9209-8708]{Patrick Sheehan}
\affiliation{National Radio Astronomy Observatory, 520 Edgemont Rd., Charlottesville, Virginia 22903, USA}

\author[0000-0003-0334-1583]{Travis J. Thieme}
\affiliation{Institute of Astronomy, National Tsing Hua University, No. 101, Section 2, Kuang-Fu Road, Hsinchu 30013, Taiwan}
\affiliation{Center for Informatics and Computation in Astronomy, National Tsing Hua University, No. 101, Section 2, Kuang-Fu Road, Hsinchu 30013, Taiwan}
\affiliation{Department of Physics, National Tsing Hua University, No. 101, Section 2, Kuang-Fu Road, Hsinchu 30013, Taiwan}

\author[0000-0001-5058-695X]{Jonathan P. Williams}
\affiliation{Institute for Astronomy, University of Hawai‘i at Mānoa, 2680 Woodlawn Dr., Honolulu, Hawai‘i 96822, USA}

\author[0000-0003-4099-6941]{Yoshihide Yamato}
\affiliation{Department of Astronomy, Graduate School of Science, The University of Tokyo, 7-3-1 Hongo, Bunkyo-ku, Tokyo 113-0033, Japan}

\author[0000-0003-1412-893X]{Hsi-Wei Yen}
\affiliation{Academia Sinica Institute of Astronomy and Astrophysics, 11F of Astronomy-Mathematics Building, AS/NTU, No. 1, Sec. 4, Roosevelt Rd., Taipei 10617, Taiwan}

%%%%%%%%%%%%%%%%%%%%%%%%%%%%%%%%%%%%%%%%%%%%%%

\begin{abstract}

While dust disks around optically visible, Class~II protostars are found to be vertically thin, when and how dust settles to the midplane are unclear. As part of the Atacama Large Millimeter/submillimeter Array (ALMA) large program, Early Planet Formation in Embedded Disks, we analyze the edge-on, embedded, Class~I protostar IRAS~04302+2247, also nicknamed the ``Butterfly Star." With a resolution of $0.05\arcsec$ (8~au), the 1.3~mm continuum shows an asymmetry along the minor axis which is evidence of an optically thick and geometrically thick disk viewed nearly edge-on. There is no evidence of rings and gaps, which could be due to the lack of radial substructure or the highly inclined and optically thick view. With $0.1\arcsec$ (16~au) resolution, we resolve the 2D snow surfaces, i.e., the boundary region between freeze-out and sublimation, for $^{12}$CO $J$=2--1, $^{13}$CO $J$=2--1, C$^{18}$O $J$=2--1, $H_{2}$CO $J$=$3_{0,3}$--$2_{0,2}$, and SO $J$=$6_{5}$--$5_{4}$, and constrain the CO midplane snow line to $\sim 130$~au. We find Keplerian rotation around a protostar of $1.6 \pm 0.4 M_{\odot}$ using C$^{18}$O. Through forward ray-tracing using RADMC-3D, we find that the dust scale height is $\sim 6$~au at a radius of 100~au from the central star and is comparable to the gas pressure scale height. The results suggest that the dust of this Class~I source has yet to vertically settle significantly. 
%In addition, the radius of the dust disk is smaller than the radius of the gas disk suggesting that radial drift occurs sooner than vertical settling.

\end{abstract}


%%% what is the broad science theme I can address? What's so valuable about this source to the broader audience

%%% broad science theme: planet formation -> infall from envelope and outflow -> decoupling of dust and gas -> dust settling 
%%% what's unclear? 
%%% - vertical structure of gas and dust. observations generally only get the radial structure 
%%% - the level of coupling with the gas (Stokes number; depends on grain size and gas density) and level of turbulence (alpha)

%%% an early source that's edge-on
%%% we can directly verify dust settling from (1) separation of gas and dust, (2) brightness asymmetry along the minor axis
%%% we can see a bunch of extended structure that is not part of the disk 

%%% dust settling happens even when the disk is chaotic!!! 
%%% there is outflow and infall 
%%% need to talk about earliest phases of disks and the importance of high angular resolution 

\section{Introduction} \label{sec:intro}
% has there been gas observations compared to dust before? 

The formation of rotationally supported circumstellar disks plays a crucial role in the star and planet formation process. As a consequence of the conservation of angular momentum, much of the material from the larger scale core is channeled to the disk and subsequently accretes onto the protostar itself \citep[e.g.][]{Terebey1984ApJ...286..529T, Li2014prpl.conf..173L, Tsukamoto2022arXiv220913765T}. The reservoir of material in the disk enables the growth of solids and serves as the birthplace of planets \citep[e.g.][]{Testi2014prpl.conf..339T, Drazkowska2022arXiv220309759D, Tu2022MNRAS.515.4780T}. Nevertheless, the process of dust evolution, from sub-micron-sized particles inherited from the core to planetesimals and planets, requires numerous mechanisms to overcome multiple growth barriers, e.g., the meter-sized barrier \citep{Weidenschilling1977MNRAS.180...57W}. One of the most favored mechanisms to overcome the meter-sized barrier is the streaming instability, which can drive rapid growth from pebbles to planetesimals, but it requires comparable densities of the dust and the gas rather than the 1:100 dust-to-gas ratio inherited from the interstellar medium \citep[e.g.][]{Youdin2005ApJ...620..459Y, Lesur2022arXiv220309821L}. One natural process to increase the dust-to-gas ratio is through dust settling \citep[e.g.][]{Gole2020ApJ...904..132G}. 

Gaseous disks are vertically extended owing to the vertical pressure support. The balance between the pressure gradient and the vertical gravitational pull sets the gas scale height. In contrast, dust particles, if decoupled from the pressure-supported gas, will inevitably descend to the midplane to form a thin dust layer. Turbulent mixing operates against dust settling by stirring up the dust and prevents the dust from becoming fully settled  \citep[e.g.][]{Nakagawa1986Icar...67..375N, Dubrulle1995Icar..114..237D}. While the tendency for settling is well established, the effectiveness of turbulence is not clear and relies on observations for constraints \citep[e.g.][]{Pinte2016ApJ...816...25P, Ohashi2019ApJ...886..103O, Villenave2022ApJ...930...11V}. However, observations that can characterize the vertical structure of disks are few in number, since it requires high angular resolution of nearly edge-on disks \citep{Tobin2010ApJ...722L..12T, Lee2017_darklane, Sakai2017MNRAS.467L..76S, Lee2020NatAs...4..142L, Villenave2020, Michel2022ApJ...937..104M, Ohashi2022ApJ...934..163O}. 
%\red{cite edisk: L1527, GSS 30, CB 68, IRAS 32}

%Recent high angular resolution observations have revealed significant radial substructures in virtually all of the Class II sources \citep[e.g.][]{Huang2018ApJ...852..122H, Andrews2018ApJ...869L..41A, Long2018ApJ...869...17L, Cieza2021MNRAS.501.2934C, Villenave2022ApJ...930...11V} and even around younger Class I sources \citep[e.g.][]{alma_partnership2015ApJ...808L...3A, Segura-Cox2020Natur.586..228S}. The existence of fine dust rings, as a result of dust trapping in local pressure maxima, should signify a certain degree of decoupling of dust from the gas. The same decoupling should 

%%% basic idea (IRAS04302 generally) and why IRAS04302 specifically
IRAS~04302+2247 (hereafter IRAS~04302) is a Class~I (bolometric temperature $T_{\text{bol}}=88$~K; \citealt{Ohashi_edisk_overview}) protostar, poetically nicknamed the ``Butterfly Star" by \cite{Lucas1997MNRAS.286..895L} for its remarkable bipolar reflection nebulae in the near-infrared. High-resolution near-infrared images from the Hubble Space Telescope/NICMOS exhibited a clear dark lane sandwiched between the reflection nebulae and depict a highly inclined system with an obscured central source and bipolar cavity walls that scatter the near-infrared photons \citep{Padgett1999AJ....117.1490P}. A molecular outflow in H$_{2}$ was detected in the same direction as the bipolar cavity walls \citep{Lucas1998MNRAS.299..723L}, and the deep absorption silicate feature in the mid-infrared requires significant inclination \citep{Furlan2008ApJS..176..184F}. Indeed, millimeter wavelength observations show an elongated continuum within the near-infrared dark lane, which is evidence of the presence of an edge-on disk \citep{Wolf2003ApJ...588..373W, Wolf2008ApJ...674L.101W, Sheehan2017ApJ...851...45S, vantHoff2020, Villenave2020}. The near edge-on disk orientation facilitates the  determination of the geometrical thickness of the dust layer \citep[e.g.][]{Villenave2020}.

Detailed models of IRAS~04302 using scattered light images and the millimeter continuum images, which trace different physical processes and regions of the circumstellar system, have ascertained an inclined system of a disk and envelope \citep[e.g.][]{Wolf2003ApJ...588..373W, Furlan2008ApJS..176..184F, Wolf2008ApJ...674L.101W, Eisner2012ApJ...755...23E, Sheehan2017ApJ...851...45S}. Intriguingly, the dust in the envelope is consistent with interstellar medium (ISM) grains \citep[e.g.][]{Lucas1997MNRAS.286..895L}, while the dust in the disk is found to have grown significantly \citep{Wolf2003ApJ...588..373W, Grafe2013A&A...553A..69G, Sheehan2017ApJ...851...45S}. Furthermore, \cite{Grafe2013A&A...553A..69G} suggested that the larger grains in the disk show evidence of radial and vertical decoupling from the small grains.

Recent molecular line observations with $\sim 0.3\arcsec$ to $0.4\arcsec$ achieved by ALMA have begun to resolve the locations where molecules trace the disk surface of IRAS~04302, making the study of its vertical structure possible \citep{vantHoff2020, Podio2020A&A...642L...7P}. \cite{vantHoff2020} identified C$^{17}$O in the midplane within 100~au and detected emission in the disk surface layers beyond 100~au which can be explained by freeze-out of CO. In addition, H$_{2}$CO ($3_{1,2}-2_{1,1}$) mainly originates from the disk surface layers with a large reduction of emission at the midplane where the continuum is located. \cite{Podio2020A&A...642L...7P} also found a similar distribution of emission for $^{12}$CO ($2-1$), H$_{2}$CO ($3_{2,1} - 2_{1,1}$), and CS ($5-4$). The pattern is consistent with results from thermochemical models that consist of a midplane freeze-out and an elevated molecular layer separated by a snow surface \cite[e.g.][]{Aikawa2002A&A...386..622A, Akimkin2013ApJ...766....8A, Dutrey2014prpl.conf..317D}. Furthermore, different isotopologues of CO trace different densities \citep{vantHoff2020, Podio2020A&A...642L...7P}.

Most of the prior continuum observations have been limited in angular resolution with $\sim 0.2\arcsec$ to $0.5\arcsec$ making it difficult to resolve the vertical structure of the dust \citep{Grafe2013A&A...553A..69G, Podio2020A&A...642L...7P, vantHoff2020}. The highest angular resolution of the continuum to date is $\sim 0.06\arcsec$ at $\lambda=2.1$~mm and hints at a flared dust disk \citep{Villenave2020}. The unique view of IRAS~04302 thus serves as a perfect laboratory to study the vertical structure of the dust and gas around a young source in detail. As part of the Early Planet Formation in Embedded Disks (eDisk) program, we present high-resolution $\lambda=1.3$~mm continuum ($\sim 0.05\arcsec$ or $8$~au) and molecular line images ($\sim 0.1\arcsec$ or 16~au) obtained from ALMA.

IRAS~04302 is located within the L1536 cloud of the Taurus star-forming region. The whole Taurus star-forming region is conventionally assumed to have a distance of 140~pc \citep{Kenyon1994AJ....108.1872K}, but recent parallax measurements found significant depth effects for each cloud. From Gaia, \cite{Luhman2018AJ....156..271L} and \cite{Roccatagliata2020A&A...638A..85R} found a distance of 161 and 160.3~pc, respectively, for the L1536 cloud. \cite{Galli2018ApJ...859...33G} inferred a distance of 162.7~pc using astrometry from the Very Long Baseline Array. For this paper, we adopt a distance of 160 pc.

The rest of the paper is organized as follows. Section~\ref{sec:observations} describes the observations and data processing, while Section~\ref{sec:results} shows the resulting dust continuum images and molecular line channel maps. We analyze the continuum and line data in more detail in Section~\ref{sec:analysis}.
%Section~\ref{sec:analysis} analyzes in more detail by providing a dust model to interpret the continuum image and by fitting the rotation curve. 
We discuss several implications in Section~\ref{sec:discussion} and conclude in Section~\ref{sec:conclusion}.

\section{Observations} \label{sec:observations}
%Sept. 30, 2021, 4302.62 sec
%Oct. 1, 2021, 4150.42s
%Oct. 1, 2021, 4303.68s
The data are obtained as part of the ALMA Large Program (2019.1.00261.L, PI: N. Ohashi). The details of the survey, including the spectral setup, calibrators, and imaging procedure, are discussed in \cite{Ohashi_edisk_overview}. We briefly describe the relevant setup for IRAS~04302. The short baseline data of IRAS~04302 were observed on Dec. 21, 2021 in configuration C-5 with baselines ranging from 15~m to 3.6~km with an on-source integration time of $\sim 35$ minutes. The long baseline data were observed on Sept.~30 and Oct.~1 in 2021 with total integration times of $\sim 2.16$~hours in configuration C-8 with baselines ranging from 70~m to 11.9~km. The spectral setup was in Band 6 with a representative wavelength of 1.3~mm (225~GHz) for the continuum. The spectral resolution for each detected line is listed in Table~\ref{tab:images_summary}. 

All calibration and imaging tasks utilized the Common Astronomy Software Applications (CASA) package \citep{McMullin2007ASPC..376..127M} version 6.2.1 and pipeline version 2021.2.0.128. From the pipeline calibrated data, we follow the self-calibration procedure presented in \cite{Ohashi_edisk_overview} which we briefly describe in the following. First, we imaged each execution block separately and aligned the peaks to a common phase center using the \texttt{fixvis} and \texttt{fixplanets} tasks. Second, to adjust for flux calibration uncertainties between each execution block, we scaled the amplitude of the visibilities that were azimuthally binned as a function of $uv$-distance. We self-calibrated the short-baseline data through three rounds of phase-only calibration. With the self-calibrated short-baseline data, we included the long-baseline data and conducted one round of phase-only calibration with a solution interval that was the length of each execution block.

We used the \texttt{tclean} task to image the self-calibrated visibilities. The continuum imaging used several Briggs robust weightings from robust=-2 to 2 \citep{Briggs1995PhDT.......238B}. Smaller robust values correspond to better angular resolution at the expense of increased noise, while larger robust values correspond to better sensitivity albeit with lower angular resolution \citep[e.g.,][]{Briggs1995PhDT.......238B, Czekala2021ApJS..257....2C}. We show the resulting images in Appendix~\ref{sec:continuum_robust}. 
We adopt the image with robust=0.5 as the representative image to compromise between spatial resolution, sensitivity, and image fidelity. 

The self-calibration solutions were applied to the measurement set used for the lines and further continuum subtracted using the \texttt{uvcontsub} task. Each line image cube used a robust=0.5 and 2 with the \texttt{uvtaper} set at 2000k$\lambda$ (or $\sim 0.09 \arcsec$). The self-calibration and imaging scripts for this source can be found at \url{http://github.com/jjtobin/edisk}. We assume a $10\%$ absolute flux calibration uncertainty, but we only consider the statistical uncertainty for the rest of this paper. The resulting resolution and noise levels for each image are listed in Table~\ref{tab:images_summary}.

The CLEAN process for lines results in 2D images as a function of frequency $\nu$ (i.e., an image cube or channel maps). From the image cube, we can define several 2D quantities to interpret the 3D data. We denote the image cube as $I(x, y, \nu)$, where $x$ and $y$ represent the sky coordinates, R.A. and Dec. With a known line transition frequency $\nu_{0}$, one can convert from the observed frequency $\nu$ to the velocity along the line-of-sight, $v$, through
\begin{equation}
    v = \big(1 - \frac{\nu}{\nu_{0}} \big) c
\end{equation}
where $c$ is the speed of light. For the rest of the paper, we express $v$ in km s$^{-1}$. The channel width in velocity units $\Delta v$ is related to the channel width in frequency units $\Delta \nu$ through
\begin{equation}
    \Delta v = \frac{ \Delta \nu }{ \nu_{0}} c
\end{equation}
where $\Delta v$ and $\Delta \nu$ are both positive quantities. $\Delta v$ is shown in Table~\ref{tab:images_summary}.

Since the image is defined on discrete pixels and spectral channels, we use $I_{i,j,k}$, where $i,j,k$ are indices, to represent the intensity value at a certain pixel $(x_{i}, y_{j})$ and at a certain velocity $v_{k}$. The integrated intensity image $M$ is defined as 
\begin{equation}
    M_{i,j} \equiv \sum_{k} I_{i,j,k} \Delta v_{k}. 
\end{equation}
The resulting two-dimensional quantity only depends on the sky coordinates and the units are in Jy beam$^{-1}$ km s$^{-1}$.

The peak intensity image $P$ results from taking the peak along the spectrum at each coordinate $(x_{i}, y_{j})$ expressed as
\begin{equation}
    P_{i,j} \equiv \text{max} [ I_{i,j,k} : k = 1,...,N_{v}]
\end{equation}
where ``$\text{max}$" represents taking the maximum along the velocity axis with $N_{v}$ points.

The intensity $I_{\nu}$ in units of Jy beam$^{-1}$ can be converted to brightness temperature in kelvins through the Planck function. Suppose the major and minor axes of the beam are $\theta_{M}$ and $\theta_{m}$, respectively. The solid angle of the beam is $\Omega=\pi \theta_{M} \theta_{m} / (4 \ln 2)$. Let $J_{\nu}$ be the intensity $I_{\nu}$ expressed in erg s$^{-1}$ cm$^{-2}$ Hz$^{-1}$ ster$^{-1}$ which are related by $J_{\nu} = I_{\nu} 10^{-23} / \Omega$. 
The peak intensity image $P$ can express in brightness temperature with
\begin{equation} \label{eq:jypbeam_to_tb}
    T_{b} = \frac{ h \nu }{k} \frac{1}{ \ln ( \frac{2 h \nu^{3}}{c^{2} J_{\nu} } + 1) }
\end{equation}
Using Eq.~(\ref{eq:jypbeam_to_tb}), we can express $P$ as a brightness temperature.

With the line-of-sight velocities known, one can analyze the velocity structure of the source by extracting a representative velocity at each pixel of the image cube. We use the ``peak velocity image," $V$, which is the velocity that corresponds to the peak of the spectrum at each pixel. The peak velocity map does not rely on an assumption on the profile shape and is shown to be less susceptible to noise compared to other methods \citep[e.g.][]{deBlok2008AJ....136.2648D, Teague2018RNAAS...2..173T}.

The images, $M$, $P$, and $V$, were created using the CASA task \texttt{immoment} setting the argument \texttt{moments} to 0, 8, and 9, respectively.\footnote{We caution that the term \textit{moment} in the \texttt{immoments} task differs from the mathematical definition of the moment, which would be defined as $\int I(v) v^{n} dv$ to express the $n$th the moment of the spectrum.} Furthermore, we only consider emission above the $3\sigma$ level to avoid ``negative" intensities from continuum oversubtraction (see Table~\ref{tab:images_summary} for the noise levels).

% name, frequency, noise level, beam size, velocity resolution
\begin{deluxetable*}{ccccccc}
    \tablenum{1}
    \tablecaption{Summary of Images \label{tab:images_summary}}
    \tablewidth{0pt}
    \tablehead{
        \colhead{Frequency} & \colhead{Image} & \colhead{Transition} & \colhead{Velocity Resolution} & \colhead{Robust} & \colhead{Noise Level} & \colhead{Beam Size} \\
        \colhead{(GHz)} & \nocolhead{dummy} & \nocolhead{dummy} & \colhead{(km s$^{-1}$)} & \nocolhead{dummy}& \colhead{(mJy beam$^{-1}$)} & \nocolhead{Name} 
        }
    \decimalcolnumbers
    \startdata
        225 & 1.3 mm continuum & - & - & 0.5 & $1.45 \times 10^{-2}$ & $0.055\arcsec \times 0.050\arcsec$ \\
        230.53800000 & $^{12}$CO & 2 -- 1 & 0.635 & 0.5 & 1.03 & $0.097\arcsec \times 0.082 \arcsec$ \\
         & & & & 2.0 & 1.26 & $0.135\arcsec \times 0.114\arcsec$ \\
        220.39868420 & $^{13}$CO & 2 -- 1 & 0.167 & 0.5 & 2.21 & $0.099\arcsec \times 0.083\arcsec$ \\
         & & & & 2.0 & 2.95 & $0.137\arcsec \times 0.118\arcsec$ \\
        219.56035410 & C$^{18}$O & 2 -- 1 & 0.167 & 0.5 & 1.64 & $0.099\arcsec \times 0.082\arcsec$ \\
        % & & & & 2.0 & 1.59 & $0.136\arcsec \times 0.116\arcsec$ \\
        219.94944200 & SO & $6_{5}$ -- $5_{4}$ & 0.167 & 0.5 & 1.96 & $0.099\arcsec \times 0.082 \arcsec$ \\
        % & & & & 2.0 & 1.86 & $0.135\arcsec \times 0.117\arcsec$ \\
        218.22219200 & H$_{2}$CO & $3_{0,3}$ -- $2_{0,2}$ & 1.34 & 0.5 & 0.554 & $0.10\arcsec \times 0.084\arcsec$ \\
        % & & & & 2.0 & 0.524 & $0.143\arcsec \times 0.117\arcsec$ \\
    \enddata
    \tablecomments{
        see \cite{Ohashi_edisk_overview} for the complete spectral setup. 
        }
\end{deluxetable*}

\section{Results} \label{sec:results}

\subsection{Continuum} \label{sec:continuum}
\begin{figure}
    \centering
    \includegraphics[width=\columnwidth,trim={0cm 0.5cm 1.5cm 0cm},clip]{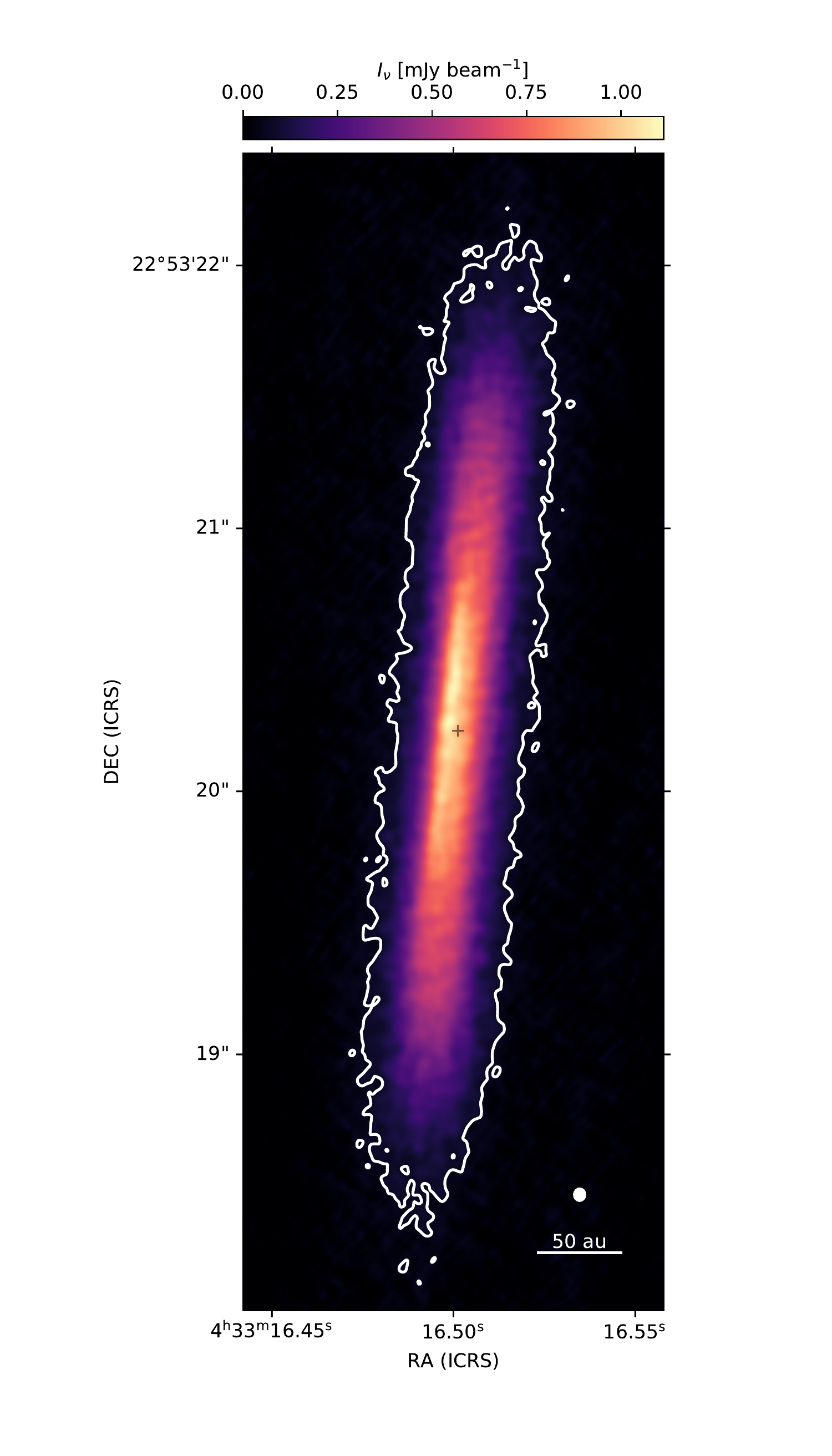}
    \caption{
        The continuum image of IRAS~04302+2247. The white contour marks the $5\sigma$ level (see $\sigma$ in Table~\ref{tab:images_summary}). The white ellipse in the lower right corner is the beam size (see Table~\ref{tab:images_summary}) and the length scale is 50~au. The black plus sign marks the center from the best-fit 2D Gaussian. 
    }
    \label{fig:big_continuum}
\end{figure}

% we can tell which is the near side
Fig.~\ref{fig:big_continuum} shows the 1.3~mm continuum image with robust=0.5 and reveals a highly elongated structure that is consistent with past low angular resolution images at millimeter wavelengths \citep{Wolf2003ApJ...588..373W, Wolf2008ApJ...674L.101W, Grafe2013A&A...553A..69G}. The image has a peak of 1.11~mJy beam$^{-1}$ with a noise level of $\sigma=$14.5~$\mu$Jy beam$^{-1}$. The total flux is $184.15$~mJy by integrating the emission above $3\sigma$. The image appears largely symmetric along the major axis, but clearly asymmetric along the minor axis in which the east side is brighter than the west side. The elongated emission is expected from an inclined disk-like structure and the kinematic analysis in Section~\ref{sec:analysis} confirms a Keplerian disk. Thus, we will refer to the elongated continuum as simply the (dust) disk. Even with the higher angular resolution compared to previous observations, there is no clear evidence of rings or gaps.

To characterize the continuum image, we fit the disk with a 2D Gaussian using the CASA task \texttt{imfit}. The coordinate center of the 2D Gaussian is one of the free parameters, and we get the best-fit value of (04:33:16.50, +22:53:20.2) in ICRS, which we set as the origin of the image hereafter unless explicitly stated otherwise. {We treat the center as the location of the star. The deconvolved full width at half maximum (FWHM) for the major and minor axes are $2.149\arcsec \pm 0.007\arcsec$ and $0.2385\arcsec \pm 0.0007\arcsec$ respectively. Assuming a completely flat disk, the ratio between the minor and major axes equals $\cos i$ where $i$ is the inclination of the disk ($i=0^{\circ}$ means face-on). With the FWHM from the 2D Gaussian fitting, we derive $i \sim 84^{\circ}$. Since the disk has a finite vertical thickness, the inclination estimation is a lower limit (see Section~\ref{sec:continuum_RT_modeling}). The position angle (PA) of the major axis of the best-fit 2D Gaussian is $174.77^{\circ} \pm 0.03^{\circ}$  which we adopt as the position angle of the major axis of the system. The total flux from the fitting is $182.6\pm0.6$~mJy ($1\sigma$ uncertainty).

Fig.~\ref{fig:cont_cuts} compares the major and minor axis cuts with the origin set at the center determined from the fitted 2D Gaussian. The cuts are produced by interpolating the image and we also calculate the brightness temperature $T_{b}$ using Eq.~(\ref{eq:jypbeam_to_tb}). The brightness temperature is low across the disk with only $\sim 14$~K at the peak. For comparison, the peak brightness temperatures at $\lambda=0.9$~mm (ALMA Band~7) and $\lambda=2.1$~mm (ALMA Band~4) are 10 and 6.7~K, respectively \citep{Villenave2020}. The slightly higher peak brightness temperature presented here is likely because the disk is better resolved. The extent of the major axis reaches up to $\sim 2\arcsec$ (320~au) from the center, which is similar to the Band 4 and 7 continuum images from \cite{Villenave2020}. The large extent implies a fairly large disk radius, which we constrain in Section~\ref{sec:continuum_RT_modeling}.

\begin{figure*}
    \centering
    \includegraphics[width=\textwidth]{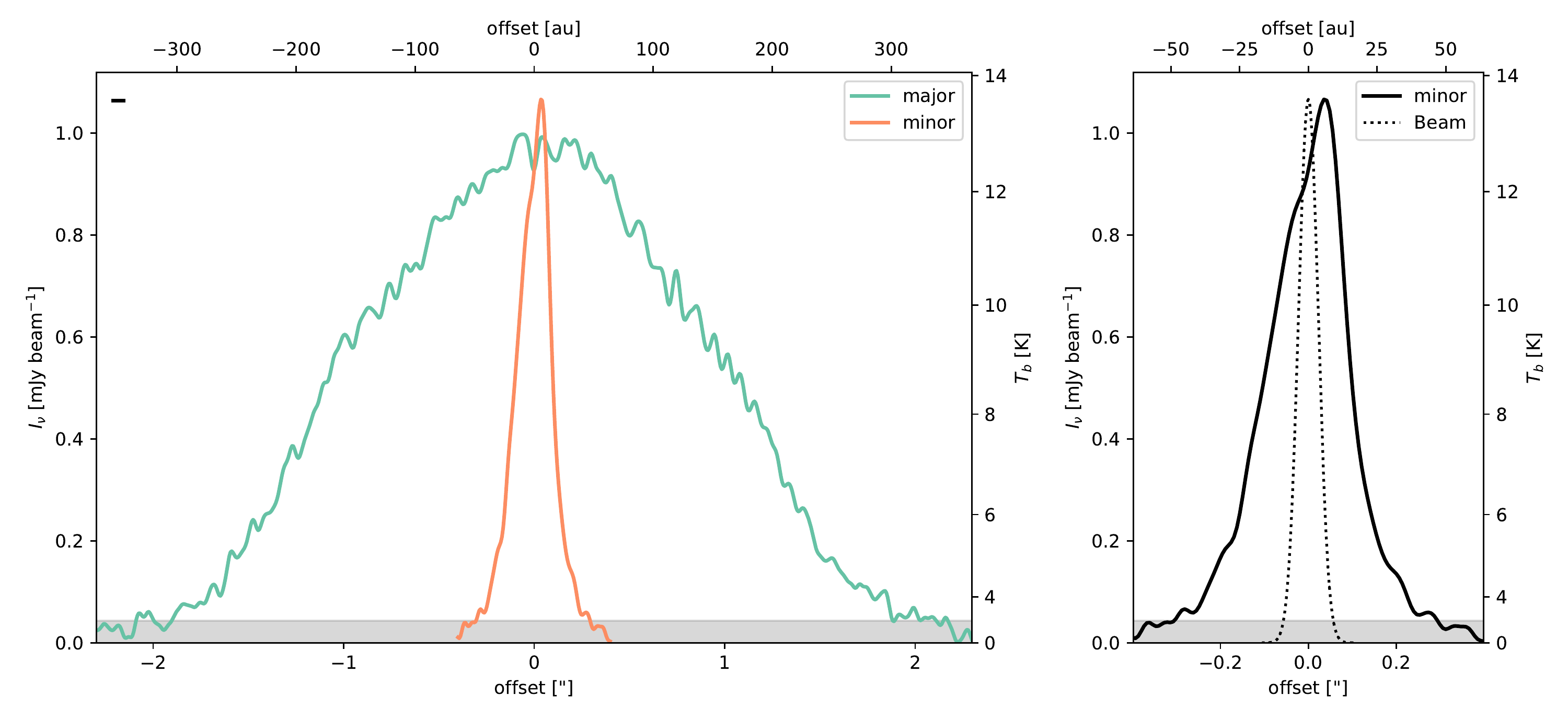}
    \caption{
        Left panel: The cuts along the major and minor axis of the continuum disk by interpolating the image. The cut along the major axis is in green and that along the minor axis is in orange. The origin is at the center of the fitted 2D Gaussian (04:33:16.5, +22:53:20.2). The bottom and top axes mark the offset from the origin along the cut in arcsec and au. The positive location for the major axis is along the northern part of the disk, while the positive location for the minor axis is along the eastern part of the disk. The left and right axes mark the intensity in mJy~beam$^{-1}$ and brightness temperature (using the full Planck function) in Kelvin respectively. The line segment to the upper left corner represents the length of the FWHM of the beam. The shaded region is the intensity below $3\sigma$. Right panel: Zoom-in comparison between the minor axis cut (solid black line) and the beam (dotted black line). 
    }
    \label{fig:cont_cuts}
\end{figure*}

To see the asymmetry along the minor axis clearly, we zoom in on the minor axis cut and show a comparison with the beam in the right panel of Fig.~\ref{fig:cont_cuts}. The FWHM of the minor axis is resolved by $\sim 3.5$ beams. The asymmetry could be due to an intrinsically asymmetric disk or due to a highly inclined axisymmetric disk that is optically thick, has a finite geometrical thickness, and is not seen exactly edge-on. We favor the latter possibility since the asymmetry occurs along the minor axis and is readily consistent with the high inclination and with the direction of the outflow (see Section~\ref{sec:lines}) Given that the emission is brighter on the east side, we can infer that the east side is the far side of the disk based on simple expectations of an optically thick disk with decreasing temperature as a function of radius \citep{Lee2017_darklane, Villenave2020, Ohashi2022ApJ...934..163O, Takakuwa_edisk} demonstrated through detailed modeling in Section~\ref{sec:continuum_RT_modeling}). In addition, the optically thinner $\lambda=2.1$~mm (ALMA Band~4) image with similar resolution ($\sim 0.06\arcsec$; 10~au) does not show a similar asymmetry \citep{Villenave2020}, which is more consistent with our picture than an intrinsically asymmetric disk.

By assuming the emission at $\nu=225$~GHz comes entirely from the dust thermal emission and is optically thin, one can estimate the total dust mass disk through
\begin{equation} \label{eq:opt_thin_mass}
    M_{\text{dust}} = \frac{ D^{2} S_{\nu} }{ \kappa_{\nu} B_{\nu}(T) }
\end{equation}
where $S_{\nu} = \int I_{\nu} d \Omega$ is the flux density, $\kappa_{\nu}$ is the mass opacity in cm$^{2}$ g$^{-1}$ of dust, $D$ is the distance to the source, $T$ is the temperature in Kelvin, and $B_{\nu}$ is the black body radiation using the Planck function. We adopt the opacity of $0.023$~cm$^{2}$ g$^{-1}$ of gas from \cite{Beckwith1990AJ.....99..924B} (see also recent evidence from \citealt{Lin2021} in support of this prescription and Section~\ref{sec:continuum_RT_modeling}) and assume a dust-to-gas mass ratio of 0.01 to obtain $\kappa_{\nu}=2.3$ cm$^{2}$ g$^{-1}$ of dust. We assume $T=20$~K which is a commonly adopted value for surveys \citep[e.g.][]{Andrews2005ApJ...631.1134A, Ansdell2016ApJ...828...46A, Tobin2020ApJ...890..130T}. Since $D=160$~pc and $S_{\nu}=184$~mJy for IRAS~04302, we have $M_{\text{dust}} \sim 140 M_{\oplus}$. Another way to estimate a representative temperature is based on the bolometric luminosity 
\begin{equation}
    T = 43 (L_{\text{bol}} / L_{\odot})^{1/4}
\end{equation}
which is optimized at a radius of $50$~au \citep{Tobin2020ApJ...890..130T}. With $L_{\text{bol}}=0.43 L_{\odot}$ \citep{Ohashi_edisk_overview}, we have $T \sim 34$~K and the dust mass is $M_{\text{dust}} \sim 70 M_{\oplus}$. Note that since the disk is clearly not optically thin (as we can see from the asymmetry from the minor axis due to optical depth effects) the estimate here is a lower limit and likely a drastic underestimation given the near edge-on view. 
%Without assuming the dust-to-gas mass ratio, the total disk mass is $\sim 2-3\%$ of the central star. 

\subsection{Lines} \label{sec:lines}

\begin{figure*}
    \centering
    \includegraphics[width=\textwidth,trim={0cm 0.5cm 0cm 0cm},clip]{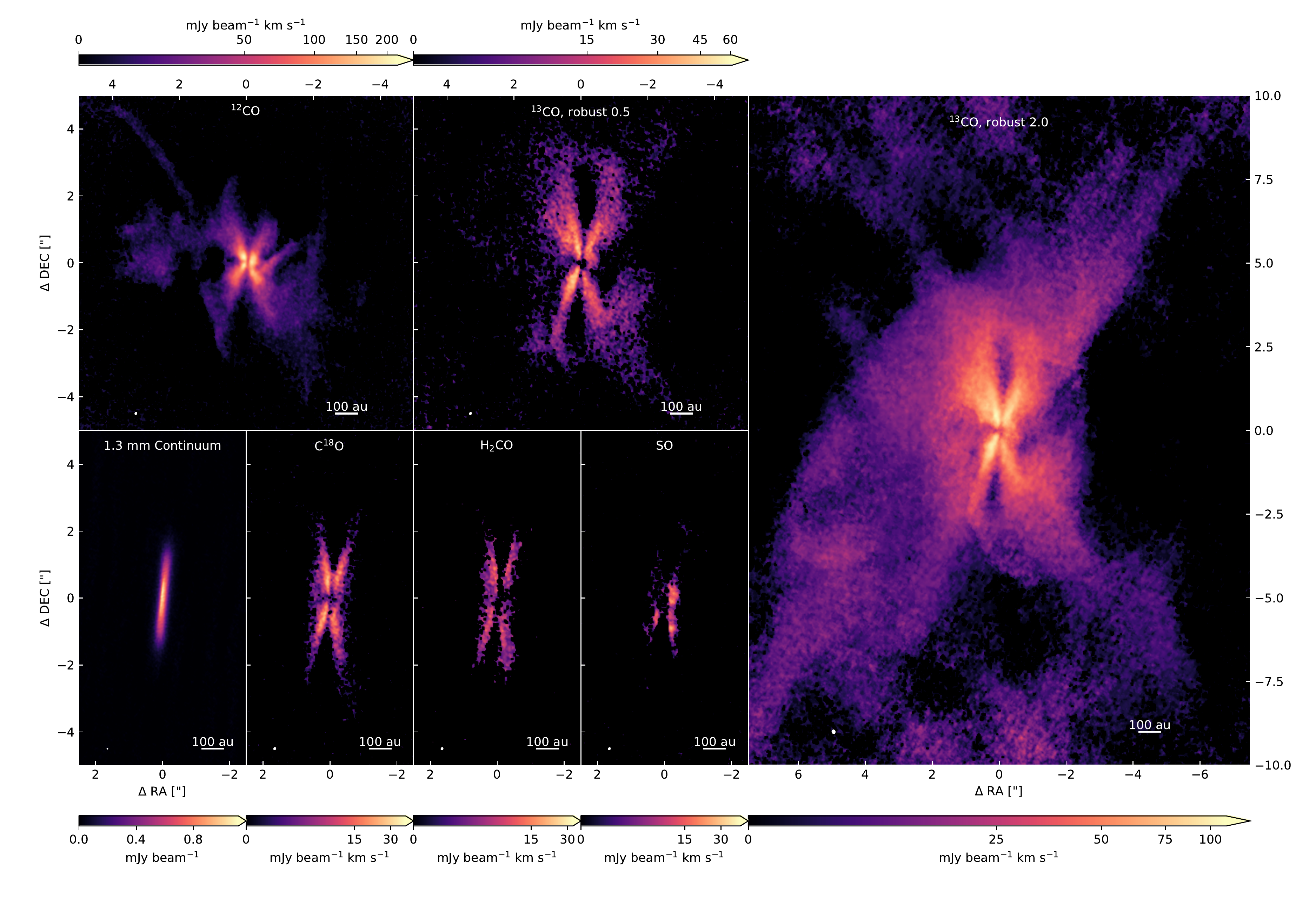}
    \caption{
        The integrated intensity images for $^{12}$CO (top left), $^{13}$CO with robust=0.5 (top center), $^{13}$CO with robust=2 (right), C$^{18}$O (bottom center-left), H$_{2}$CO 3$_{0,3}$-2$_{0,2}$ (bottom center-right), and SO (bottom right). The continuum is shown in the bottom left. The color scale starts from 0. The horizontal bar represents 100~au which is the same across all panels. The white ellipse in the lower right corner of each plot represents the beam size.
    }
    \label{fig:image_collection}
\end{figure*}

% talk about the different scales and snow surface
Fig.~\ref{fig:image_collection} shows the integrated intensity images for $^{12}$CO~2--1, $^{13}$CO~2--1, C$^{18}$O~2--1, H$_{2}$CO~3$_{0,3}$--2$_{0,2}$, and SO~$6_{5}$--$5_{4}$ with robust=0.5. We also show $^{13}$CO with robust=2 which captures more large-scale emission that is resolved out from robust=0.5.
The different molecules trace different spatial scales of the edge-on disk and their images also differ from the continuum image.

In the direction parallel to the disk major axis, which is described by the impact parameter, the extents of C$^{18}$O, H$_{2}$CO, and $^{12}$CO (robust=0.5) appear comparable to the continuum image, while SO clearly spans a smaller range in impact parameter. The $^{13}$CO image with robust=0.5 is the most extended and even more so with robust=2. The large extent of $^{13}$CO with robust=0.5, which is on the order $\sim 4\arcsec$ (640~au) from the center, suggests a gas disk that is larger than the dust disk (see Section~\ref{sec:outer_cap} for more detail).

In the direction parallel to the disk minor axis, which corresponds to the ``vertical" direction of an edge-on disk, the more optically thin lines, C$^{18}$O, H$_{2}$CO, and SO, are more confined to regions just to the east and west of the continuum. This suggests that these molecules trace the disk surface and not the outer edges of the disk in the radial direction. In contrast, the more optically thick lines, $^{12}$CO and $^{13}$CO, are far more vertically extended. The emission traced by $^{13}$CO with robust=0.5 appears to reach $\sim 2\arcsec$ (320~au) in the vertical direction (see Section~\ref{sec:snow_line_and_surface}). 
%For $^{12}$CO, the emission in the east side is along the disk rotation axis which is where we would expect associated outflow or jet emission. 

%%% V-shape snow surface
A common feature across all images is the lack of emission near the supposed disk midplane and the emission appears to form a V-shaped pattern to the north and south. The same feature was observed in previous lower angular resolution observations in C$^{17}$O 2--1 and H$_{2}$CO 3$_{1,2}$--2$_{1,1}$ by \cite{vantHoff2020} and in $^{12}$CO 2--1, CS 5--4, H$_{2}$CO 3$_{1,2}$--2$_{1,1}$ by \cite{Podio2020A&A...642L...7P}. The lack of emission is largely due to freeze-out, especially at larger impact parameters, and the V-shape is a natural result of the snow surface given the typical 2D temperature structure of an irradiated disk \citep[e.g.,][]{Aikawa1999A&A...351..233A, Dutrey2017A&A...607A.130D, vantHoff2018A&A...615A..83V, Qi2019ApJ...882..160Q, Zhang2019ApJ...883...98Z, Flores2021AJ....161..239F}. Indeed, an absorption feature due to CO ice is also detected for this source in the infrared \citep{Aikawa2012A&A...538A..57A}. At smaller impact parameters where the continuum could be optically thick, it could also be due to dust extinction (see Section~\ref{sec:dust_extinction}). 
Intriguingly, the $^{13}$CO image with robust=0.5 (and, similarly, the case with robust=2) not only has the V-shape where the emission diverges, but the emission converges at large impact parameters beyond $\sim 3\arcsec$ (480~au), enclosing a dark cavity and resembling the shape of the number ``8" overall. We can infer that in the midplane, $^{13}$CO is frozen out at small radii, but reappears at large radii (see Section~\ref{sec:snow_line_and_surface}). We discuss the cause for the re-emergence in Section~\ref{sec:outer_cap}. 

%... cite Bosman+2018 somewhere

Apart from the major axis, there is a lack of emission directly along the minor axis of the disk for $^{12}$CO, $^{13}$CO, C$^{18}$O, and H$_{2}$CO (it is less clear for SO). This can be explained by optical depth effects (see \citealt{vantHoff2018A&A...615A..83V} for a visualization) through the following. For a rotating disk seen edge-on, 
only the material with projected speeds near the systemic velocity, which is along the minor axis, can contribute. At those channels, we trace regions further from the disk where it is colder since there is more material along the line-of-sight (see also Section~\ref{sec:snow_line_and_surface} for the channel maps). For optically thicker lines, like $^{12}$CO and $^{13}$CO, much of the emission can even be resolved out and the depression is even more pronounced. Another factor that can decrease the brightness at small impact parameters is beam dilution \citep{Flores2021AJ....161..239F} though the high resolution images here are likely less susceptible.

%%% brighter sides
Another common feature seen in Fig.~\ref{fig:image_collection} is that the lines are all brighter on the east side compared to the west side, except for SO which appears brighter on the west side.
The brighter east side can be interpreted as an inclination effect for a disk with a two-dimensional temperature distribution \citep{Dutrey2017A&A...607A.130D, Flores2021AJ....161..239F}. We can infer that the brighter eastern side is the far side of the disk and the western side is the near side. The orientation is consistent with the orientation inferred from the continuum. The opposite behavior of SO, however, is puzzling and it could be due to other reasons, like chemical effects, rather than inclination effects \citep{Sakai2014Natur.507...78S}. Intriguingly, SO$_{2}$ was also detected to be brighter on the same side as SO \citep{Garufi2022A&A...658A.104G}. 

%%% look at moment 9
Fig.~\ref{fig:moments_C18O} expands upon the C$^{18}$O in Fig.~\ref{fig:image_collection} and compares the intensity integrated image, peak intensity image, and the peak velocity image. To distinguish the redshifted and blueshifted halves, the peak velocity images are shown relative to a systemic velocity, $\vsys$, of $5.7$~km s$^{-1}$ (see Section~\ref{sec:rotation_curve} for the measurement of $\vsys$). The velocity gradient of C$^{18}$O shows a clear signature of rotation at the disk surface. The other two optically thin tracers, H$_{2}$CO and SO, are shown in Fig.~\ref{fig:moments_H2CO} and Fig.~\ref{fig:moments_SO}. Similar to C$^{18}$O, both tracers also follow the disk surface and show similar velocity features.

\begin{figure*}
    \centering
    \includegraphics[width=\textwidth]{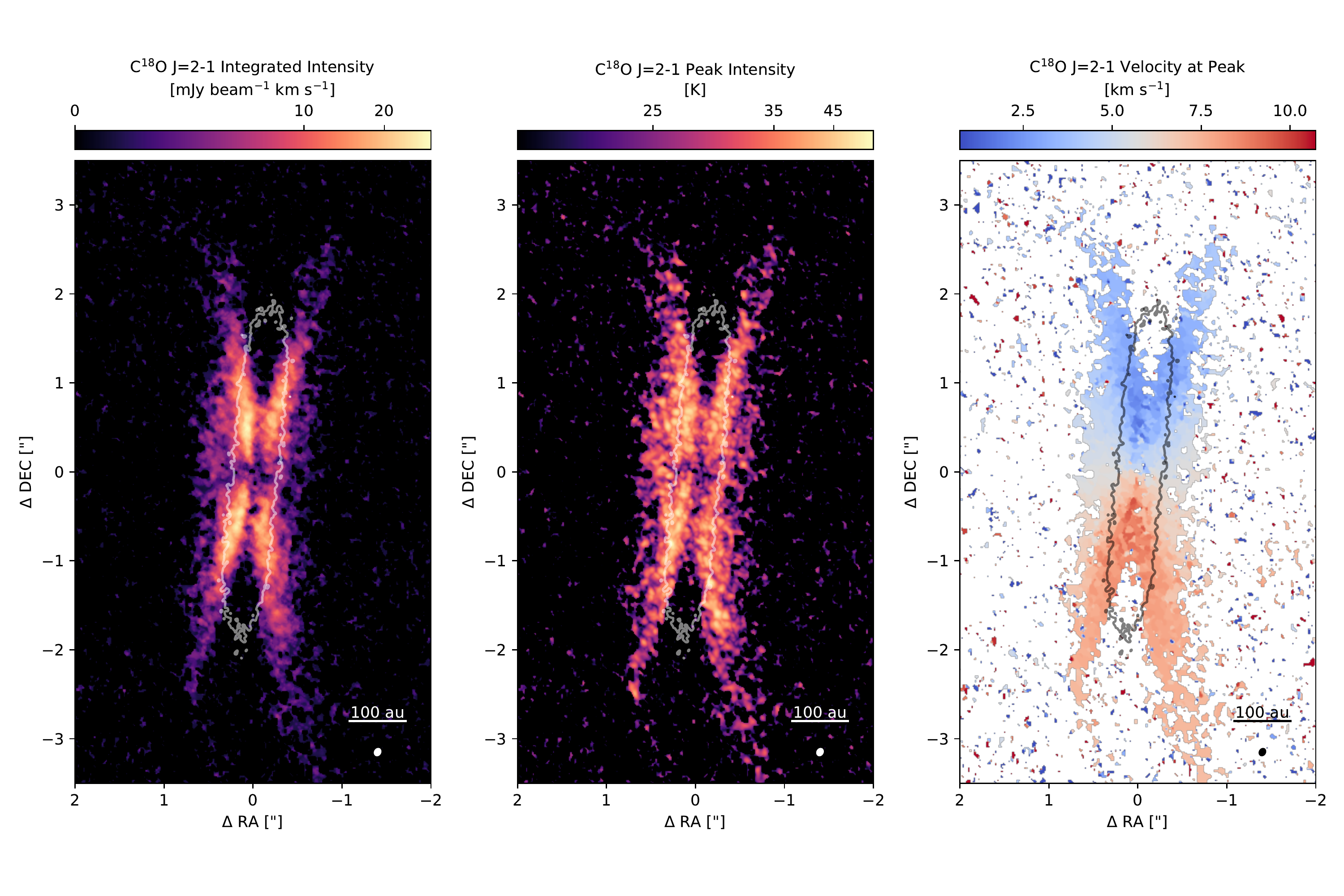}
    \caption{
        The integrated intensity image (left), the peak brightness temperature map (center), and the peak velocity image (right) for C$^{18}$O with robust=0.5. The velocity image is plotted relative to $\vsys=5.7$ km s$^{-1}$ (see Section~\ref{sec:rotation_curve}). The grey contour outlines the $5\sigma$ level of the continuum. The horizontal bar represents the 100~au length scale. The ellipse to the lower right is the beam size. 
        }
    \label{fig:moments_C18O}
\end{figure*}

\begin{figure*}
    \centering
    \includegraphics[width=\textwidth]{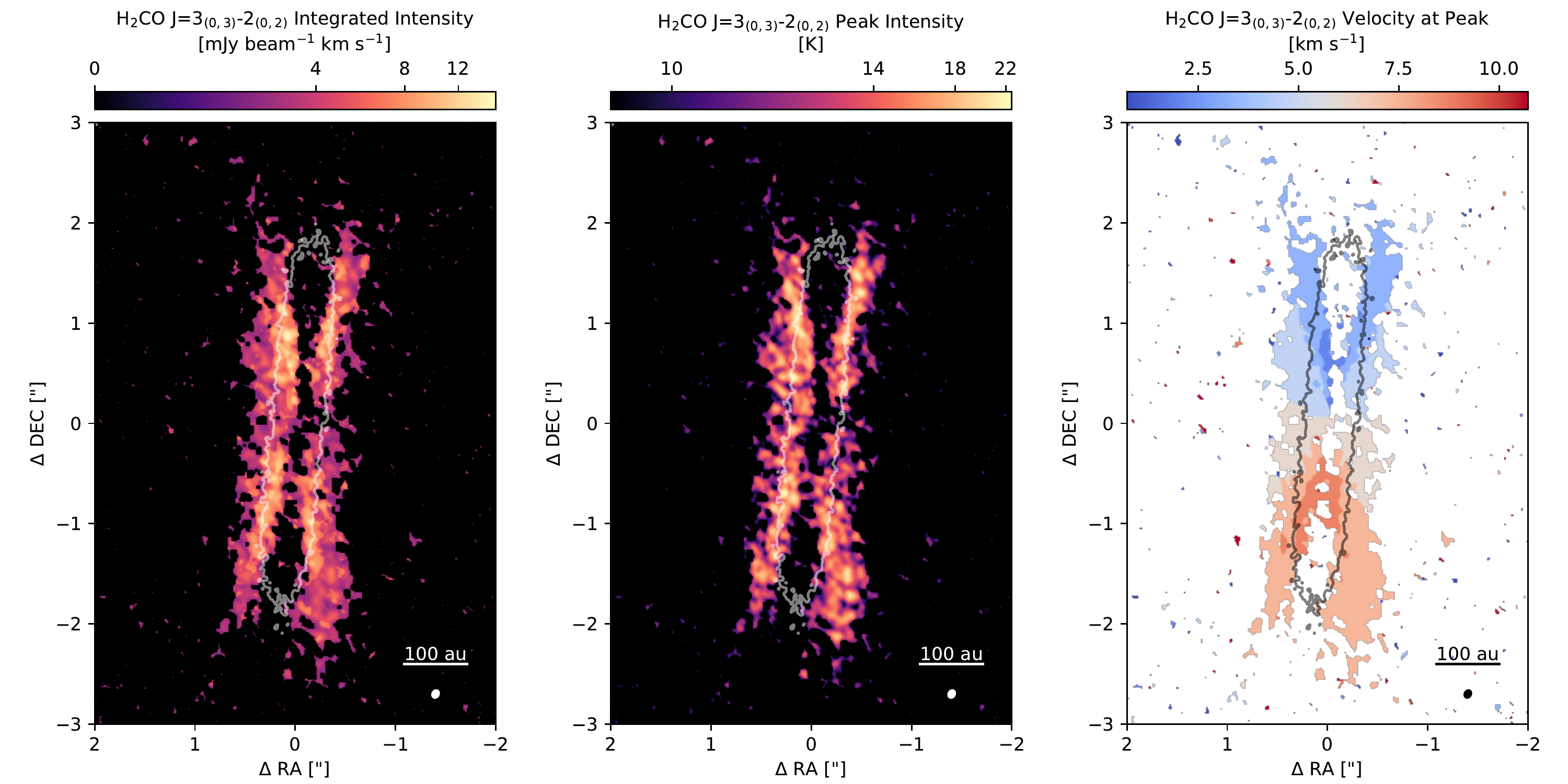}
    \caption{
        Similar to Fig.~\ref{fig:moments_C18O} but for H$_{2}$CO with robust=0.5. 
        }
    \label{fig:moments_H2CO}
\end{figure*}

\begin{figure*}
    \centering
    \includegraphics[width=\textwidth]{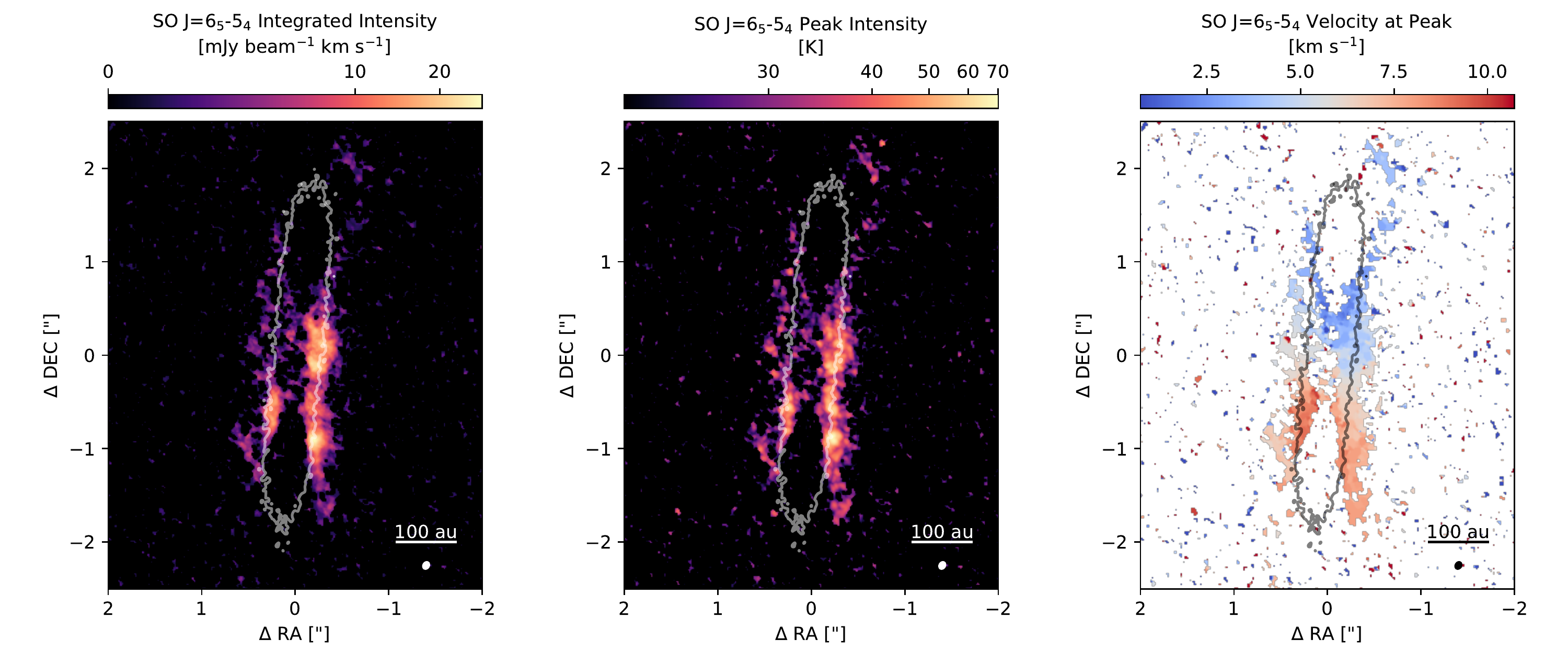}
    \caption{
        Similar to Fig.~\ref{fig:moments_C18O} but for SO with robust=0.5. 
        }
    \label{fig:moments_SO}
\end{figure*}

From Fig.~\ref{fig:moments_12CO} and Fig.~\ref{fig:moments_13CO}, we also see the same blueshifted and redshifted halves for $^{12}$CO and $^{13}$CO near the dust continuum which are similar to the C$^{18}$O, H$_{2}$CO, and SO, but there are additional extensions that do not follow what is expected from rotation. Notably, the extension towards the east side of the $^{12}$CO image at $\sim 3\arcsec$ from the center is blueshifted. The level of blueshift increases with increasing distance from the center which is consistent with a Hubble-type outflow\citep[e.g][]{Arce2007prpl.conf..245A}. A blueshifted outflow to the east is also consistent with the orientation of the disk where the east side is the far side. In addition, $^{12}$CO with robust=2 appears more extended than its robust=0.5 version, but it is still less extended than $^{13}$CO with robust=2. This is likely because much of $^{12}$CO remains resolved out even with robust=2.

\begin{figure*}
    \centering
    \includegraphics[width=\textwidth]{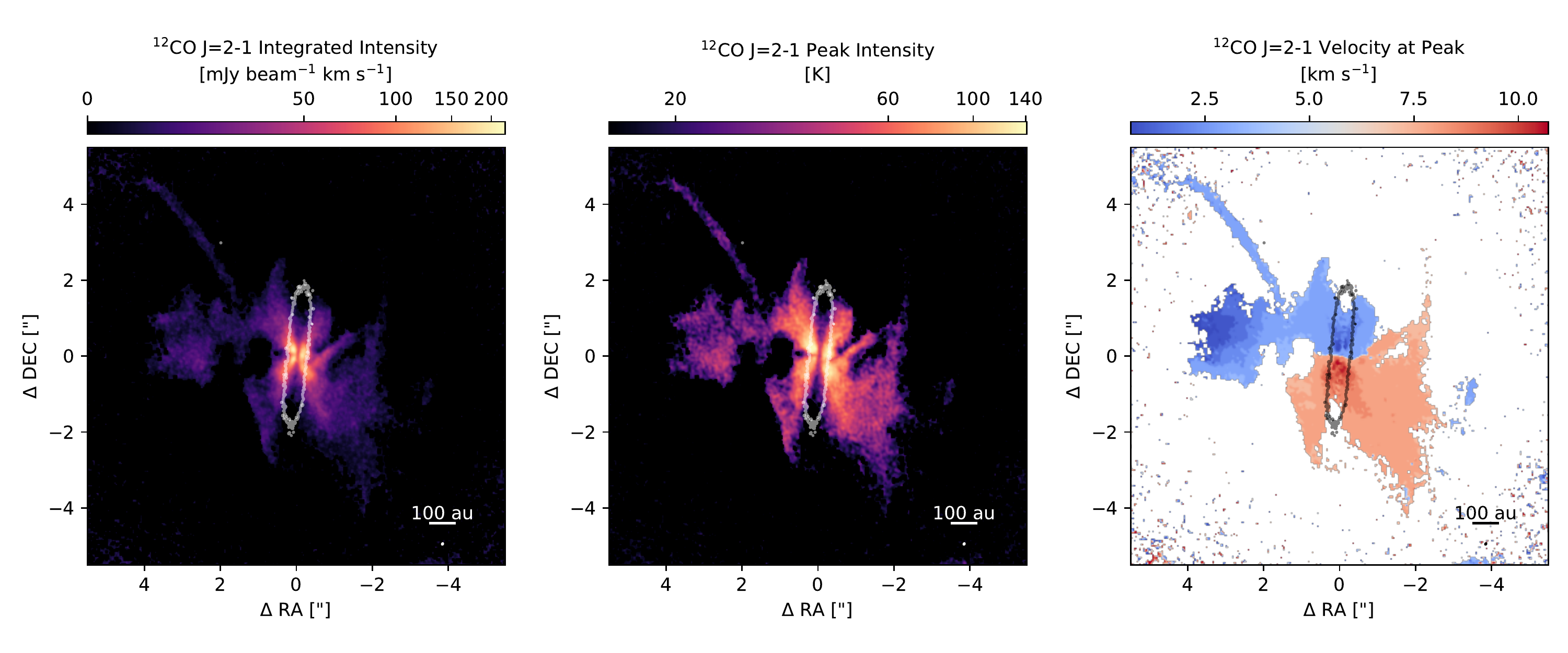}
    \includegraphics[width=\textwidth]{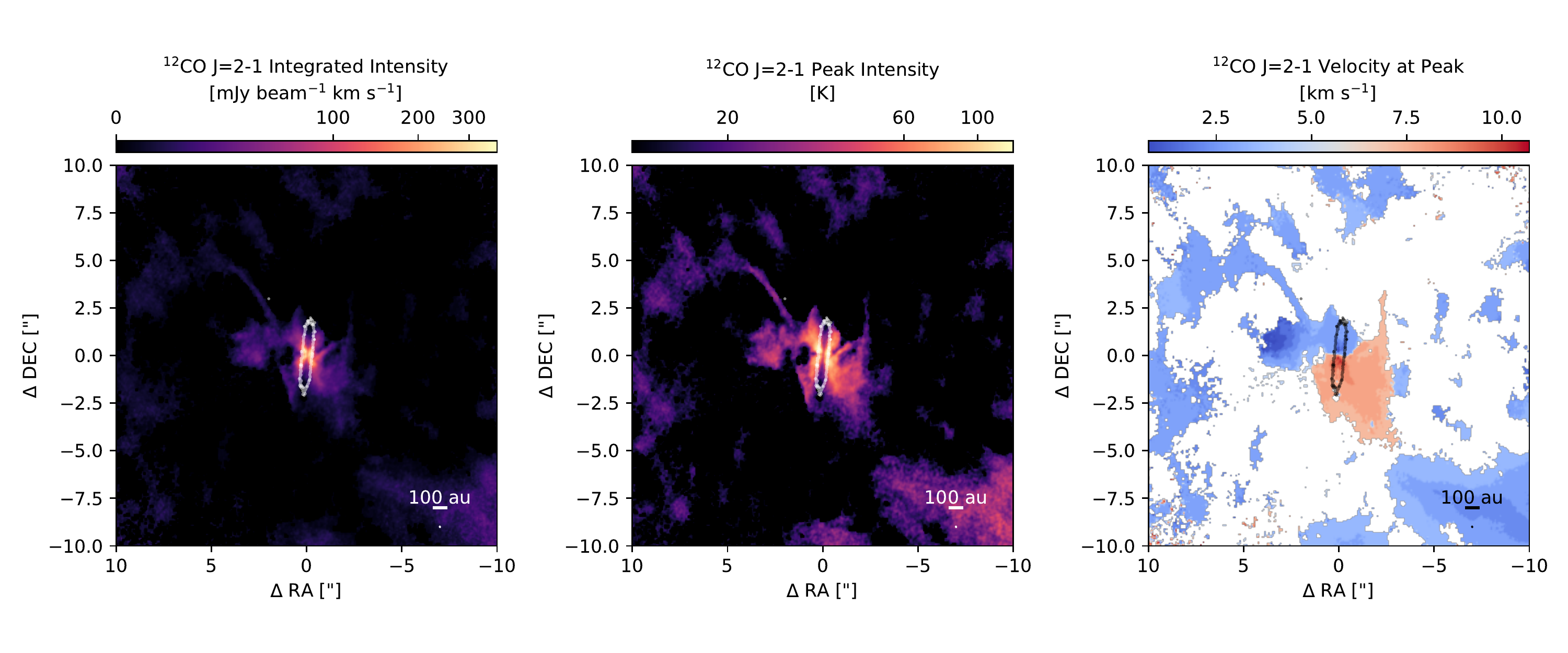}
    \caption{
        Top row: Similar to Fig.~\ref{fig:moments_C18O} but for $^{12}$CO with robust=0.5. Bottom row: The moment images for $^{12}$CO with robust=2.0. Note that the image covers a larger region than the top row to show the larger scale structure.
        }
    \label{fig:moments_12CO}
\end{figure*}

\begin{figure*}
    \centering
    \includegraphics[width=\textwidth]{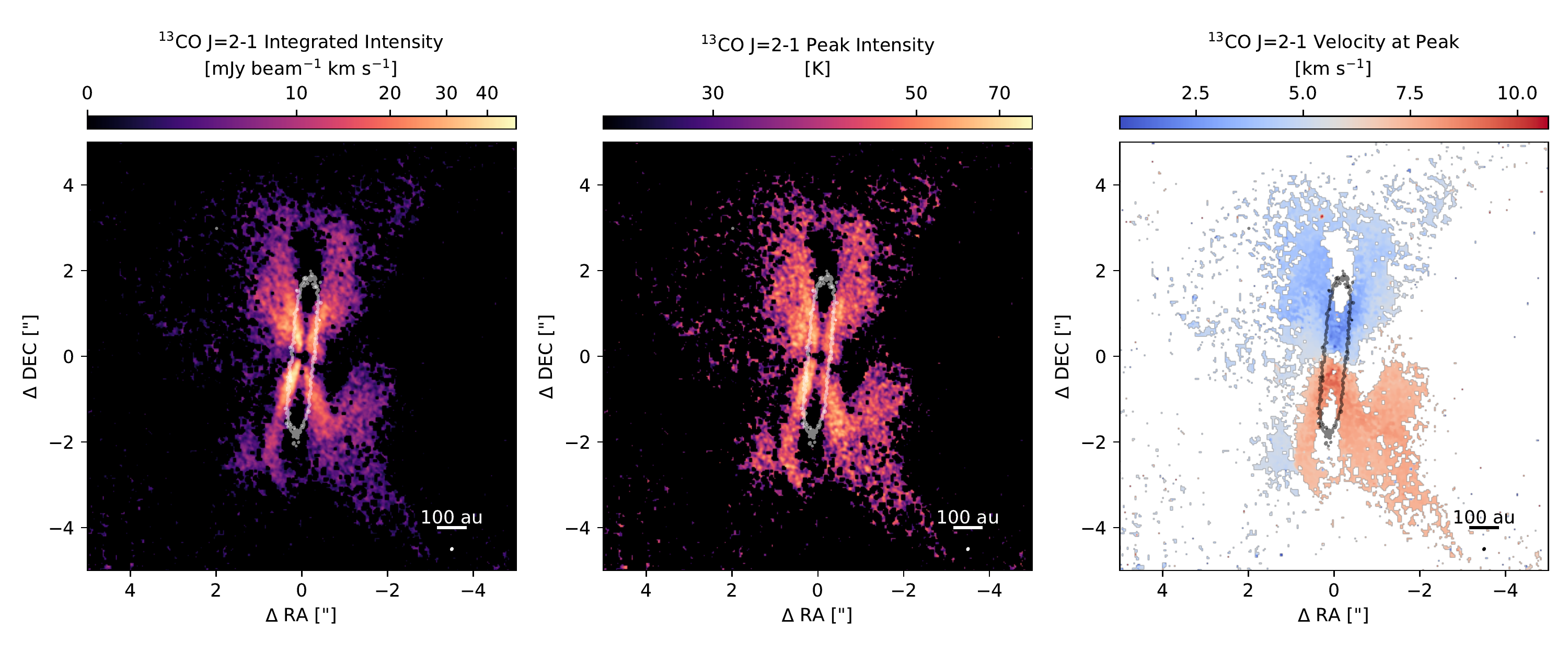}
    \includegraphics[width=\textwidth]{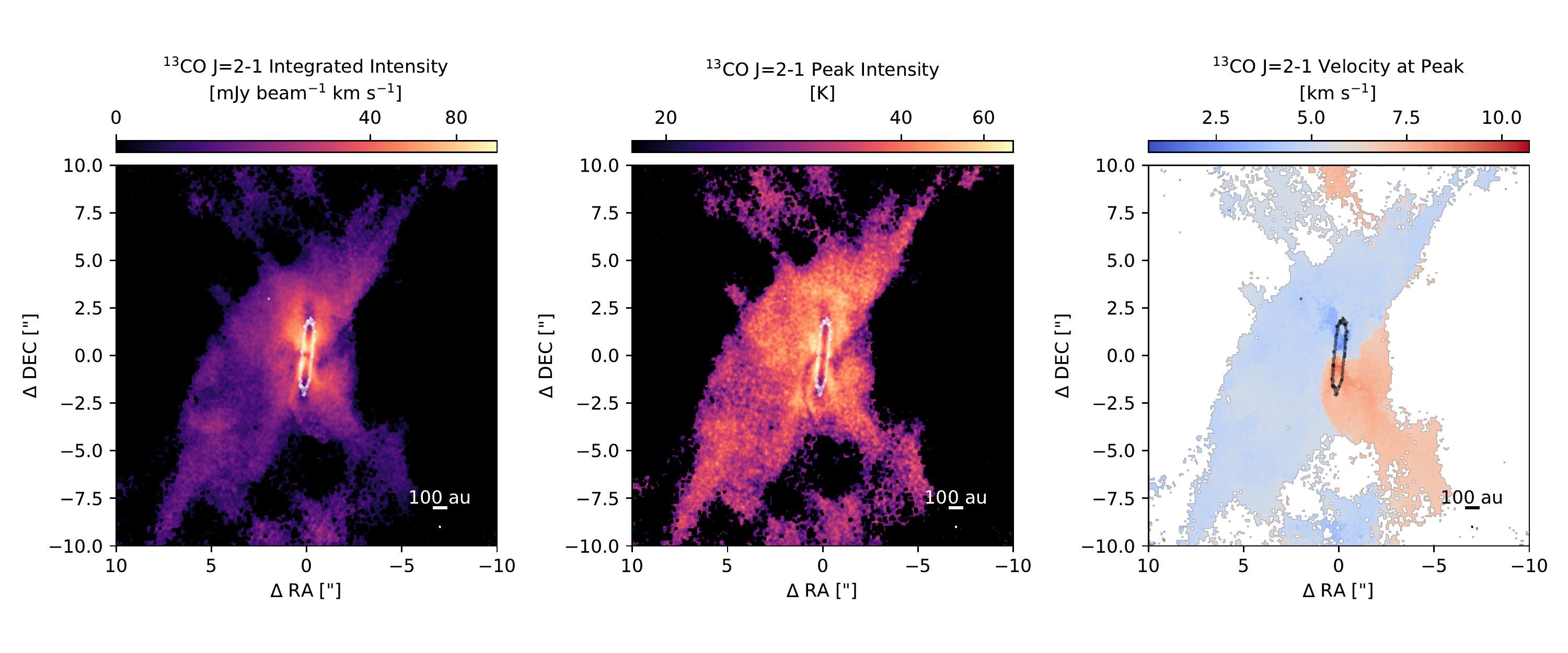}
    \caption{
        Top row: Similar to Fig.~\ref{fig:moments_C18O} but for $^{13}$CO with robust=0.5. Bottom row: The moment images for $^{13}$CO with robust=2.0. Note that the image covers a larger region than the top row to show the larger scale structure. 
        }
    \label{fig:moments_13CO}
\end{figure*}

\subsection{Tracing the CO Snow Line and Snow Surface} \label{sec:snow_line_and_surface}

The optically thinner tracer C$^{18}$O probes the snow line and the immediate snow surface better. For the convenience of the discussion from this section, we define $x$ as the impact parameter along the disk major axis, where positive $x$ lies in the northern part of the major axis and $y$ as the location along the disk minor axis where positive $y$ lies along the blueshifted side of the jet axis to the east of the disk midplane. The origin, $x=0$ and $y=0$, corresponds to the center of the fitted 2D Gaussian from Sec.~\ref{sec:continuum}. 

% take the point with outermost detection in the midplane
A key feature of the C$^{18}$O is a region near the midplane that clearly lacks emission which can be attributed to freeze-out \citep[e.g.][]{Dutrey2017A&A...607A.130D, vantHoff2020, Villenave2022ApJ...930...11V}. Given the fine resolution, we can trace the snow surface to $\sim 0.1\arcsec$. We show selected channel images of C$^{18}$O with robust=0.5 in Fig.~\ref{fig:channels_c18o_freezeout} and focus only on the northern half of the disk, since the freeze-out zone appears symmetric to the southern half (see Fig.~\ref{fig:moments_C18O}). To increase the signal-to-noise ratio and to limit the number of channel images, we averaged every 3 channels and the noise level used for the figure is correspondingly decreased by $\sqrt{3}$. 

To outline the snow surface, we give a simple prescription (motivated by a similar prescription in \citealt{Lee2021ApJ...910...75L}):
\begin{equation} \label{eq:snow_surface}
    y(x) = H_{v}
    \begin{cases}
        \bigg( \dfrac{x}{ R_{s} } \bigg)^{1.5} & R_{s} \leq x < R_{v} \\
        \sqrt{1 - \bigg(\dfrac{ x - R_{s} }{ R_{e} - R_{s} } \bigg)^{2} } & R_{v} \leq x < R_{e}
    \end{cases}. 
\end{equation}
This describes an increasing surface that begins from the snow line $R_{s}$ to some transition radius $R_{v}$ after which the V-shaped snow surface begins to close and ends at $R_{e}$ where the gas re-emerges. $H_{v}$ is the height at $R_{s}$. The closing of the snow surface at large impact parameters is less clear in C$^{18}$O, but more obvious in $^{13}$CO which we show in Fig.~\ref{fig:channel_13co_freezeout} and discuss later. We estimated the parameters to be $R_{s}=0.8\arcsec$ (130~au), $H_{v}=0.3\arcsec$ (48~au), $R_{v}=2\arcsec$ (320~au), and $R_{e}=2.8\arcsec$ (448~au) by eye. Although the disk is not perfectly edge-on, we assume a symmetric outline across $y=0$ for simplicity.

The mid-velocity channels ($\sim 2-4$ km s$^{-1}$ for the northern half) show the iconic V-shaped emission expected for a snow surface. Under the simple expectation that C$^{18}$O should exist from the center to the snow line, the emission should appear from the disk center at high velocities and emerge away from the center with decreasing velocity until the emission begins to concentrate along the disk minor axis at velocities near $\vsys$ (see other edge-on sources with Keplerian rotation, e.g., \citealt{Dutrey2017A&A...607A.130D}, \citealt{Teague2020MNRAS.495..451T}, \citealt{Flores2021AJ....161..239F}). Thus, the snow line $R_{s}$ is based on the maximum impact parameter with emission that exists between the east and west surfaces. Note that the location could be an upper limit due to contamination from finite beam averaging of the east and west snow surfaces. Nevertheless, the $R_{s}$ of $0.8\arcsec$ (130~au) appears consistent with previous constraints using C$^{17}$O \citep{vantHoff2020}. 
%The distance between the two snow surfaces is $\sim 0.15 \arcsec$ at $R_{b}$, which is only $\sim 1.7$ times the FWHM of the beam. 

%This could be due to simultaneous effects of dust extinction, continuum absorption, and contamination from the snow surface... can be better understood through modeling. 

\begin{figure*}
    \centering
    \includegraphics[width=\textwidth, trim={1.5cm 0.5cm 0.5cm 0cm},clip]{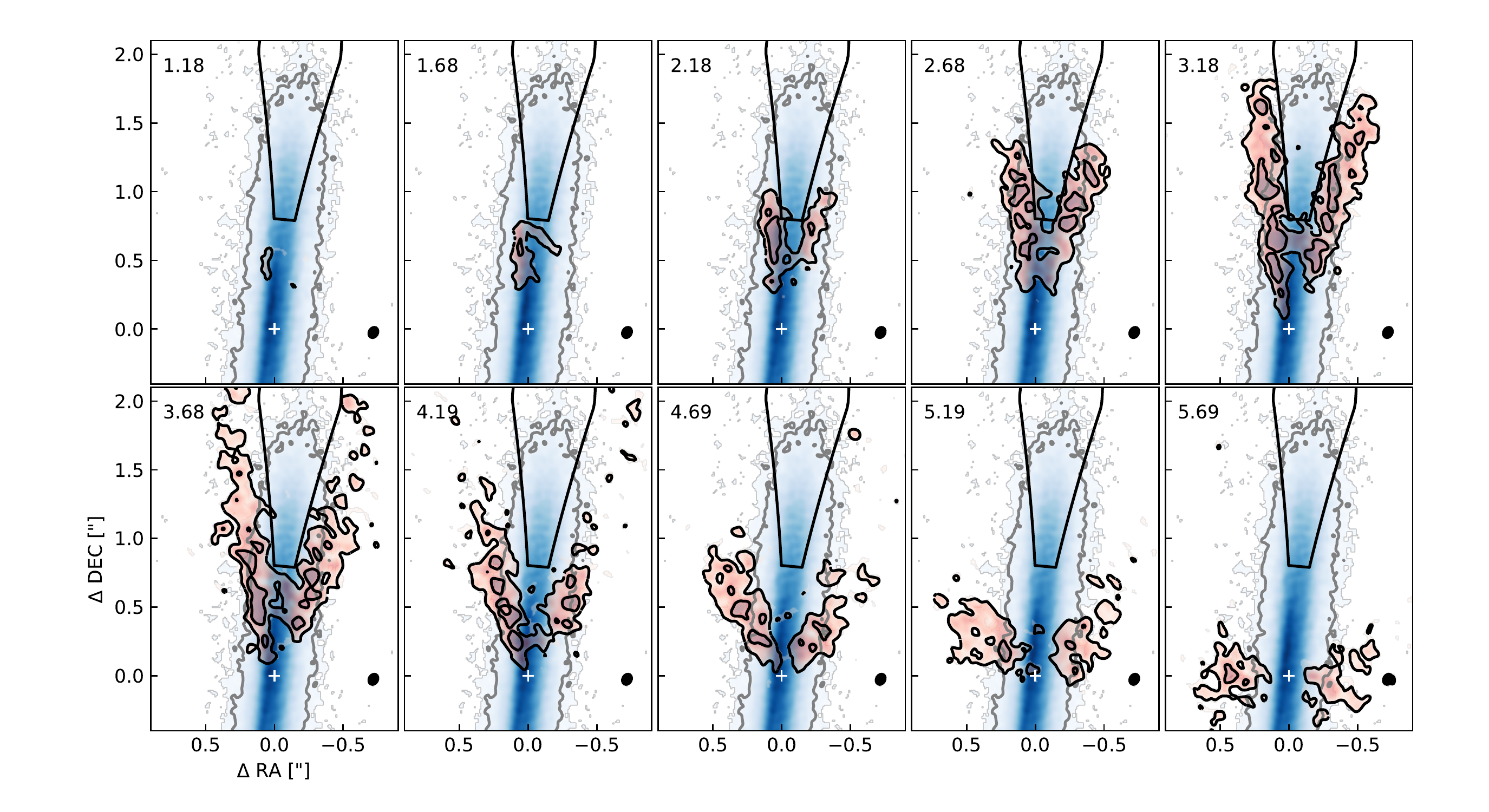}
    \caption{
        Selected channel images of C$^{18}$O with robust=0.5. Each image here is first averaged by 3 channels to increase the signal-to-noise ratio. The red color map is the C$^{18}$O emission and the black contours mark the 3 and 5$\sigma$ levels. The underlying blue color map is the continuum, while the grey contour is the 5$\sigma$ level of the continuum. The black ellipse in the lower right of each image represents the beam for C$^{18}$O. The text in the upper left corner of each plot is the velocity in km s$^{-1}$. The inferred snow surface is outlined in black. The white cross marks the origin of the image. 
    }
    \label{fig:channels_c18o_freezeout}
\end{figure*}

%% there is an outer ``cap" to the freeze-out 
%% interesting blue atmosphere in SE
While the optically thinner C$^{18}$O probes the snow line and the immediate snow surface at small impact parameters, $^{13}$CO reveals a complete freeze-out zone explained by the following. Fig.~\ref{fig:channel_13co_freezeout} shows selected channels for the northern half of the $^{13}$CO with robust=2 for better detection of the larger scale structure. The first three high velocity channels also show the distinct V-shaped snow surface that extends further than C$^{18}$O. At the low velocity channels, the east side and the west side of the emission appear to connect at large impact parameters ($\sim 3\arcsec$) forming an apparent ``cap" to the V-shaped emission that closes the opening. The difference between C$^{18}$O and $^{13}$CO is likely due to optical depth and sensitivity. Since $^{13}$CO can be detected more easily at lower column densities, we can identify the full spatial extent (or complete) freeze-out zone of CO, while C$^{18}$O can only reveal the partial freeze-out zone. From the cap, it appears that CO is no longer frozen-out on grains at the larger radii even though one may expect that the temperature is lower than the inner radii. We extend this discussion in Section~\ref{sec:outer_cap}.

\begin{figure*}
    \centering
    \includegraphics[width=\textwidth, trim={0.5cm 0.5cm 1.5cm 0cm},clip]{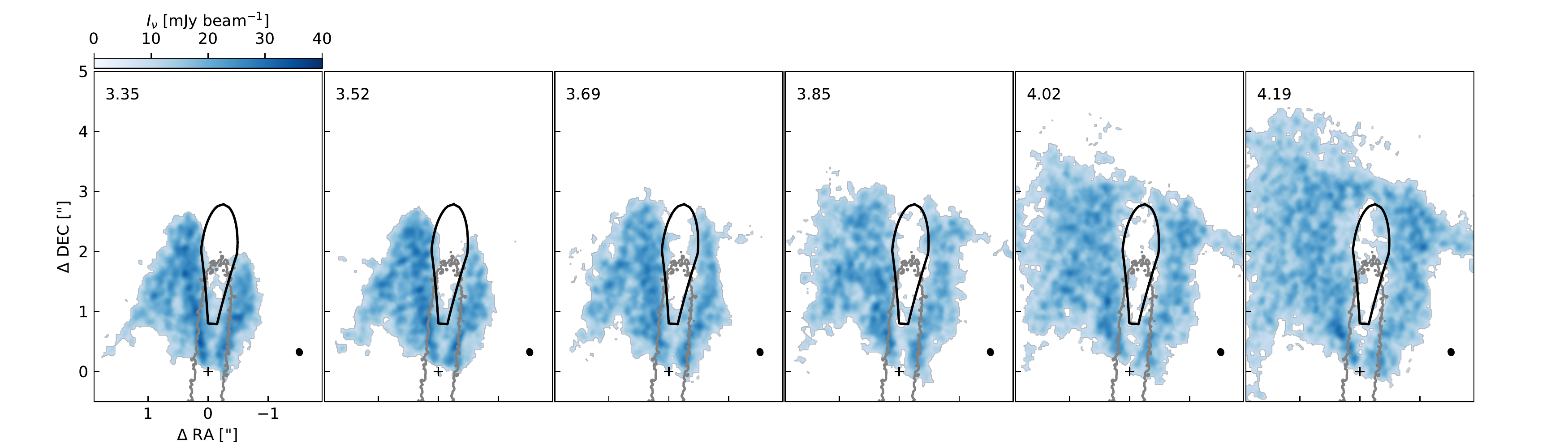}
    \caption{
        Channel images of $^{13}$CO (robust=2.0) with emission above $3\sigma$. The grey contour is the continuum at $5\sigma$. The snow surface based on Eq.~(\ref{eq:snow_surface}) is marked in black. In each image, the black  ellipse in the lower right represents the beam and the text in the upper left is the velocity in km s$^{-1}$. The black cross marks the center of the disk. 
    }
    \label{fig:channel_13co_freezeout}
\end{figure*}

Another intriguing feature is the non-Keplerian, blueshifted feature in the southeast atmosphere of the disk. Fig.~\ref{fig:channel_13co_blue_SE_atmosphere} shows selected $^{13}$CO channel map with robust=2 focusing on the southern half. At redshifted channels (bottom row of Fig.~\ref{fig:channel_13co_blue_SE_atmosphere}), the disk near the midplane shows the typical Keplerian rotation for an edge-on disk (like that of C$^{18}$O in Fig.~\ref{fig:channels_c18o_freezeout}, but for the blueshifted half). In addition, we see the freeze-out zone and the outer cap that is similar to the northern half (Fig.~\ref{fig:channel_13co_freezeout}). However, at blueshifted channels (top row of Fig.~\ref{fig:channel_13co_blue_SE_atmosphere}), the southern half is not void of emission as expected from Keplerian rotation, but hosts large extensions to the east. The extension is along the disk minor axis at the most blueshifted channel of Fig.~\ref{fig:channel_13co_blue_SE_atmosphere} and extends to the south when closer to the system velocity. Intriguingly, the edge of the extension closest to the disk appears to match the eastern edge of the redshifted Keplerian part in shape (the emission in the bottom row of Fig.~\ref{fig:channel_13co_blue_SE_atmosphere}). Thus, it appears that the blue extension is aware of the atmosphere of the southeast part of the disk and forms an interface. With a simple modification to Eq.~(\ref{eq:snow_surface}), we outline the interface by:
\begin{equation} \label{eq:13co_interface}
    y(x) = 
    \begin{cases}
        H_{0} + (H_{a} - H_{0}) \bigg( \dfrac{x}{ R_{a} } \bigg) & x < R_{a} \\
        H_{a} \sqrt{1 - \bigg(\dfrac{ x - R_{a} }{ R_{c} - R_{a} } \bigg)^{2} } & R_{a} \leq x < R_{c}
    \end{cases}
\end{equation}
where $H_{0}$ is the distance from the center along the minor axis, $H_{a}$ is the height at some transition radius $R_{a}$, and $R_{c}$ is the outer radius of the cap. We find that $H_{0}=0.5\arcsec$ (80~au), $H_{a}=0.9\arcsec$ (140~au), $R_{a}=2.2\arcsec$ (350~au), and $R_{c}=3.9\arcsec$ (620~au) by eye. We show the outline symmetric across $x=0$ and $y=0$ for convenience of discussion.

From the difference in the kinematics within and outside the southwest interface, we can distinguish the disk component and the envelope component. The southeast blue extension outside the interface is connected to even larger distances at $\sim 10\arcsec$ shown in the $^{13}$CO with robust=2 (bottom row of Fig.~\ref{fig:moments_13CO}). Given that the extension is closest to the disk at high blueshifted channels (e.g., $\sim 4.7$~km s$^{-1}$) and becomes more extended at lower blueshifted channels (e.g., $\sim 5.4$~km s$^{-1}$), the nature of the extension can be explained by infalling material from behind the plane-of-sky that lands onto the southern half of the disk. In such a scenario, it would make sense for an interface to form, since the infalling material moving from behind the plane-of-sky has to collide with the disk material moving into the plane-of-sky. The existence of infalling material is not too surprising given evidence in other younger Class 0/I disks \citep{Pineda2020NatAs...4.1158P, Alves2020ApJ...904L...6A, ValdiviaMena2022A&A...667A..12V, Garufi2022A&A...658A.104G} or even late-stage infall onto Class~II disks (e.g., \citealt{Tang2012A&A...547A..84T, Ginski2021ApJ...908L..25G, Huang2020ApJ...898..140H, Huang2021ApJS..257...19H, Gupta2023A&A...670L...8G}; see also \citealt{Kuffmeier2020A&A...633A...3K}).

Though the outline of the interface was determined from the kinematic difference in the southeast part of the disk, the southwest part of the outline appears to also separate the disk from a broad extension to the west. Different from the southeast blue extension, the west extension is redshifted as the southern part of the Keplerian disk should be, making it indistinguishable kinematically and thus the outline from Eq.~(\ref{eq:13co_interface}) may not mark a clear interface. However, given that there is no symmetric counterpart across $y=0$ on the east side at the same channels, it is morphologically distinct from the material within the outline. It is unclear what the nature of the west red extension is.

%%% is it infall? 
% - does the morphology make sense?
% - does the value make sense? 
%%% calculate sqrt(2) difference from Keplerian

\begin{figure*}
    \centering
    \includegraphics[width=\textwidth]{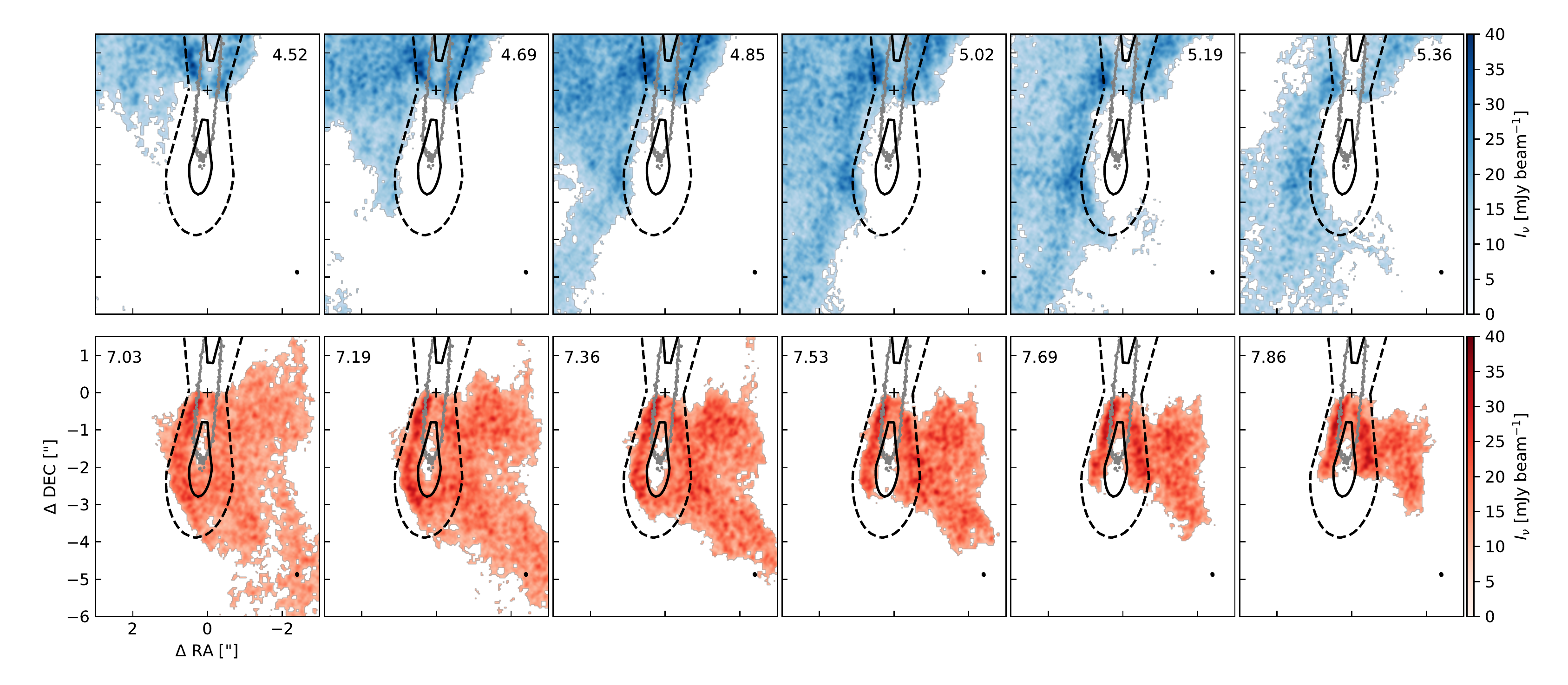}
    \caption{
        Selected channel maps of $^{13}$CO (robust=2.0) showing emission above $3\sigma$. The images in the top row are blueshifted and those in the bottom row are redshifted with respect to the system velocity. The grey contour is the $5\sigma$ level of the continuum. The black cross marks the center of the image. The black solid contour is the snow surface (from Eq.~(\ref{eq:snow_surface})), and the black dashed contour represents the ``interface" (from Eq.~(\ref{eq:13co_interface}); see Section~\ref{sec:snow_line_and_surface} for more detail). The black ellipse to the lower right corner is the beam. 
    }
    \label{fig:channel_13co_blue_SE_atmosphere}
\end{figure*}

\section{Analysis} \label{sec:analysis}

In this section, we analyze the data presented in Section~\ref{sec:results} in more detail. Section~\ref{sec:continuum_RT_modeling} analyzes the continuum image through forward ray-tracing of the dust and provides constraints on the dust scale height and inclination. Section~\ref{sec:rotation_curve} analyzes the position-velocity (PV) diagram of C$^{18}$O and measures the stellar mass. 

\subsection{Continuum Forward Ray-Tracing} \label{sec:continuum_RT_modeling}

Although a 2D Gaussian captures the overall features, such as the position angle and the overall shape, certain deviations stand out. Fig.~\ref{fig:Gaussian_and_Qconstant}a shows the original continuum and the fitted 2D Gaussian, while Fig.~\ref{fig:Gaussian_and_Qconstant}b shows the residuals, which are defined as the observed image subtracted by the 2D Gaussian. The largest deviation is the significant positive residual extending parallel to the disk major axis that is slightly offset from the center to the east. This corresponds to the asymmetry along the minor axis where the east side is brighter.

\begin{figure}
    \centering
    \includegraphics[width=0.75\columnwidth, trim={0cm 0.2cm 0.5cm 0cm},clip]{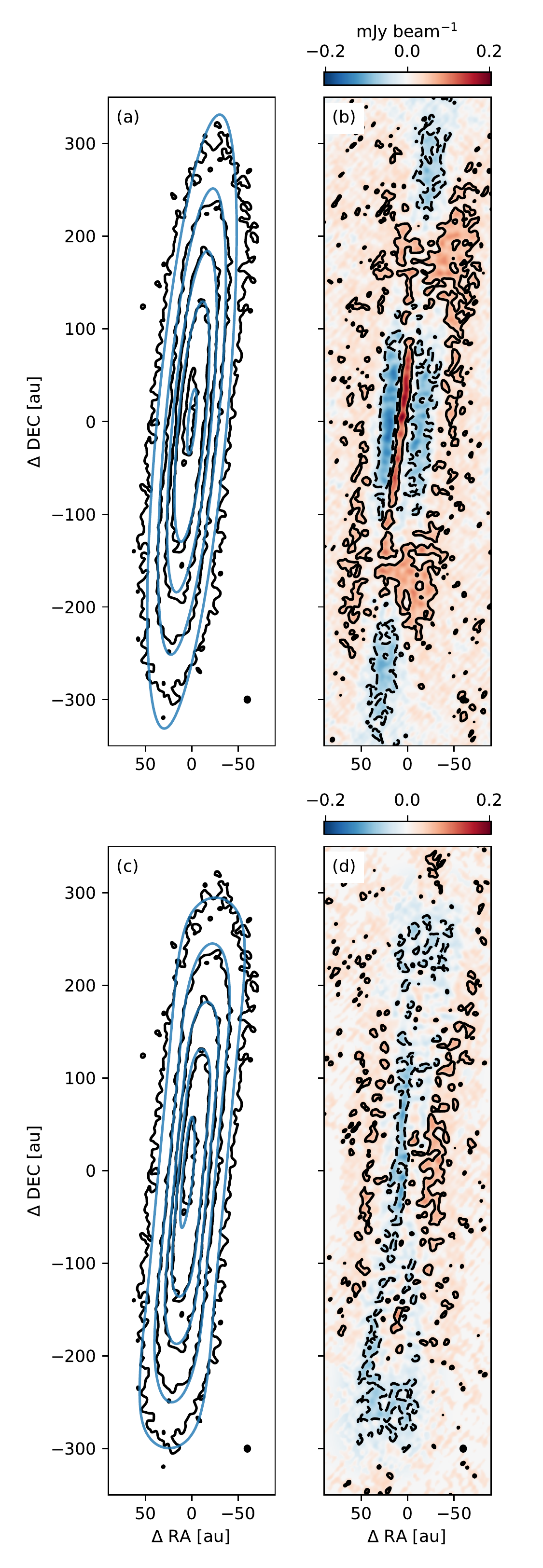}
    \caption{
        Comparisons between the observed continuum and models. The top row shows the 2D Gaussian model, while the bottom row shows the model with radiative transfer (see Section~\ref{sec:continuum_RT_modeling}). The plots in the left column show the model (in blue contours) plotted against the observed continuum (in black contours). The color maps in the right column show the residuals (the observed subtracted by the model) in mJy~beam$^{-1}$. The solid and dashed contours trace the $3\sigma$ and $-3\sigma$ levels respectively. 
    }
    \label{fig:Gaussian_and_Qconstant}
\end{figure}

In this section, we demonstrate that the asymmetry along the minor axis is due to the inclination effect of an optically thick disk. We use a parameterized disk model and use RADMC-3D\footnote{RADMC-3D is available at \url{https://www.ita.uni-heidelberg.de/~dullemond/software/radmc-3d/}.} to conduct the ray-tracing \citep{Dullemond2012ascl.soft02015D}. We refrain from conducting the heating/cooling calculations from RADMC-3D given the large computational cost and complexities regarding the dust opacity spectrum \citep[e.g.][]{Birnstiel2018ApJ...869L..45B}. The calculation is beyond the scope of this paper and we leave it to a future paper. The parameterized disk model is a similar version of the disk model from \cite{Lin2021} which is suited for a disk viewed near edge-on.\footnote{The usefulness of the model comes from the characteristic optical depth defined later in Eq.~(\ref{eq:tau0_definition}). The optical depth for an edge-on (or nearly edge-on) source depends on the \textit{radial} extent of the disk, while the optical depth for a face-on source relies on the \textit{vertical} extent.} The model was applied to a Class 0 edge-on source, HH~212~mms, and successfully reproduced the asymmetry along the minor axis across ALMA Bands 3, 6, and 7. In the following, we briefly describe the key parts of the model and include modifications.

We parameterize the disk using the Toomre $Q$ parameter \citep{Toomre1964ApJ...139.1217T}
\begin{equation} \label{eq:Toomre_Q_parameter}
    Q \equiv \frac{ c_{s} \Omega_{k} }{ \pi G \Sigma }
\end{equation}
where $c_{s}$ is the isothermal velocity, $\Omega_{k}\equiv \sqrt{GM_{*} /R^{3}}$ is the Keplerian frequency, $R$ is the cylindrical radius, and $\Sigma$ is the gas surface density. For a gravitationally stable disk, $Q$ must be greater than a value of order unity \citep[e.g.][]{Kratter2016ARA&A..54..271K}. The pressure scale height of the gas is 
\begin{equation} \label{eq:gas_scale_height}
    H_{g} \equiv \dfrac{c_{s}}{\Omega_{k}} \text{ .}
\end{equation}
From basic arguments of vertical hydrostatic equilibrium, the gas density in the midplane is $\rho_{g, \text{mid}} = \Sigma / \sqrt{2\pi} / H_{g}$ and thus, when combined with Eq.~(\ref{eq:Toomre_Q_parameter}), we have
\begin{equation} \label{eq:midplane_density}
    \rho_{g, \text{mid}}(R) = \dfrac{ M_{*} }{ \pi \sqrt{2 \pi} R_{0}^{3} Q } \bigg( \dfrac{R}{R _{0}} \bigg)^{-3} \text{,}
\end{equation}
where $R_0$ is a characteristic radius, which we take to be the outer radius of the disk. For illustrative purposes, we assume that $Q$ is a constant in the disk, and introduce a characteristic density $\rho_0\equiv M_*/(\pi \sqrt{2\pi} R_0^3 Q)$, which is the density at the disk outer edge. 

Since the dust disk appears vertically thin, we approximate the temperature with just a vertically isothermal prescription: 
\begin{equation} \label{eq:dust_temperature_profile}
    T(R) = T_{0} \bigg( \dfrac{R}{R_{0}} \bigg)^{-q}
\end{equation}
where $T_{0}$ is the temperature at the outer edge of the disk and $q$ specifies the temperature gradient. Note that the whole gas disk should have a vertical temperature gradient (warmer temperature in the atmosphere), which is needed for the existence of the clear snow surface (Fig.~\ref{fig:image_collection}, \ref{fig:channel_13co_freezeout}). However, since the bulk of the dust disk appears to lie below the snow surface and there is no continuum dark lane (such as that found for HH~212~mms), the effect of a vertical temperature gradient is likely marginal, and thus, we only use a vertically isothermal profile for the dust disk. 

As a further simplification, we fix $q=0.5$ which is expected from passively irradiated disks in radiative equilibrium \citep[e.g.][]{Chiang1997ApJ...490..368C, DAlessio1998ApJ...500..411D}. This assumption may not be entirely applicable to embedded protostars, which can have additional accretion heating or warming from the envelope \cite[e.g.][]{Butner1994ApJ...420..326B, AgurtoGangas2019A&A...623A.147A}. Accretion heating should lead to a steeper temperature gradient, usually $q=0.75$ \citep{Armitage2015arXiv150906382A}, and dominate the inner regions of the disk \citep{Takakuwa_edisk}. Envelope warming prevails in the outer regions and should make the temperature gradient shallow (e.g., $q\leq0.4$ from \citealt{Whitney2003ApJ...591.1049W}). The Class~I designation of IRAS~04302 motivates a smaller $q$, however, \cite{vantHoff2020} found $q=0.75$ based on the location of snow lines of H$_{2}$CO and C$^{17}$O though the resolution is not ideal. We also refrain from fitting $q$ directly, since a single wavelength image of an edge-on disk probes a limited range in radius due to the high optical depth \citep{Lin2021}. Longer wavelength observations are necessary to probe the temperature of the inner regions and using multiwavelength observations that probe different radii will better constrain $q$. Thus, given the uncertainties, we fix $q=0.5$ as a compromise for this paper and leave the exploration of $q$ for a future study.\footnote{We have tried $q=0.75$ and found that the qualitative results remain the same, while the best-fit parameters only differ slightly. We found smaller residuals with $q=0.5$ though we caution that our hand search may not be comprehensive and a more sophisticated parameter search could be done in the future. }

\cite{Lin2021} assumed that the dust and the gas are well-coupled and thus the dust also follows the gas in hydrostatic equilibrium (qualitatively, this means the dust scale height is equal to the gas scale height if the disk is vertically isothermal). However, to directly explore the dust scale height independent of what the gas scale height should be, we parameterize the dust scale height by
\begin{equation} \label{eq:dust_scale_height}
    H_{d}(R) = H_{100} \bigg( \dfrac{R}{ 100~\text{au} } \bigg)^{1.25}
\end{equation}
where the power-law index is the same as that from the gas scale height, i.e., $1.5-q/2$. Eq.~(\ref{eq:dust_scale_height}) allows us to easily explore the effects of height with one parameter $H_{100}$.

By assuming that the midplane density of the dust is related to the midplane density of the gas (Eq.~(\ref{eq:midplane_density})) through a dust-to-gas mass ratio $\eta$, the complete dust density as a function of radius and height is 
\begin{equation} \label{eq:dust_density_distribution}
    \rho_{d}(R,z) = \rho_{d,0} \bigg( \dfrac{R}{R_{0}} \bigg)^{-3} \exp{ \bigg[ - \dfrac{1}{2} \bigg( \dfrac{z}{H_{d}} \bigg)^{2} \bigg] }
\end{equation}
where $z$ is the vertical height and $\rho_{d,0} \equiv \rho_{0} / \eta$ is the midplane dust density at the outer edge of the disk. 

Instead of prescribing the dust opacity $\kappa_{\nu}$ (in units of cm$^{2}$ per gram of dust) explicitly, we use the characteristic optical depth $\tau_{0,\nu}$ defined as
\begin{equation} \label{eq:tau0_definition}
    \tau_{0,\nu} \equiv \rho_{d,0} R_{0} \kappa_{\nu}. 
\end{equation}
The definition makes sense because the characteristic length scale along the line-of-sight for an edge-on disk is $R_{0}$. This parameter reflects the fact that opacity and density are degenerate and it is the optical depth (proportional to the product of opacity and density) that controls how an image appears (see \citealt{Lin2021} for detailed derivation and for exploration of how $\tau_{0,\nu}$ controls the image of an edge-on disk). In other words, $\tau_{0,\nu}$ is a free parameter that we can fit from the image. 

As an initial exploration for this paper, we conduct the parameter search by hand. To limit the parameter space, we fix the position angle to $174.77^{\circ}$ obtained from the 2D Gaussian fit. The free parameters include $\tau_{0,\nu}$, $T_{0}$, $i$, $R_{0}$ and $H_{100}$ in addition to the location of the star ($\delta_{\text{RA}}$, $\delta_{\text{DEC}}$). The parameters for the best-fit model are listed in Table~\ref{tab:Qconstant_model_parameters}.

\begin{deluxetable}{lcc}
    \tablenum{2}
    \tablecaption{Adopted parameters for the dust model \label{tab:Qconstant_model_parameters} }
    \tablewidth{0pt}
    \tablehead{
        \colhead{Parameter} & \colhead{Variable} & \colhead{Value} \\
        }
    \decimalcolnumbers
    \startdata
        Inclination     & $i$               & 87$^{\circ}$ \\
        Disk Edge       & $R_{0}$           & 310~au \\
        Temperature at $R_{0}$ & $T_{0}$    & 7.5~K \\
        Dust Scale Height at 100~au    & $H_{100}$           & 6~au \\
        Characteristic Optical Depth & $\tau_{0,\nu}$ & 0.35 \\
        RA offset of star        & $\delta_{\text{RA}}$           & -0.03$\arcsec$ \\
        DEC offset of star        & $\delta_{\text{DEC}}$           & -0.04$\arcsec$ \\
    \enddata
    \tablecomments{
        These are the parameters from the search by hand that appear to match best and provide the model in Fig.~\ref{fig:Gaussian_and_Qconstant}. The RA and DEC offset are relative to the center based on the 2D Gaussian fitting in Section~\ref{sec:results}. 
        }
\end{deluxetable}

Fig.~\ref{fig:Gaussian_and_Qconstant}c shows that the model compares quite well with the observations. The dust model can easily reproduce the shift along the minor axis towards the far side of the disk (towards the east for the case of IRAS~04302) since the disk is optically thick and highly inclined \citep{Villenave2020, Takakuwa_edisk}. The residuals are shown in Fig.~\ref{fig:Gaussian_and_Qconstant}d and are evidently much lower than that from the simple 2D Gaussian fit (Fig.~\ref{fig:Gaussian_and_Qconstant}b).

We find that the $H_{100}$ is 6~au. The dust scale height from past modeling efforts based on lower resolution mm-images varies in the literature and ranges from $\sim 2$~au to 15~au at a radius of 100~au \citep{Wolf2003ApJ...588..373W, Wolf2008ApJ...674L.101W, Grafe2013A&A...553A..69G, Sheehan2017ApJ...851...45S} though it depends on the exact prescription of each model. By resolving the asymmetry along the disk minor axis, the new high-resolution image presented here offers a strong constraint on the dust scale height. In addition, the value is consistent with an independent study that modeled another high-resolution image at Band~4 \citep{Villenave2023ApJ...946...70V}. On the other hand, the derived radius of $R_{0}=310$~au is consistent with past modeling efforts based on lower resolution mm-images in which case the major axis of the disk was well resolved \citep{Wolf2003ApJ...588..373W, Grafe2013A&A...553A..69G}. 

The inferred inclination of $i=87^{\circ}$ provides the necessary deviation from being perfectly edge-on ($i=90^{\circ}$) which would not produce an asymmetry along the minor axis since both halves across the midplane would be perfectly symmetric \citep[e.g.][]{Wolf2003ApJ...588..373W}. The value is also consistent with the lower limit of $\sim 84^{\circ}$ assuming the disk is completely flat (see Section~\ref{sec:continuum}). It is not surprising that the actual inclination is larger than the inclination inferred just from the ratio between the minor and major axes, or $\arccos(\text{minor} / \text{major})$. Using the ratio assumes that only the radial extent contributes to the projected length along the minor axis which is indeed the case for a geometrically thin disk. However, for a highly inclined geometrically thick disk, the vertical thickness contributes to the projected width along the minor axis which decreases $\arccos(\text{minor} / \text{major})$. 
% explain the boxiness
% asymmetry becomes less obvious 

The inferred $T_{0}$ of 7.5~K appears to be lower than necessary when compared to what is expected from the estimated snow line of CO. The low temperature profile is necessary because the peak brightness temperature is only $\sim 14$~K and yet the disk has to be optically thick to produce the minor axis shift of the continuum. Based on the fitted $T_{0}$, the snow line for CO, assuming a freeze-out temperature of 20~K, should be at $\sim 44$~au ($0.275 \arcsec$). However, this appears inconsistent with the observed location of the snow line which is $\sim 130$~au ($\sim 0.8 \arcsec$) from C$^{18}$O (also similar to what was derived in \citealt{vantHoff2018A&A...615A..83V} from lower angular resolution observations of C$^{17}$O). One possibility is that the dust temperature profile is correct and the observed C$^{18}$O emission beyond the inferred snow line location (of $44$~au) is contaminated by emission from the warmer surface layers due to the finite beam.

%%% Li: Again, at tau=1 surface, the dust brightness temperature should be close to (but somewhat smaller than) the dust temperature. Please check whether this is consistent. 
%%% From the best-fit dust model, the tau=1 surface for impact parameter=0 has radius of 120au
%%% the temperature is 7.5 * (120 / 310)**(-0.5) = 12.1 K 
%%% indeed similar to the peak of the major axis cut

Another possibility to alleviate the above discrepancy is through scattering. Scattering makes objects appear dimmer, which means the actual temperature should be higher than what is inferred when assuming no scattering \citep[e.g.][]{Birnstiel2018ApJ...869L..45B}. Interestingly, radiation transfer calculations for this source including scattering of 100~$\mu$m grains infer a temperature of 20~K at 100~au \citep{Grafe2013A&A...553A..69G} which is higher than the 13~K at 100~au based on the model prescribed here. Given that scattering only scales the image intensities and does not alter the relative shape of the image much \citep{Lin2021}, the inferred low temperature could be evidence of scattering, but we leave the incorporation of scattering to a future study.

Intriguingly, the outermost contour of the model appears systematically less extended than the observations along the minor axis (Fig.~\ref{fig:Gaussian_and_Qconstant}c). This is also seen as two lanes of generally positive residuals to the east and west of the disk in Fig.~\ref{fig:Gaussian_and_Qconstant}d which suggests a more extended upper layer. However, increasing $H_{c}$ to broaden the image along the minor axis leads to even broader widths at the endpoints of the major axis of the disk. Thus, it appears that the dust scale height should not be too flared at the outer radius compared to the inner radius. This is in fact what we would expect from dust settling of a given grain size, where the outer region should be more settled than the inner region because the Stokes number of the grains increases as the density decreases towards larger radii \citep{Dullemond2004A&A...421.1075D}. We leave also this possibility for future exploration.

We found that the characteristic optical depth is $\tau_{0, \nu}=0.35$ which can be related to the opacity.\footnote{As demonstrated in \cite{Lin2021}, if there is scattering, the intensity of the image decreases, but the relative shape of the image does not change much, and $\tau_{0,\nu}$ is mainly determined by the extinction opacity. } From Eq.~(\ref{eq:tau0_definition}) and the definition of $\rho_{0}$ from Eq.~(\ref{eq:midplane_density}), we can explicitly solve for $\kappa_{\nu}$ through 
\begin{equation} \label{eq:solve_kappa}
    \kappa_{\nu} = \frac{\pi \sqrt{2\pi} Q \eta \tau_{0} R_{0}^{2} }{ M_{*} }. 
\end{equation}
Using the best-fit $R_{0}$ and $\tau_{0, \nu}$ from this section, the $M_{*}$ derived based on the rotation curve of C$^{18}$O (see Section~\ref{sec:rotation_curve}), the opacity is $\kappa_{\nu} = 0.019 Q \eta$ in units of cm$^{2}$ g$^{-1}$ of gas. If the disk is gravitationally stable, $Q$ should be greater than of order unity. Otherwise, the disk should fragment \citep{Kratter2016ARA&A..54..271K}. Thus, taking $Q=1$ gives a lower limit to $\kappa_{\nu}$. We note that the lower limit to the opacity is per mass of gas since it is the gas that contributes most of the mass, and that limits the amount of material. However, theoretical dust models calculate dust opacity with respect to the mass of the dust \citep[e.g.][]{Ossenkopf1994A&A...291..943O} and thus we have to assume a $\eta$ to directly compare the dust opacity calculations to the observationally constrained opacity presented here. By assuming the standard $\eta=100$, we get $\kappa_{\nu} = 1.9 Q$ cm$^{2}$ g$^{-1}$ of dust. The uncertainty of $\kappa_{\nu}$ is $0.5$~cm$^{2}$ g$^{-1}$ of dust based on error propagation from the uncertainty of $M_{*}$ derived in Section~\ref{sec:rotation_curve}. We add the caveat that the opacity can vary spatially which is not captured through the model and thus, the value measured here is an effective opacity of the region observable at Band~6.

The conventional \cite{Beckwith1990AJ.....99..924B} opacity at $\lambda$=1.3~mm is $\kappa_{\nu} = 2.3$ cm$^{2}$ g$^{-1}$ of dust (also constrained observationally and assumed $\eta=100$) and the opacity based on HH~212~mms is $\kappa_{\nu} = 1.33$ cm$^{2}$ g$^{-1}$ of dust \citep{Lin2021}. By taking $Q=1$, it appears that the lower limit from IRAS~04302 lies right in between the two previous studies as shown in Fig.~\ref{fig:opacity}. For completeness, we have included opacity constraints at other wavelengths for HH~212~mms \citep{Lin2021} and also another commonly adopted dust opacity model from \cite{Ossenkopf1994A&A...291..943O} with calculations adopted for low and high densities. The lower limit from IRAS~04302 disfavors the opacity model from \cite{Ossenkopf1994A&A...291..943O} and is more consistent with the \cite{Beckwith1990AJ.....99..924B} prescription. 

The proximity of the lower limit from IRAS~04302 to the opacity from HH~212~mms is intriguing, given that HH~212~mms is vastly different compared to IRAS~04302 in class, size of the disk, and stellar mass. While HH~212~mms is likely to be marginally gravitationally unstable given the small stellar mass, bright continuum, and early stage \citep{Tobin2020ApJ...890..130T}, IRAS~04302, as a Class~I source, is less certain. Even if grains have a universal opacity, the lower limit from IRAS~04302 need not be similar, since from Eq.~(\ref{eq:solve_kappa}), taking $Q=1$ is only a lower limit after all and $Q$ can take on any value greater than 1 if the disk is not marginally gravitationally unstable.

If not purely coincidental, a possible physical explanation is that the grains could be similar between these two systems \textit{and} both systems are marginally gravitationally unstable which fixes $Q$ to a value of order unity \citep[e.g.][]{Lodato2007NCimR..30..293L, Kratter2016ARA&A..54..271K, Xu2021MNRAS.502.4911X}. It may not be too surprising if IRAS~04302 can also be marginally gravitationally unstable given the large disk, an available reservoir of envelope material, and cold midplane temperature. There is growing evidence of other Class~0/I sources that are marginally graviationally unstable \citep[e.g.][]{Kwon2011ApJ...741....3K, Tobin2020ApJ...890..130T, Xu2022ApJ...934..156X}. Furthermore, from an evolutionary standpoint, this is in line with evidence of Class~II sources with $Q$ that largely falls within $1$ to $10$ \citep[e.g.][]{Kwon2015ApJ...808..102K, Cleeves2016ApJ...832..110C, Booth2019ApJ...882L..31B, Veronesi2021ApJ...914L..27V,  PanequeCarreno2021ApJ...914...88P, Ueda2022ApJ...930...56U, Schwarz2021ApJS..257...20S, Sierra2021ApJS..257...14S, Yoshida2022ApJ...937L..14Y, Lodato2022arXiv221103712L}.

\begin{figure}
    \centering
    \includegraphics[width=\columnwidth]{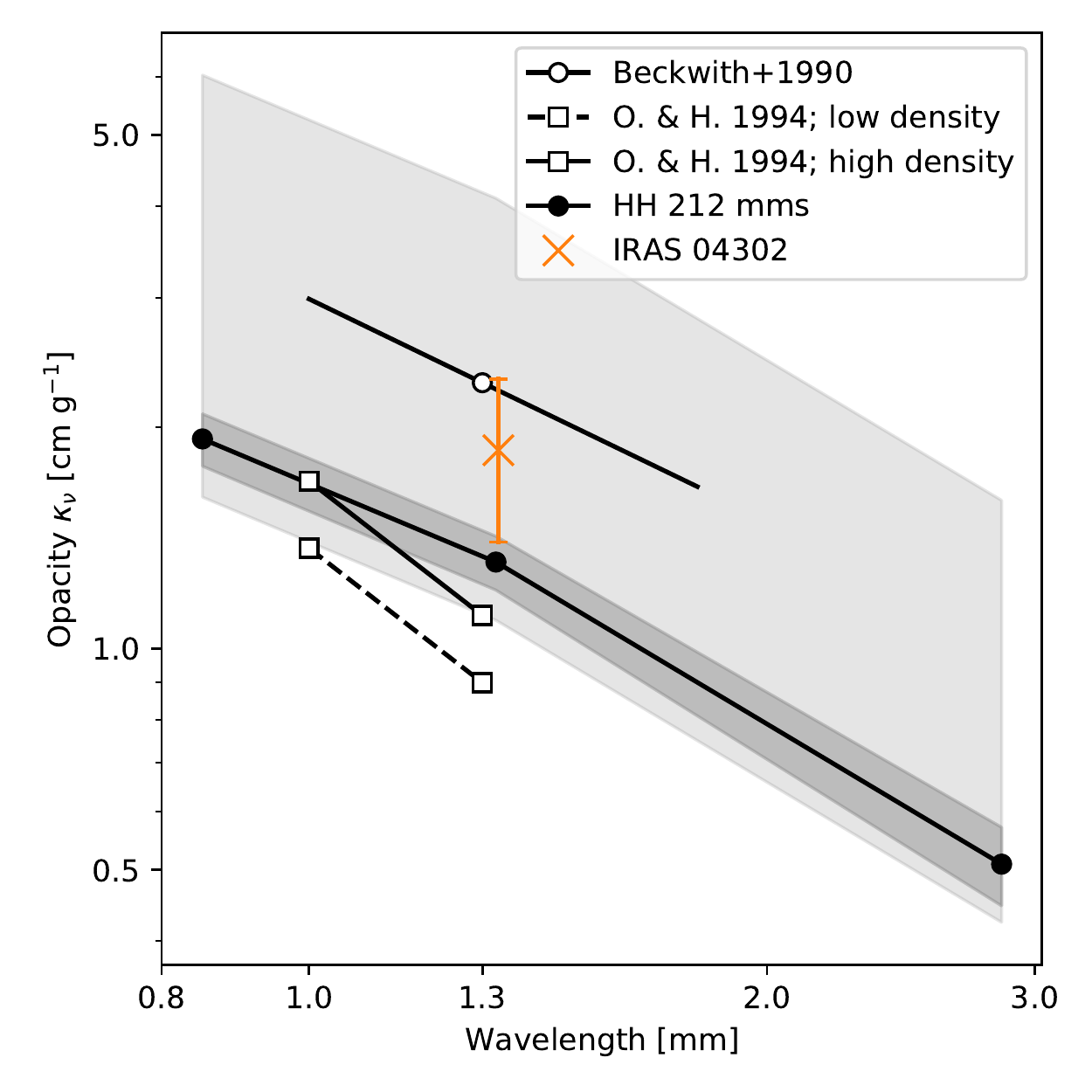}
    \caption{
        The lower limit to the dust opacity (absorption cross section per gram of dust assuming a dust-to-gas ratio of 0.01) inferred from the IRAS~04302 disk (marked as an orange cross) in comparison to other millimeter dust opacities from the literature. The error bar is the uncertainty associated with the uncertainty from the stellar mass. The filled circles with solid lines are the lower limit to the dust opacity for the HH~212~mms disk from \cite{Lin2021}. The corresponding lighter, shaded region is the uncertainty associated with the stellar mass and also the Toomre~$Q$ parameter (ranging $1$ to $2.5$) and the darker, shaded region is the uncertainty from the noise. The open circle is the \cite{Beckwith1990AJ.....99..924B} opacity at 1.3~mm and its line segment represents the opacity index of $1$. The open squares are opacities from \cite{Ossenkopf1994A&A...291..943O} at 1 and 1.3~mm. 
    }
    \label{fig:opacity}
\end{figure}

% add reference to Villenave's current work with the VLA

\subsection{Deriving the Stellar Mass from Disk Rotation} \label{sec:rotation_curve}

Given the clear evidence of rotation (e.g., right panel of Fig.~\ref{fig:moments_C18O}), we further analyze the rotation curve for IRAS~04302 using the position-velocity (PV) diagram along the major axis of the disk. Fig.~\ref{fig:slamfit} shows the PV diagram for C$^{18}$O with robust=0.5. We choose C$^{18}$O since it is optically thinner and only traces the disk as opposed to $^{12}$CO and $^{13}$CO which are more susceptible to surrounding envelope material. Also, C$^{18}$O is better detected than the other optically thin lines. To create the PV diagram, we use the position angle derived from the Gaussian fit of the continuum (see Section~\ref{sec:continuum}) and we use a slit with a width of $\sim 2$ beams to increase the signal-to-noise necessary for the analysis below. 

We use the Spectral Line Analysis/Modeling (SLAM) code \citep{yusuke_aso_2023_7783868} \footnote{The SLAM code is available at \url{https://github.com/jinshisai/SLAM}} to extract the rotation curve from the PV diagram \citep{Aso2015, Sai2020}. Inferring the rotation properties relies on first assigning pairs of radius and velocity points based on the PV diagram and later fitting the points to a rotation curve. Details of SLAM are described in \cite{Ohashi_edisk_overview}, but we describe the essential steps and parameters adopted here. 

For the first step, we aim to trace the ``outer" edge of the PV diagram (the top of the second quadrant and the bottom of the fourth quadrant). We use the 5$\sigma$ level for each spectrum along the position as the representative pairs of radius and velocity, which corresponds to the ``edge" method in SLAM. The reason is as follows. For an edge-on disk, the line-of-sight at a particular impact parameter $x$ (i.e., the position along the major axis) crosses several radii. The spectra is simply the collective emission of material along that line-of-sight each with varying levels of projected velocities (without considering any complications from finite line width). Along the line-of-sight, there is a minimal radius that contributes the maximal velocity and that is the location in plane-of-sky which equals the impact parameter $x$ (see e.g., \citealt{Dutrey2017A&A...607A.130D} for an illustration). Thus, in the spectra, we would expect that the maximum velocity where we have detection is precisely the representative velocity for the radius that equals (absolute value of) the impact parameter. Complications arise when considering finite line width, temperature effects, inclination, and detection levels, which can be addressed through modeling. However, as a working expectation, we use the 5$\sigma$ level for each spectrum along the position to fit for the Keplerian rotation \citep{Seifried2016MNRAS.459.1892S}. To assess how sensitive the parameters are to the chosen level, we also use the 3$\sigma$ level.

Another common way to extract representative pairs of radius and velocity from the PV diagram is to take the mean of the intensity profile, which corresponds to the ``ridge" method in SLAM \citep{Aso2015, Yen2017ApJ...834..178Y, Sai2020}. This extraction usually underestimates the true stellar mass \citep[e.g.][]{Maret2020A&A...635A..15M}, but we use it to complement the ``edge" method described above to assess the systematic uncertainty. Conventionally, there are two ways to take the mean of the intensity profile, either along the velocity axis (i.e., the spectra at a certain impact parameter) or along the position axis (i.e., the profile of the image of a certain channel). From experimentation, we find that using both was necessary to trace the PV diagram.

The noise level used here is assessed in regions of the PV diagram where no emission is expected. We have $\sigma=0.976$~mJy~beam$^{-1}$ which is $\sim \sqrt{2}$ smaller than the channel map noise level (Table~\ref{tab:images_summary}) as expected from our adopted slit width. In addition, we avoid fitting the spectra within $0.5\arcsec$ (80~au), since the PV diagram is even qualitatively different from the typical Keplerian rotation curve. The lack of high-velocity emission could be due to the lack of material at inner radii \citep{Dutrey2017A&A...607A.130D} or dust extinction, but we leave the verification for future exploration and focus on fitting the Keplerian parts in practice. 

The next step involves fitting a rotation curve to the data points which we use
\begin{equation} \label{eq:rotation_curve}
    v = - \text{sign}(x) v_{b} \bigg( \frac{|x|}{r_{b}} \bigg)^{-p} + \vsys
\end{equation}
where $r_{b}$ is a characteristic radius, $v_{b}$ is the rotational velocity at $r_{b}$, and $p$ is the power-law index of the rotation profile. The sign of $v$ has been adjusted to account for the definition of $x$ in this paper (positive along the northern part of the major axis which is consistent throughout the paper). If the disk is in Keplerian rotation, we should retrieve $p=0.5$ and one can infer the stellar mass $M_{*}$ from $v_{b} = \sqrt{GM_{*} / r_{b}} \sin i$ where $i$ is the inclination.

The left panel of Fig.~\ref{fig:slamfit} shows the assigned pairs of position (radius) and velocity from the edge method using the 5$\sigma$ level of the spectra (which corresponds to the intensity in the vertical direction of Fig.~\ref{fig:slamfit}) at the largest (absolute) velocity with respect to $\vsys$. The inferred rotation curve (plotted as a white curve) follows the outer edge of the PV diagram reasonably well. Considering only statistical uncertainty, we find that $p=0.52 \pm 0.02$ which verifies that the disk is consistent with Keplerian rotation. The systemic velocity $\vsys$ is $5.72 \pm 0.02$ km s$^{-1}$ and the stellar mass is $M_{*} = 1.65 \pm 0.02$ M$_{\odot}$ assuming an inclination of $i=87^{\circ}$ derived from the dust continuum (see Section~\ref{sec:continuum_RT_modeling}).

To assess the systematic uncertainty, we compare the edge method with 5$\sigma$ to the edge method with 3$\sigma$ and the ridge method. The edge method using data points at the 3$\sigma$ level (not shown in Fig.~\ref{fig:slamfit} for brevity) yielded a stellar mass of $M_{*} \sim 2 M_{\odot}$ (see Table~\ref{tab:slam_best_fit_parameters} for a comparison of the results). Measuring a larger stellar mass is not too surprising, since adopting a lower threshold adds to the range of the measured velocity which could be due to the line width and it can artificially increase the measured stellar mass. In the other extreme, the extracted points for the ridge method (right panel of Fig.~\ref{fig:slamfit}) yielded a smaller stellar mass of $M_{*} \sim 1.2 M_{*}$ as expected \citep[e.g.][]{Maret2020A&A...635A..15M}. Given the large spread in measurements depending on the assumed method, we adopt $M_{*}=1.6\pm0.4$~$M_{\odot}$ and $\vsys = 5.7 \pm 0.1$~km s$^{-1}$.

The measured stellar mass is similar to the adopted mass of $1.7$~M$_{\odot}$ from \cite{Grafe2013A&A...553A..69G} although the detailed derivation was not described in the literature. Otherwise, as far as we know, there are no other published measurements of the stellar mass through dynamical measurements.

\begin{deluxetable}{lccc}
    \tablenum{3}
    \tablecaption{Results from SLAM \label{tab:slam_best_fit_parameters} }
    \tablewidth{0pt}
    \tablehead{
        \colhead{Method} & \colhead{$p$} & \colhead{$M_{*}$} & \colhead{ $\vsys$ } \\
        \colhead{} & \colhead{} & \colhead{[$M_\odot$]} & \colhead{[km s$^{-1}$]}
        }
    \decimalcolnumbers
    \startdata
        edge ($5\sigma$)  & $0.52 \pm 0.02$  & $1.65 \pm 0.02$  & $5.72 \pm 0.02$ \\
        edge ($3\sigma$)  & $0.49 \pm 0.01$  & $2.07 \pm 0.03$  & $5.60 \pm 0.01$ \\
        ridge   & $0.540 \pm 0.008$ & $1.225 \pm 0.005$  & $5.821 \pm 0.007$ \\
    \enddata
    \tablecomments{
        The $M_{*}$ was derived assuming an inclination $i=87^{\circ}$. 
        }
\end{deluxetable}

%%% there are issues like: dust extinction, contamination from the warmer surface
%%% we're not sure what we're probing 
%%% we need complete forward modeling
%We caution that the dynamical measurement measured here could be susceptible to contamination from the warmer surface and we are unclear how dust extinction can affect the results. The slit width of $\sim 2$ beams is necessary to increase the signal-to-noise. However, such a width inevitably includes contribution from the warmer surface, since there are fitted  inferred snow line of $\sim 0.8\arcsec$ (128~au) 

\begin{figure*}
    \centering
    \includegraphics[width=\textwidth]{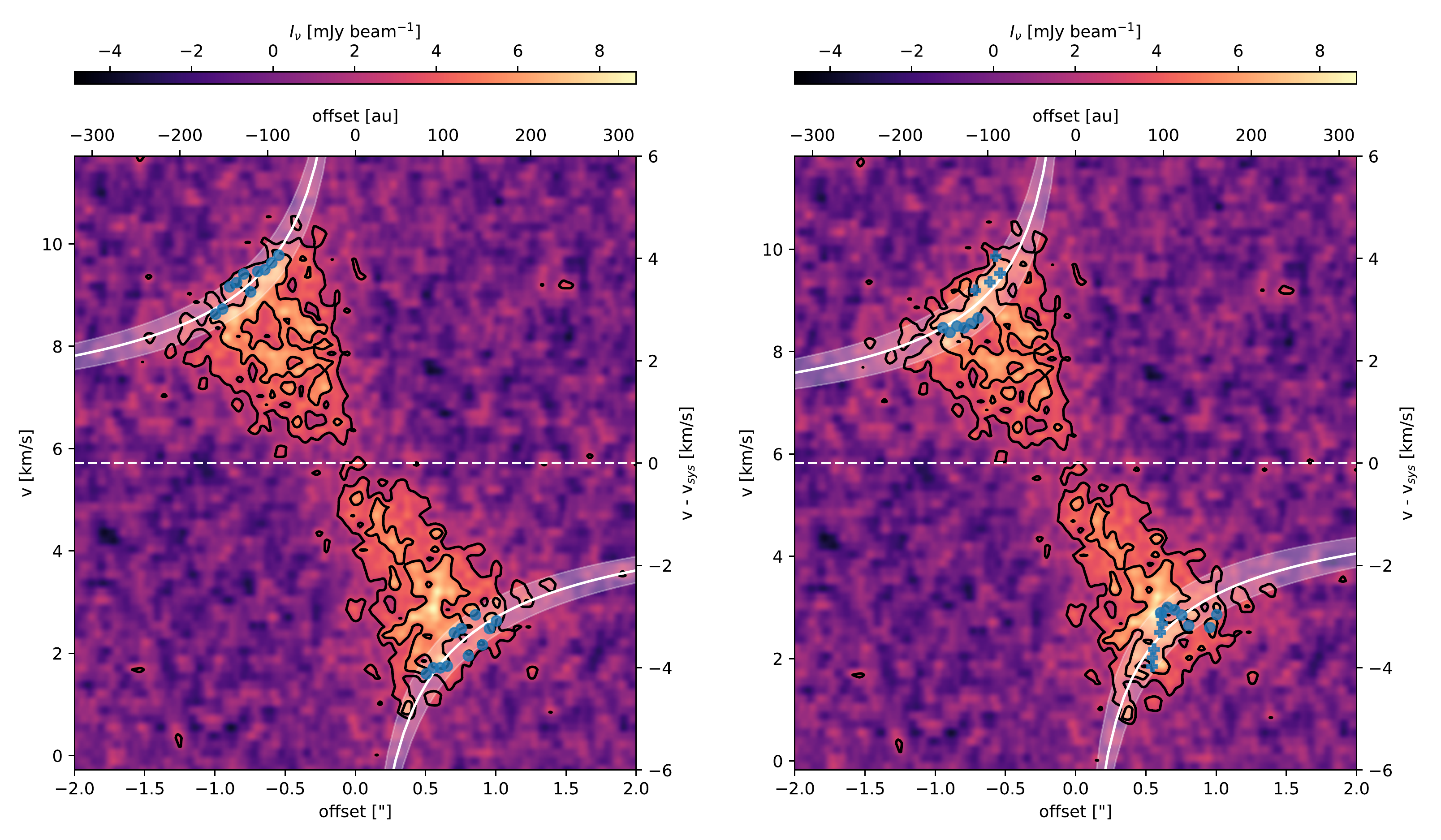}
    \caption{
        For both plots, the color scale is the pv diagram of C$^{18}$O and the black contour shows the 5$\sigma$ level. The left plot includes the fitted points from the ``edge" method along the velocity axis, while the right plot includes the fitted points from the ``ridge" method along the velocity axis (circles) and position axis (crosses). The best-fit rotation curve, using $M_{*}=1.6 M_{\odot}$, for each is in solid white line and the best-fit $\vsys$ is shown as a horizontal dashed white line. The shaded region is the range of the rotation curve using $M_{*}=1.6 \pm 0.4 M_{\odot}$. The bottom and top axes are the location along the major axis in units of arcseconds and au respectively. The left and right axes show the velocity and that relative to $\vsys$. 
    }
    \label{fig:slamfit}
    
\end{figure*}

\section{Discussion} \label{sec:discussion}

\subsection{Evidence for Dust Extinction} \label{sec:dust_extinction}
%\red{need to mention the opposite case which is line extinction: for edge-on disks, the line emission should be lower than the continuum. Thus we would expect negative emission due to continuum (over)subtraction}
%% the lack of emission is also evident for HH212
%% 

As described in Section~\ref{sec:lines}, the lack of line emission along the major axis of the disk at large impact parameters is due to freeze-out which gives the iconic V-shaped emission \citep{vantHoff2020}. However, at small impact parameters where we do not expect freeze-out, a depression is shared across all lines and is especially obvious from the moment 0 images of H$_{2}$CO and SO (Fig.~\ref{fig:image_collection}, \ref{fig:moments_H2CO}, \ref{fig:moments_SO}).

The depression along the innermost parts of the major axis can be explained by dust extinction. The lack of emission for SO and H$_{2}$CO is likely because their snow lines lie well within the $\tau=1$ surface of the dust, i.e., the location where the optical depth to the observer is $1$. In such a scenario, the dust essentially buries the emission behind the $\tau=1$ surface. 
For example, from the dust model shown in Section~\ref{sec:continuum_RT_modeling}, the impact parameters where the total optical depth equals 1 and 5 are $\sim 215$~au and $\sim 105$~au, respectively. H$_{2}$CO has a larger freeze-out temperature at $\sim 70$~K \citep{Noble2012A&A...543A...5N} and we would expect the snowline to be at impact parameters much less than $105$~au well into the optically thick regions of the dust disk. On the other hand, the observed CO snowline from imaging of $\sim 130$~au is roughly in the translucent region between the two limits which makes it possible to see the emission from the midplane.

Dust extinction can also explain the asymmetry of the high velocity emission from $^{12}$CO. We show the channel maps focused on the innermost region of the disk in Fig.~\ref{fig:channel_highvel_12co}. At high blueshifted channels (top row of Fig.~\ref{fig:channel_highvel_12co}), $^{12}$CO initially emerges as a single point in the east side ($v=-2.85$~km s$^{-1}$) until a second point appears on the west side ($v=-2.21$~km s$^{^-1}$). The same is true for the high redshifted channels (bottom row of Fig.~\ref{fig:channel_highvel_12co}) where the single point in the east side at the highest velocity channel ($v=14.30$~km s$^{-1}$) and the second point in the west appears at a lower redshifted channel ($v=13.03$~km s$^{-1}$). A similar behavior is evident for C$^{18}$O in Fig.~\ref{fig:channels_c18o_freezeout}. 
With the nearly edge-on view, the emission from the front side of the disk can be seen unobstructed, while the emission from the back side of the disk must travel through the dust disk to reach the observer. 

%Assuming the points at high velocity channels come from the same distance away from the midplane, we can infer the level of dust extinction from the asymmetry from $I_{2} = I_{1} e^{-\Delta \tau}$ where $I_{1}$ is the emission without extinction (the east point for our case), $I_{2}$ is the emission after extinction (the west point), and $\Delta \tau$ is the optical depth due to the dust. Using the channel at $v=-1.58$ km s$^{-1}$, which is at a velocity far away from $\vsys$ and since the west point is clearly detected and separate from the east point, we have $I_{1} \sim 0.029$ and $I_{2} \sim 0.012$ which gives $\Delta \tau \sim 0.88$. \red{John: Where are these measurements being made (needs coordinates) and over how much area is being averaged? Text is not clear on this}

The existence of dust extinction is consistent with the requirement that the dust must be optically thick to produce the continuum asymmetry along the minor axis (see Section~\ref{sec:continuum_RT_modeling}; unless it is due to an intrinsic asymmetry in the density distribution of the disk). It is also not too surprising as other sources also have examples of dust extinction, for example, the rings of HD~163296 \citep{Isella2018ApJ...869L..49I} and DG Tau B \citep{Garufi2020A&A...636A..65G}.

We note that an inclined disk with a two-dimensional temperature structure of a warmer surface and colder midplane could also contribute to the asymmetry in the brightness between the near- and far-sides \citep[e.g.][]{Flores2021AJ....161..239F}. The brightness asymmetry further away from the major axis of the disk is more likely from the inclination effect. A complete radiation transfer including both dust and gas would be required to identify the separate contributions. 

% other examples
% HD 163296 rings, DG Tau B, TMC1A? 

%Using $\vsys=5.72$ and $M_{s}=1.67 M_{\odot}$ (derived in Sec.~\ref{sec:rotation_curve}), the velocity of the channel before the second point emerges is $\sim 8.5$ km s$^{-1}$ away from $\vsys$. Thus, assuming Keplerian rotation, the emission should come from a radius of $\sim 20$ au ($\sim 0.13\arcsec$). 

\begin{figure*}
    \centering
    \includegraphics[width=\textwidth]{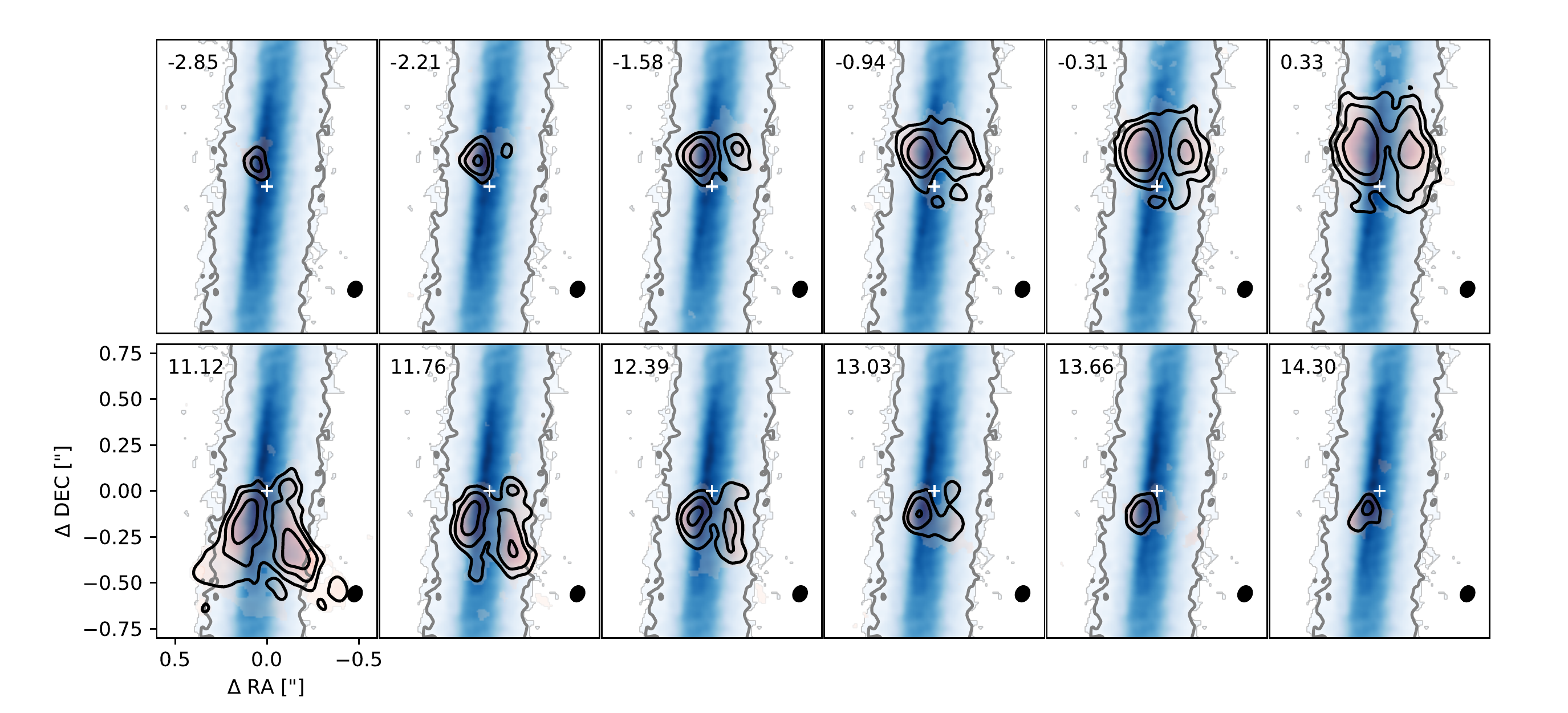}
    \caption{
        Selected channel maps of $^{12}$CO in comparison to the continuum. The top row correspond to the high blueshifted channels and the bottom row are the high redshifted channels. The red color map is the $^{12}$CO J=2-1 emission above 3$\sigma$ (see Table~\ref{tab:images_summary}), while the black contours mark the 5, 10, and 20$\sigma$ levels. The underlying blue color map is the continuum, while the grey contour is the 5$\sigma$ level of the continuum. The black ellipse in the lower right of each image represents the beam for the $^{12}$CO. The text in the upper left corner of each plot is the velocity in km s$^{-1}$. 
    }
    \label{fig:channel_highvel_12co}
\end{figure*}

\subsection{The outer cap of $^{13}$CO} \label{sec:outer_cap}
% reverse snowline? 
% does this have implications for composition of planets? Jovian planets can only form in the freeze-out zone? 
% implications for JWST? 
% ... could Law+2022 outer surface be probing the cap? 

% we can easily tell the edge of the Keplerian disk simply because the material becomes radial and stream-like 

% ice absorption detected in Aikawa+2012

An intriguing part of the $^{13}$CO morphology is the detection at $\sim 4\arcsec$ from the center along the disk major axis even though the molecule is not seen from $\sim 1\arcsec$ to $3\arcsec$ along the disk major axis. We interpret the lack of emission due to freeze-out and the transition to the freeze-out zone extends into the atmosphere resembling the shape of ``V." However, we detect $^{13}$CO at $\sim 4\arcsec$ in the form of a cap that closes off the freeze-out zone which means the molecule is no longer frozen-out and somehow ``re-emerges" at larger radii where the temperature is usually expected to be lower. The cap also exists for $^{12}$CO, but is only visible when robust=2.0 (Fig.~\ref{fig:moments_12CO} bottom row), which suggests that much of the emission along with the cap is mostly resolved out for $^{12}$CO. The C$^{18}$O cap is not evident with robust=0.5 and also not clear with robust=2.0 which could be due to the lack of sufficient signal-to-noise for the optically thinner isotopologue.

A natural question is whether the cap belongs to the disk or envelope. Fig.~\ref{fig:pv_13co} shows the PV diagram along the major axis of $^{13}$CO with robust=2.0 with a slit width of $\sim 1$ beam. For comparison, we show the Keplerian curve with $M_{*}=1.6 M_{\odot}$ and $\vsys$ measured from C$^{18}$O in Section~\ref{sec:rotation_curve}. At impact parameters within $\sim \pm 1.5\arcsec$ (240~au), the Keplerian curve follows the outer extent of the PV diagram quite well which is similar to the case of C$^{18}$O (Fig.~\ref{fig:slamfit}). Within $\pm [1.5\arcsec, 2.8\arcsec]$, there is a lack of material that follows Keplerian rotation which corresponds to the freeze-out zone. The snow line from C$^{18}$O is $0.8\arcsec$ (Section~\ref{sec:snow_line_and_surface}) which is less than the inner boundary of $1.5\arcsec$ here due to significant contamination from the warmer surface from larger beam averaging. The cap begins at $\sim \pm 2.8\arcsec$ and appears to follow the Keplerian rotation curve up to $\sim \pm 3.9\arcsec$ (620~au). For the southern part (negative $x$-axis of Fig.~\ref{fig:pv_13co}), the emission stops and we can directly identify the same edge of the emission in the channel maps in Fig.~\ref{fig:channel_13co_blue_SE_atmosphere}. For the norther part (positive $x$-axis), there appears to be a sharp break in the PV diagram in which case much of the emission appears more redshifted than Keplerian. Given the consistency with Keplerian rotation, we reason that the cap belongs to the Keplerian rotating disk and the sharp deviations from Keplerian rotation at $\sim 3.9\arcsec$ correspond to the edge of the gas disk outside of which is a part of the envelope.

\begin{figure}
    \centering
    \includegraphics[width=\columnwidth]{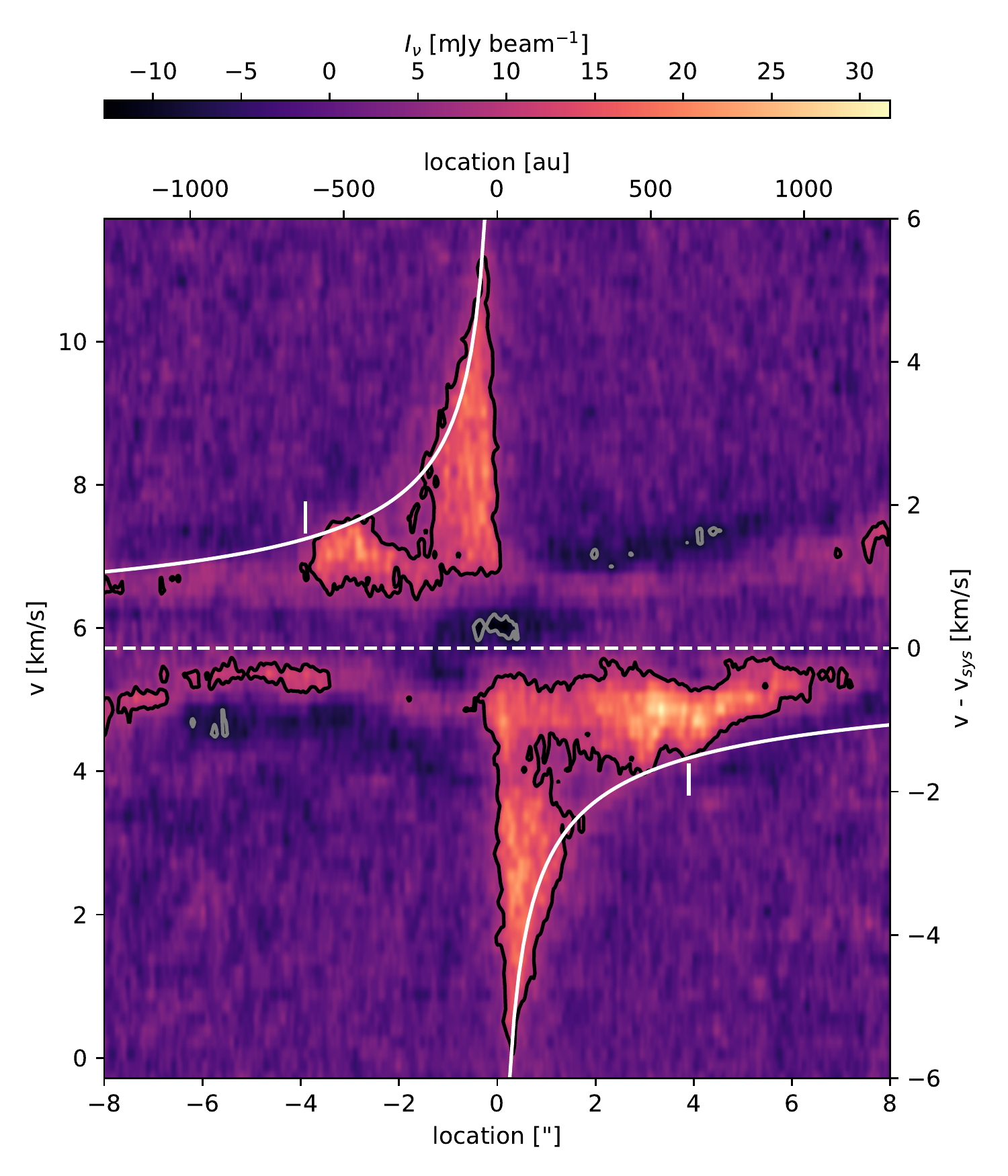}
    \caption{
        The colormap is the PV diagram of $^{13}$CO (robust=2.0). The black and grey contours show the 3$\sigma$ and -3$\sigma$ levels respectively. The white line is the Keplerian rotation curve and the white horizontal dashed line is $\vsys$. The two white vertical line segments marks $\pm 3.9\arcsec$ to denote the sharp deviation from Keplerian rotation (see Section~\ref{sec:outer_cap} for more detail). The bottom and top axes are the location along the major axis in units of arcseconds and au respectively. The left and right axes show the velocity and that relative to $\vsys$.
    }
    \label{fig:pv_13co}
\end{figure}

There are two other edge-on sources with an apparent cap, namely 2MASS~J16281370-2431391 (so called ``Flying Saucer"; \citealt{Dutrey2017A&A...607A.130D}) and SSTC2D~J163131.2-242627 (Oph~163131 for short; \citealt{Villenave2022ApJ...930...11V}) which are both Class II sources. \cite{Dutrey2017A&A...607A.130D} showed that $^{12}$CO also emerges beyond the freeze-out zone. They found that the transition coincided with a change in grain properties and proposed that the behavior was expected if an efficient rise of UV penetration was re-heating the disk.

The disk around Oph~163131 also showed a $^{12}$CO cap \citep{Flores2021AJ....161..239F, Villenave2022ApJ...930...11V}. Since the transition from the inner ``V"-shaped region to the outer cap region roughly coincided where the millimeter-continuum disk ends and the disk's scattered light stops, \cite{Flores2021AJ....161..239F} also interpreted the behavior as external UV radiation providing an additional source of heating to the outer part of the disk where dust particles are not present.

In contrast to the two sources with $^{12}$CO caps, the cap of IRAS~04302 is seen in the optically thinner $^{13}$CO which suggests that there is much more material. Another difference is the location of the cap region. Unlike the two sources whose caps begin at the extent of their millimeter-continuum, the continuum disk of IRAS~04302 clearly ends (at $\sim 1.9\arcsec$; 310~au) well before $^{13}$CO emerges (at $\sim 3\arcsec$; 500 ~au). In other words, the freeze-out zone extends beyond the millimeter-continuum disk. A potential explanation is that the freeze-out temperature could decrease in the low density region in the outer disk \citep{Harsono2015A&A...582A..41H}. Another possibility is that external UV irradiation still impacts the disk and heats up the outer region, but the UV photons are efficiently blocked by the smaller grains which are invisible at mm-wavelengths. This scenario may suggest significant radial drift of the larger $\sim$ mm-grains which is not too surprising given the much smaller radius of the dust disk (310~au; see Section~\ref{sec:continuum_RT_modeling}) compared to the radius of the gas disk (620~au). In fact, when modeling the scattered light and lower resolution mm-continuum simultaneously, \cite{Grafe2013A&A...553A..69G} required one population of large grains with a smaller radius and another population of small grains with a larger radius. Another related possibility is also warming of the outer disk, but from the envelope \citep[e.g.][]{Whitney2003ApJ...591.1049W}. 

%In contrast to the Class 0 sources, ... L1527, HH212?... the radial extent of the dust is usually comparable to the edge of the Keplerian part of the disk... there may be evolutionary difference where significant radial drift occurs before the Class I stage

%Other than edge-on sources, emission surfaces of mid-inclination Class~II sources appear to increase with radius, but decrease at the outermost radii \citep{Law2021ApJS..257....4L, Law2022ApJ...932..114L, PanequeCarreno2022A&A...666A.168P}. Whether the cap seen edge-on could 
%% IM Lup: Cleeves 2016, Pinte 2018

\subsection{Dust Settling in the Class I stage}

One of the most striking features of the IRAS~04302 disk is the shift of the intensity peak along the minor axis of the continuum image which is a tell-tale sign of dust with finite vertical extent, i.e., non-settled dust. This feature exists for several other sources among the eDisk sample, including CB~68 \citep{Kido_edisk}, L1527 IRS \citep{vantHoff_edisk}, IRS~7B \citep{Ohashi_edisk_overview, Takakuwa_edisk}, GSS~30~IRS3 \citep{Santamaria_Miranda_edisk}, IRAS~32 \citep{Encalada_edisk}, BHR~71 \citep{Gavino_edisk}, IRAS~04169+2702 \citep{Han_edisk}, and IRAS~16253-2429 \citep{Aso_edisk}.

In one extreme, dust settled into an infinitely thin sheet should appear symmetric across the minor axis and for disks with rings, the rings and gaps should not show azimuthal variation \citep[e.g.][]{Pinte2016ApJ...816...25P, Doi2021ApJ...912..164D}. Several observations of Class II sources show that the dust is predominantly well settled \citep[e.g.][]{Andrews2018ApJ...869L..41A, Long2018ApJ...869...17L, Villenave2020, Doi2021ApJ...912..164D, Liu2022SCPMA..6529511L, Villenave2023ApJ...946...70V}. One of the clearest case is SSTC2D~J163131.2-242627 (or Oph~163131 for short) whose gaps are resolved even though the disk is near edge-on (\citealt{Villenave2022ApJ...930...11V}, $i \sim 84^{\circ}$). The inferred dust scale height is $\leq 0.5$~au at 100~au, which is an order of magnitude smaller than that of IRAS~04302. Furthermore, the significant difference in the vertical extent of the gas and dust also shows that dust is decoupled from the gas over most of the disk volume away from the midplane \citep[e.g.][]{Villenave2020, Law2021ApJS..257....4L, Law2022ApJ...932..114L}.

In the other extreme, the Class 0 source, HH~212~mms, hosts a clear dark lane sandwiched between two bright lanes in the dust continuum at $\sim 1$~mm, which is evidence that the dust is elevated high enough to trace the warm surface layers. The dust scale height is $\sim 12$~au at a radius of $\sim 36$~au and the dust was shown to follow the gas in hydrostatic equilibrium \citep{Lee2017_darklane, Lin2021}.

From Section~\ref{sec:continuum_RT_modeling}, we found that the dust scale height is $6$~au at a radius of 100~au. For comparison, the gas pressure scale height from Eq.~(\ref{eq:gas_scale_height}) is $H_{g}=5.8\pm0.7$~au at a radius of 100~au after adopting $M_{*} = 1.6\pm0.4 M_{\odot}$ from Section~\ref{sec:rotation_curve} and the dust isothermal temperature profile of Eq.~(\ref{eq:dust_temperature_profile}) with the best-fit $T_{0}$ (only the stellar mass uncertainty is included here). The effectively equivalent scale heights given the uncertainties suggest that the dust has not separated from the gas vertically.

We caution that there is ambiguity in the midplane temperature, since the temperature derived from dust modeling appears different from the temperature inferred from the freeze-out location of CO. Using the snow line of 130~au (see Section~\ref{sec:snow_line_and_surface}) and assuming a freeze-out temperature of $20$~K with $q=0.5$, the temperature at $100$~au is $\sim 23$~K and results in $H_{g}=7.6\pm1.0$~au. 
%Thus, taken at face value, the measured dust scale height $H_{d}$ appears only slightly smaller than the gas scale height and dust settling has barely progressed. 
Considering the ambiguity of the temperature profile from the two scenarios, we have $\sim 0.8 \leq H_{d}/ H_{g} \leq \sim 1$. We also note that $H_{d}$ inferred from Section~\ref{sec:continuum_RT_modeling} assumes a mixed, single population of grains. However, if grain growth has occurred, we may expect grains of different sizes to settle at various characteristic heights \citep[e.g.][]{Dubrulle1995Icar..114..237D}. Nevertheless, the inferred $H_{d}$ represents the characteristic height of the bulk of the material that is responsible for the $\lambda=1.3$~mm emission which is already different from the Class~II sources where the dust responsible for the emission at the same wavelength has already settled to a much smaller scale height as mentioned above. At face-value, the non-significant level of dust settling may pose difficulties for the streaming instability to produce planetesimals \citep{Gole2020ApJ...904..132G} and thus delay planet formation.
% Nevertheless, adopting the temperature profile inferred from the snow line of CO still gives a comparable $H_{g}$. 

Although the dust traced by 1.3~mm continuum is non-settled, the dust in general appears very distinct from the distribution of gas molecules (demonstrated in Fig.~\ref{fig:image_collection}) and also very distinct from the scattered light images of IRAS~04302. Fig.~\ref{fig:hst_with_ALMA} shows a comparison between the 1.3~mm continuum, $^{13}$CO, and scattered light images from the Hubble Space Telescope (HST) at $1.6$~$\mu$m \citep{Padgett1999AJ....117.1490P}. We describe the correction for proper motion in Appendix~\ref{sec:vla_bandka}. Strikingly, each image traces a spatially distinct location. The 1.3~mm continuum appears only near the midplane, while the scattered light only exists in the bipolar cavities. $^{13}$CO fills the atmospheric regions of the disk and reaches beyond the radial extent of the 1.3~mm continuum and scattered light. Nevertheless, the gas pressure scale height $H_{g}$ of $6\sim 7$~au may not be too surprising, since the line emission can typically be at several pressure scale heights above the midplane \citep[e.g.][]{Dullemond2004A&A...421.1075D, Wolff2021AJ....161..238W, Flores2021AJ....161..239F, Villenave2022ApJ...930...11V, Law2022ApJ...932..114L, PanequeCarreno2022arXiv221001130P}, and small dust grains are present in the bipolar nebula to scatter optical/IR light. Detailed modeling using the high-angular resolution observations of the molecular lines with the dust could give a more robust view on the level of dust settling.

Another distinction between the gas and mm-continuum is the radial extent. The edge of the dust disk has a radius of $\sim 310$~au (see Section~\ref{sec:continuum_RT_modeling}), while the edge of the gas disk has a radius of $\sim 620$~au (see Section~\ref{sec:outer_cap}). In light of the disparity in the dust and gas radii, but similarity in the dust and gas scale heights (see Section~\ref{sec:continuum_RT_modeling}), IRAS~04302 demonstrates that radial settling occurs sooner than vertical settling. Nevertheless, proper forward ray-tracing including both the dust and gas will make the disparity more definitive.

IRAS~04302 is formally a Class~I source based on the SED \citep{Ohashi_edisk_overview}. Although an object with the Class~I designation could actually be a Class~II source if viewed edge-on, there is additional evidence that IRAS~04302 is indeed younger than formal Class II sources. First, the scattered light image of IRAS~04302 is noticeably irregular which indicates potential interactions with its envelope. In contrast, scattered light images of Class II sources tend to be well-ordered \citep{Villenave2020}. Second, IRAS~04302 has clear evidence of extended $^{13}$CO and $^{12}$CO emission beyond the Keplerian disk surface with kinematics inconsistent with Keplerian rotation which is likely part of the envelope (Fig.~\ref{fig:moments_12CO} bottom row and \ref{fig:moments_13CO} bottom row). Thus, it is quite clear that this Class~I source is a case where there is relatively little dust settling amid infall and outflow. Given that most Class~II sources appear settled, we speculate that substantial dust settling should happen between the Class~I stage and Class~II stage.

It is curious whether IRAS~04302 has any radial substructure given its Class~I stage. Rings and gaps are ubiquitous around Class~II protostars \citep[e.g][]{Andrews2018ApJ...869L..41A, Long2018ApJ...869...17L} and these structures could be signposts of planets \citep[e.g.][]{Zhang2018ApJ...869L..47Z}. Gaps from a highly inclined disk like Oph~163131 were resolved \citep{Villenave2022ApJ...930...11V}, but the order of magnitude larger dust scale height of IRAS~04302 can easily obscure the gaps if there exists any.

%%% Jake Simon, Daniel Carrero group: 
%%% streaming instability cannot work with large turbulence, because disk puffed up -> dust-to-gas ratio too low

%From a theoretical standpoint, planetesimal formation through streaming instability may be difficult, since the midplane dust-to-gas ratio is not high enough without dust settling \citep{Gole2020ApJ...904..132G}. We speculate that planetesimal formation 

%we suspect that the large scale infall material can play a role in dust settling? 

% do rings form first or do dust settle first? 
% past multiwavelength modeling have found grain growth though! Grains have grown without settling!??!?! or resolution effect?

\begin{figure}
    \centering
    \includegraphics[width=\columnwidth]{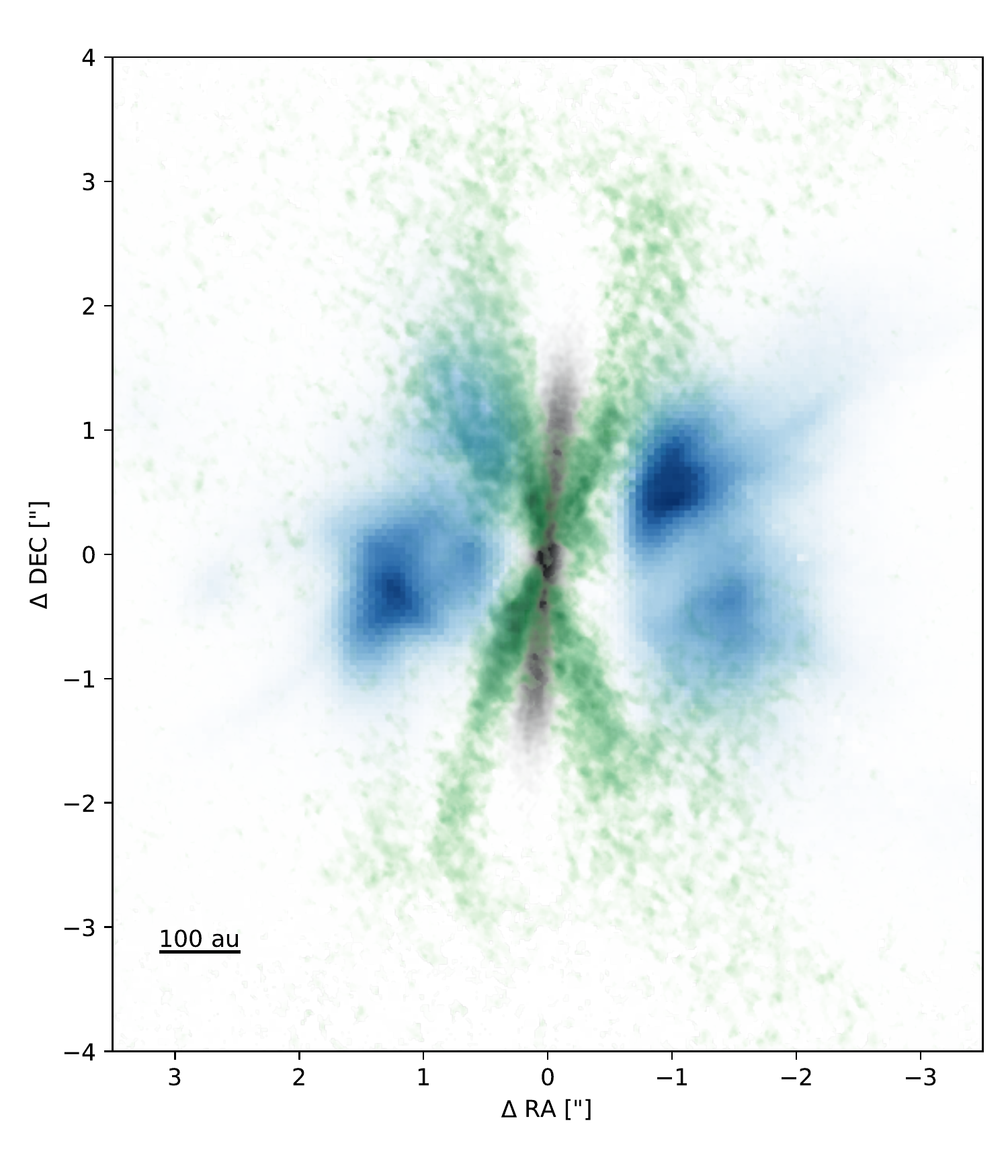}
    \caption{
        Composite image comparing the 1.3~mm continuum (black), 1.6~$\mu$m from the HST (blue), and $^{13}$CO with robust=0.5 (green). 
    }
    \label{fig:hst_with_ALMA}
\end{figure}

\subsection{Stellar Mass-Luminosity Tension}
From the rotation curve of C$^{18}$O, we find that the stellar mass is $\sim 1.65 M_{\odot}$. Given the stellar mass and depending on the age of the protostar, we should expect a luminosity that is greater than $\sim 2 L_{\odot}$ \citep[e.g.][]{Iben1965ApJ...141..993I, Kippennhahn1994sse..book.....K, Hillenbrand2004ApJ...604..741H}. However, the bolometric luminosity which is estimated to be $\sim 0.43 L_{\odot}$ \citep{Ohashi_edisk_overview} is much smaller. 

The diminished level of luminosity is likely because of the edge-on view. Most of the stellar photons along the line-of-sight are removed by extinction and not replenished by scattering causing an underestimation of the total luminosity \citep{Whitney2003ApJ...591.1049W}. Indeed, \cite{Grafe2013A&A...553A..69G} relied on a larger input stellar luminosity of $5 L_{\odot}$ to explain both the scattered light and mm-continuum of IRAS~04302. In contrast, younger protostars, like L1527 IRS, do not see such a difference between the bolometric and expected protostellar luminosity \citep{Tobin2012Natur.492...83T}. We speculate that the small envelope size, low density, and wide outflow cavities for IRAS~04302 results in lots of emission not being reprocessed and simply escape along the polar regions. The much larger envelope in size and mass of L1527 IRS \citep{Tobin2008ApJ...679.1364T}, on the other hand, could help capture and reprocess the photons. Whether the observationally inferred low bolometric luminosity of IRAS~04302 is consistent with the newly obtained stellar mass remains to be determined quantitatively. In principle, a fully consistent physical modeling including stellar irradiation can constrain the stellar luminosity since the disk temperature is constrained through imaging \citep[e.g.][]{Grafe2013A&A...553A..69G, Sheehan2017ApJ...851...45S}, but we leave it as a future effort.

%\subsection{Infall Material}
%several other sources also show this infall material
%which features are likely infall vs outflow? 

%do we expect infall? 

%there are also spiral features? elevated spirals? Contrast to Huang's sources with spirals 

%with all this infall, the small grains are obviously elevated based on the HST image ... the mm-continuum looks much thinner and is obviously not as settled as other Class II sources 

%infall is not from the midplane 

%is the infall different from Class II sources? 

\section{Conclusion} \label{sec:conclusion}

As part of the ALMA large program, eDisk, we presented high resolution ALMA Band 6 dust continuum and line emission of the nearly edge-on Class~I disk IRAS~04302. Our main results are as follows:

%treasure trove of complex interactions between envelope and disk.

% disk is nearly edge-on: inclination
% Keplerian rotation, stellar mass
% freeze-out zone
% large scale infall material
% outflow features 
\begin{enumerate}
    \item The dust continuum image has an angular resolution of $\sim 0.05\arcsec$ ($\sim$ 8~au) and shows a nearly edge-on disk with a clear brightness asymmetry along the disk minor axis. 
    %The brighter side is to the east which corresponds to the far side of the disk and the orientation is consistent with the blueshifted outflow to the east. 
    By fitting the disk with a 2D Gaussian, we find that the lower limit to the inclination is $\sim 83.6^{\circ}$ using the ratio of the major and minor axis FWHM. Through forward ray-tracing of the dust, we find that the inclination is $\sim 87^{\circ}$ and that the disk needs to be optically thick and geometrically thick to produce minor axis asymmetry. There is no evidence of rings and gaps, which could be due to the lack of radial substructure or because the highly inclined and optically thick view obscures the gaps. 
    
    \item We detect $^{12}$CO, $^{13}$CO, C$^{18}$O, H$_{2}$CO, and SO and find that all five exhibit V-shaped integrated intensity images which can be explained by freeze-out near the midplane. From C$^{18}$O, we estimate by eye that the CO snow line is located at $\sim 0.8\arcsec$ (130~au). However, the frozen-out midplane only extends to $\sim 2.8\arcsec$ (450~au) after which we detect the optically thicker tracer $^{13}$CO emission out to $\sim 3.9\arcsec$ (620~au) which forms a ``cap" of emission closing the V-shaped opening and produces a well-defined ``8"-shaped CO depletion region along the disk major axis (see Fig.~\ref{fig:image_collection}). 
    
    \item By fitting the position-velocity diagram of C$^{18}$O along the disk major axis, we find that the disk is in Keplerian rotation and that the stellar mass is $1.6 \pm 0.4 M_{\odot}$ (see Fig.~\ref{fig:slamfit}). The mass is in tension with the low observationally inferred bolometric luminosity of $\sim 0.43 L_{\odot}$. 
    
    \item The optically thick lines, $^{12}$CO and $^{13}$CO, trace significant amounts of complex extended structures outside of the Keplerian rotating disk. We find $^{12}$CO outflow along the blueshifted jet axis to the east which is consistent with the orientation of the continuum disk in which the far side of the disk is also to the east. In addition, we find blueshifted $^{13}$CO emission next to the redshifted part of the Keplerian disk surface which we suggest as material infalling onto the disk (see Fig.~\ref{fig:channel_13co_blue_SE_atmosphere}). 
    
    \item Our most important conclusion is that the dust has yet to settle significantly in the Class I IRAS~04302 disk. We find a dust scale height $\sim 6$ au at a radius of 100 au, which is comparable to the gas scale at the same radius. This result, coupled with the lack of dust settling in Class~0 disks, such as HH~212~mms, indicates that substantial dust settling should happen between the Class~I stage and Class~II stage. In addition, the radial extent of the dust disk is likely smaller than the gas disk which suggests that radial drift occurs sooner than vertical settling.
\end{enumerate}

\section*{Acknowledgments}
We thank the reviewer for the constructive comments. 
ZYDL acknowledges support from NASA 80NSSC18K1095, the Jefferson Scholars Foundation, the NRAO ALMA Student Observing Support (SOS) SOSPA8-003, the Achievements Rewards for College Scientists (ARCS) Foundation Washington Chapter, the Virginia Space Grant Consortium (VSGC), and UVA research computing (RIVANNA). ZYL is supported in part by NASA 80NSSC18K1095 and NSF AST-1910106. J.J.T. acknowledges support from NASA 21-XRP21-0064 and XRP 80NSSC22K1159. N.O. and C.O. acknowledges support from National Science and Technology Council (NSTC) in Taiwan through the grants NSTC 109-2112-M-001-051 and 110-2112-M-001-031. JKJ acknowledges support from the Independent Research Fund Denmark (grant No. 0135-00123B). LWL acknowledges support from NSF AST-2108794. S.T. is supported by JSPS KAKENHI grant Nos. 21H00048 and 21H04495, and by NAOJ ALMA Scientific Research grant No. 2022-20A. Y.A. acknowledges support by NAOJ ALMA Scientific Research Grant code 2019-13B, Grant-in-Aid for Scientific Research (S) 18H05222, and Grant-in-Aid for Transformative Research Areas (A) 20H05844 and 20H05847. M.L.R.H. acknowledges support from the Michigan Society of Fellows. IdG acknowledges support from grant PID2020-114461GB-I00, funded by MCIN/AEI/10.13039/501100011033. FJE acknowledges support from NSF AST-2108794. S.G. acknowledge support from the Independent Research Fund Denmark (grant No. 0135-00123B). PMK acknowledges support from NSTC 108-2112- M-001-012, NSTC 109-2112-M-001-022 and NSTC 110-2112-M-001-057. SPL and TJT acknowledge grants from the National Science and Technology Council of Taiwan 106-2119-M-007-021-MY3 and 109-2112-M-007-010-MY3. W.K. was supported by the National Research Foundation of Korea (NRF) grant funded by the Korea government (MSIT) (NRF-2021R1F1A1061794). C.W.L. is supported by the Basic Science Research Program through the National Research Foundation of Korea (NRF) funded by the Ministry of Education, Science and Technology (NRF- 2019R1A2C1010851), and by the Korea Astronomy and Space Science Institute grant funded by the Korea government (MSIT; Project No. 2022-1-840-05). JEL was supported by the National Research Foundation of Korea (NRF) grant funded by the Korean government (MSIT) (grant number 2021R1A2C1011718). R.S. acknowledge support from the Independent Research Fund Denmark (grant No. 0135-00123B). PDS acknowledges support from NSF AST-2001830 and NSF AST-2107784. JPW acknowledges support from NSF AST-2107841. Y.Y. is supported by the International Graduate Program for Excellence in Earth-Space Science (IGPEES), World-leading Innovative Graduate Study (WINGS) Program of the University of Tokyo. H.-W.Y. acknowledges support from the National Science and Technology Council (NSTC) in Taiwan through the grant NSTC 110-2628-M-001-003-MY3 and from the Academia Sinica Career Development Award (AS-CDA-111-M03).

This paper makes use of the following ALMA data: ADS/JAO.ALMA\#2019.1.00261.L. ALMA is a partnership of ESO (representing its member states), NSF (USA) and NINS (Japan), together with NRC (Canada), MOST and ASIAA (Taiwan), and KASI (Republic of Korea), in cooperation with the Republic of Chile. The Joint ALMA Observatory is operated by ESO, AUI/NRAO and NAOJ. The National Radio Astronomy Observatory is a facility of the National Science Foundation operated under cooperative agreement by Associated Universities, Inc.

\restartappendixnumbering

\appendix

\section{Continuum} \label{sec:continuum_robust}

% discuss resolved out structure? 
% potential for dark lane? 

Fig.~\ref{fig:continuum_robust} compares the continuum images for different robust weightings from -1, -0.5, 0, 0.5, and 1. The asymmetry along the disk minor axis is more evident with robust weightings smaller than $1$, but robust weightings less than 0 begins to resolve out the large scale major axis. The robust = 0.5 is a good compromise between resolving the minor axis asymmetry and not resolving out the large scale major axis. 

\begin{figure*}
    \centering
    \includegraphics[width=\textwidth]{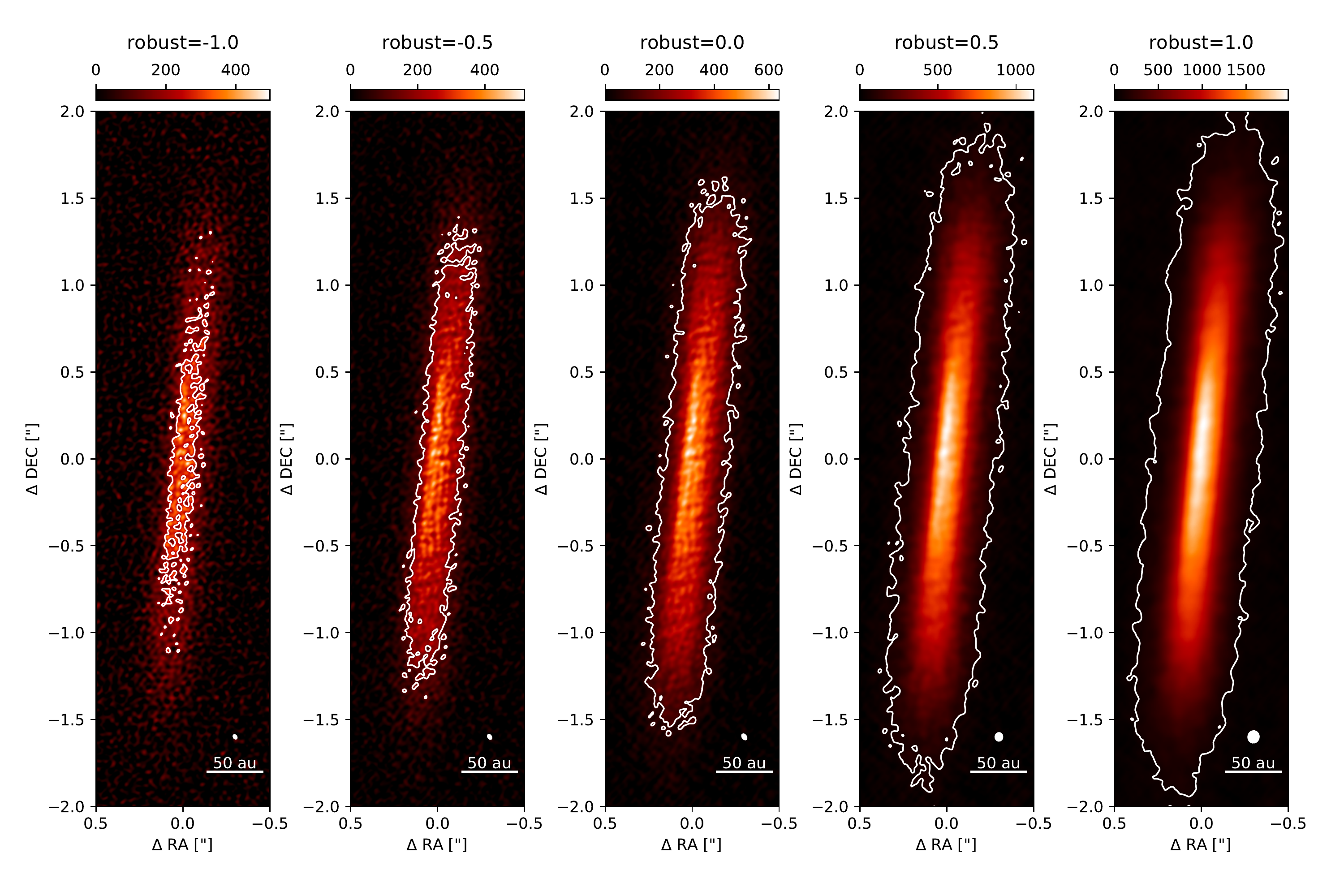}
    \caption{
        The continuum images for different robust weightings from -1, -0.5, 0, 0.5, and 1 (left to right). The color scale is the flux density in $\mu$Jy beam$^{-1}$. The white contour shows the 5$\sigma$ region. The white ellipse to the lower right of each image is the corresponding beam size. 
    }
    \label{fig:continuum_robust}
\end{figure*}

\section{Dust Modeling Uncertainty} \label{sec:dust_model_uncertainty}

Section~\ref{sec:continuum_RT_modeling} presented model image with parameters that are able to capture most of the features of the data. In this section, we provide a simple verification that the adopted parameters are the local best-fit by varying each individual parameter. The simple approach assumes that the parameters are not too correlated, which can be verified with better parameter space sampling techniques \citep[e.g.][]{Foreman-Mackey2013PASP..125..306F}. However, the complete exploration of the multi-dimensional parameter space is beyond the scope of this first-look paper.

For interferometric data, the visibility plane is where one should consider the goodness-of-fit between the model and observation to include the effects of finite sampling in the visibility plane and also it is where the true native uncertainty of each measurement resides. However, as an initial assessment and for easier comprehension of the image, we simply compare the model and observation in the image plane and leave the more complete post processing to a future exploration. 
To assess the goodness-of-fit, we calculate the reduced $\chi^{2}$, defined as $\Tilde{\chi}^{2}$, through
\begin{equation}
    \Tilde{\chi}^{2} \equiv \frac{1}{N} \sum_{i} \frac{ ( O_{i} - M_{i} )^{2} }{ \sigma^{2} }
\end{equation}
where $i$ iterates through each pixel of the image in steps of the beam size. $O_{i}$ and $M_{i}$ are the intensities the data and that of the model, respectively, at the $i$th pixel and $N$ is the total number of the selected pixels.

Considering the 7 free parameters ($i$, $R_{0}$, $T_{0}$, $H_{100}$, $\tau_{0,\nu}$, $\delta_{\text{RA}}$, and $\delta_{\text{DEC}}$), we create a series of models by varying each parameter. The range and step size for each parameter are listed in Table~\ref{tab:parameter_gridspace}. 

%An estimation of the uncertainty relies on looking for the value of the parameter that corresponds to the $\Tilde{\chi}^{2}+1$ level. We caution that the estimation works under the assumption that the error distribution of the measured data is Gaussian which may not be exactly the case when assessing the goodness-of-fit in the image plane. 

%we caution the chi+1 level as an estimate to the uncertainty 
%Note that it's not perfect, but it at least gives the relative sensitivity of each parameter. 
%On the assumption that the parameters are not too correlated.... 

\begin{deluxetable}{lcccccc}
    \tablenum{4}
    \tablecaption{The grid of parameters considered for the dust model \label{tab:parameter_gridspace} }
    \tablewidth{0pt}
    \tablehead{
        \colhead{Parameter} & \colhead{Units} & \colhead{Variable} & \colhead{Minimum} & \colhead{Maximum} & \colhead{Step} & \colhead{Adopted Value} \\
        }
    \decimalcolnumbers
    \startdata
        Inclination & $^{\circ}$     & $i$  & 85 & 89 & 1 & 87 \\
        Disk Edge & au  & $R_{0}$ & 270 & 350 & 20  & 310 \\
        Temperature at $R_{0}$ & K & $T_{0}$  & 6.5 & 8.5 & 0.5 & 7.5 \\
        Dust Scale Height at 100~au & au & $H_{100}$ & 4 & 8 & 1 & 6 \\
        Characteristic Optical Depth & & $\tau_{0,\nu}$ & 0.25 & 0.45 & 0.5 & 0.35 \\
        RA offset of star & $\arcsec$ & $\delta_{\text{RA}}$ & -0.05 & 0.05 & 0.01  & -0.03 \\
        DEC offset of star & $\arcsec$ & $\delta_{\text{DEC}}$ & -0.15 & 0.15 & 0.01 & -0.04 \\
    \enddata
    \tablecomments{
        Column (1): The name of the parameter. Column (2): The units of the parameter. Column (3): The variable used to represent the parameter. Column (4) and (5): The minimum and maximum range considered. Column (6): The step size of the parameter. Column (7): The final adopted value which is consistent with Table~\ref{tab:Qconstant_model_parameters}.
        }
\end{deluxetable}

The right column of Fig.~\ref{fig:chisq1d} shows the $\Tilde{\chi}^{2}$ by varying each parameter around the best-fit values shown in Section~\ref{sec:continuum_RT_modeling}. The intensities along the major and minor axes of the model with each varying parameter are also shown in the left and middle column of Fig.~\ref{fig:chisq1d} as a comparison to the observation to identify the effects of each parameter. We note that while we only show the major and minor axes profiles, $\Tilde{\chi}^{2}$ is evaluated across the image and not only along the major and minor axes.

\begin{figure}
    \centering
    \includegraphics[width=0.85\textwidth]{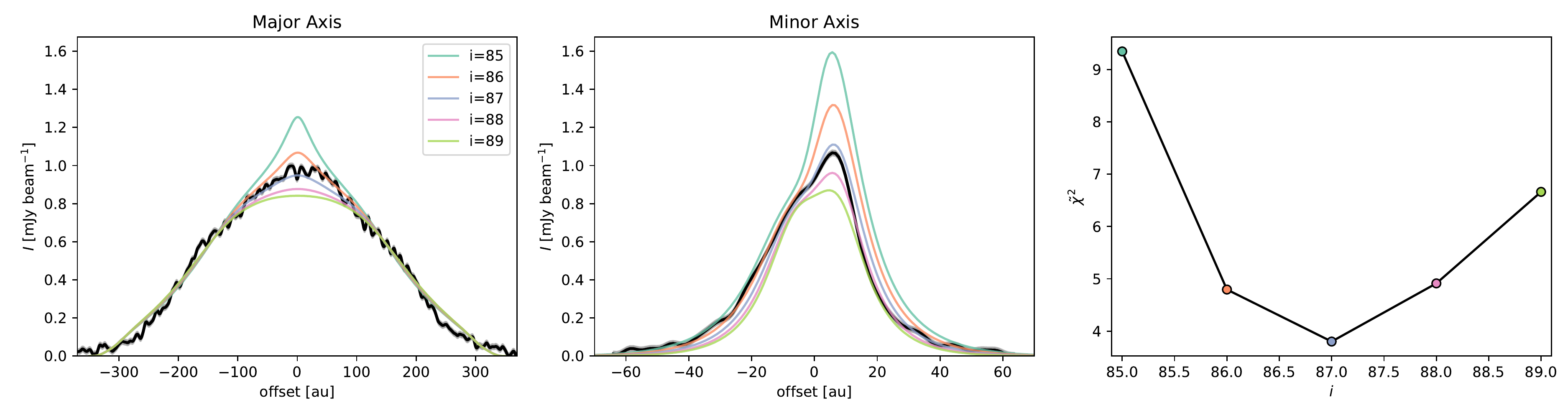}
    \includegraphics[width=0.85\textwidth]{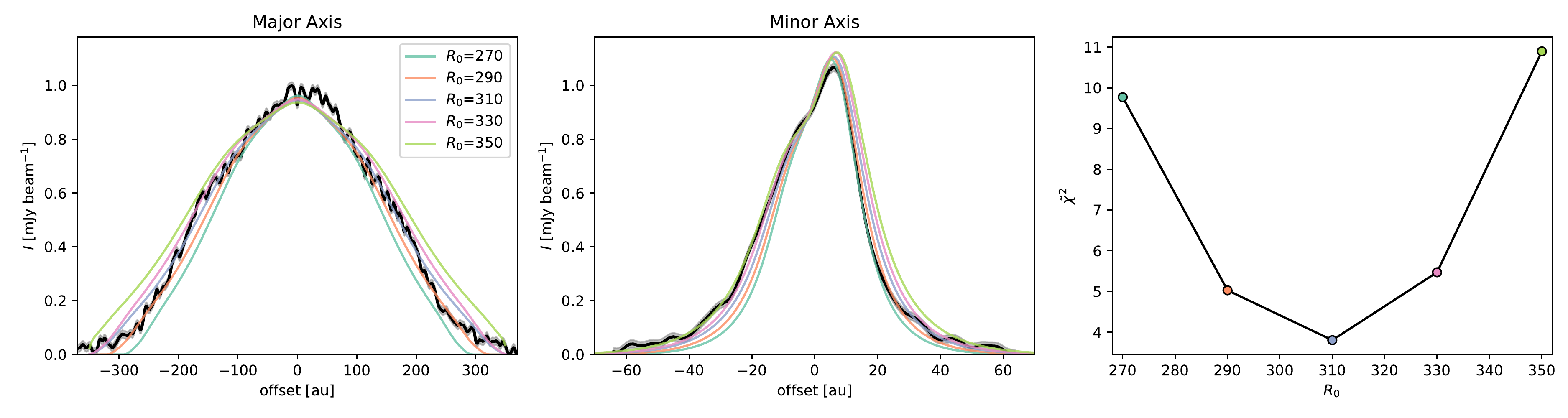}
    \includegraphics[width=0.85\textwidth]{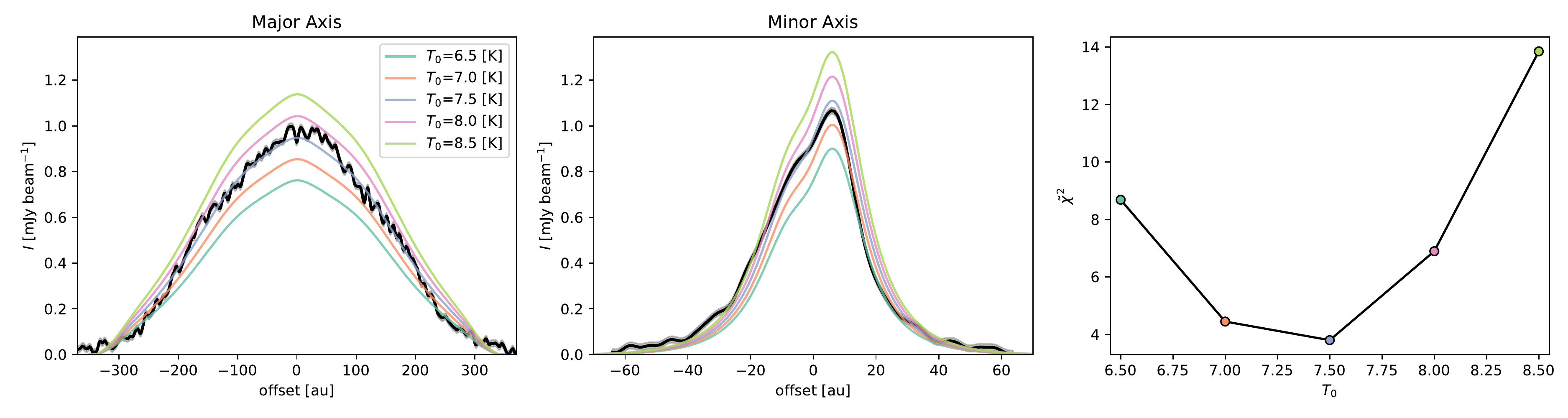}
    \includegraphics[width=0.85\textwidth]{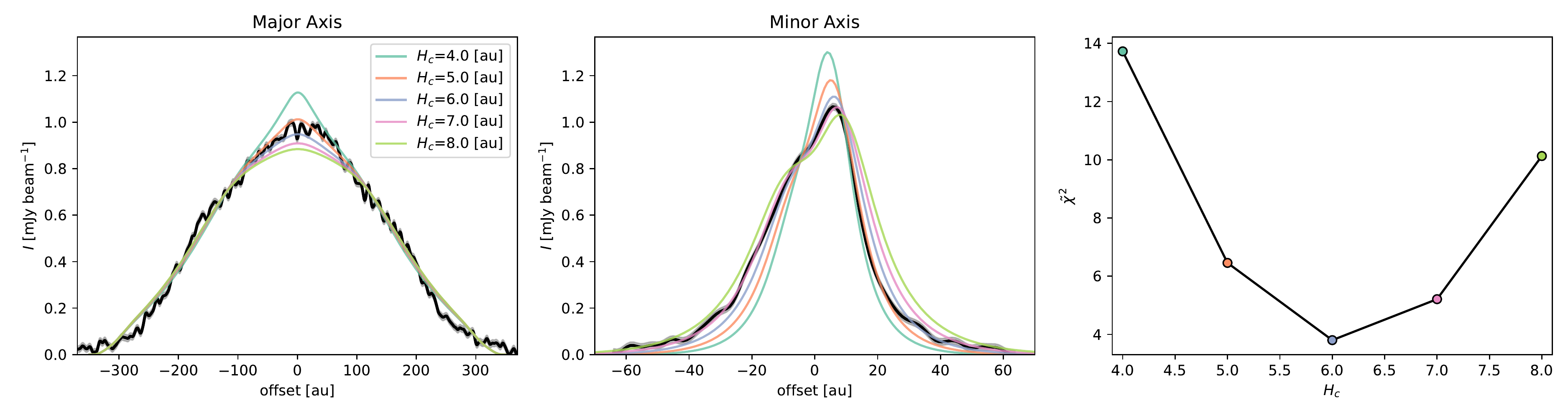}
    \includegraphics[width=0.85\textwidth]{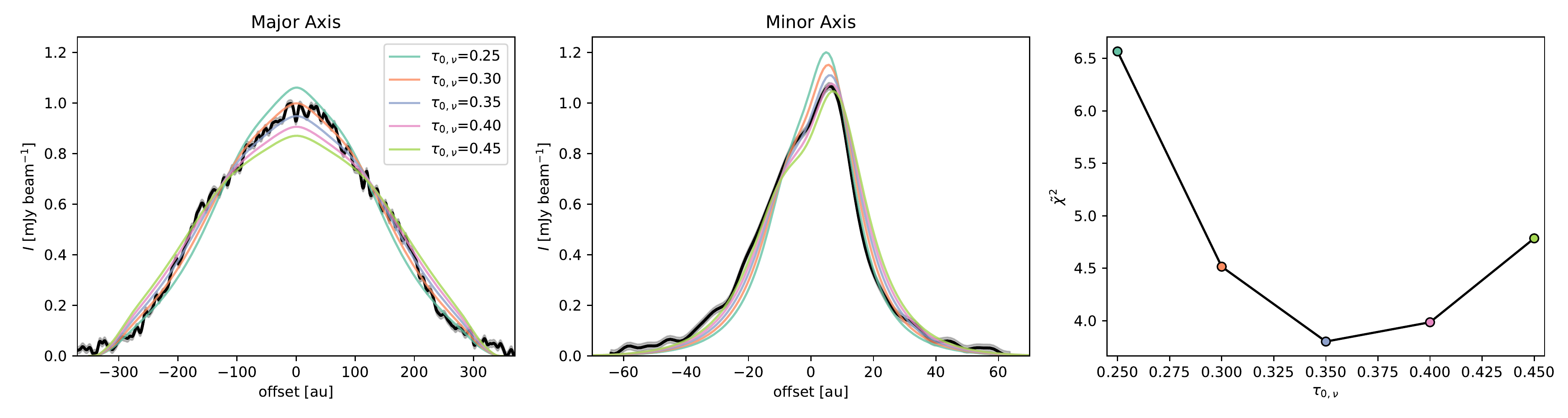}
    \caption{
        Left column: The comparison between the major axis of the observed continuum and that of the models with various parameters. The grey shaded region of the observed cut is the noise uncertainty. Middle column: The same comparison but along the minor axis. Right column: The $\Tilde{\chi}^{2}$ distribution for each value of the considered parameter. The color for each data point corresponds to the color used for each model in the major and minor axes cuts in the left and right columns. The parameters from top to bottom are: $i$, $R_{0}$, $T_{0}$, $H_{c}$, and $\tau_{0,\nu}$. 
    }
    \label{fig:chisq1d}
\end{figure}

Fig.~\ref{fig:chisq2d_cen} shows the $\Tilde{\chi}^{2}$ in two-dimensions from varying both $\delta_{\text{RA}}$ and $\delta_{\text{DEC}}$, since both simply describes a translation of the image in the plane-of-sky. 

\begin{figure}
    \centering
    \includegraphics[width=\textwidth]{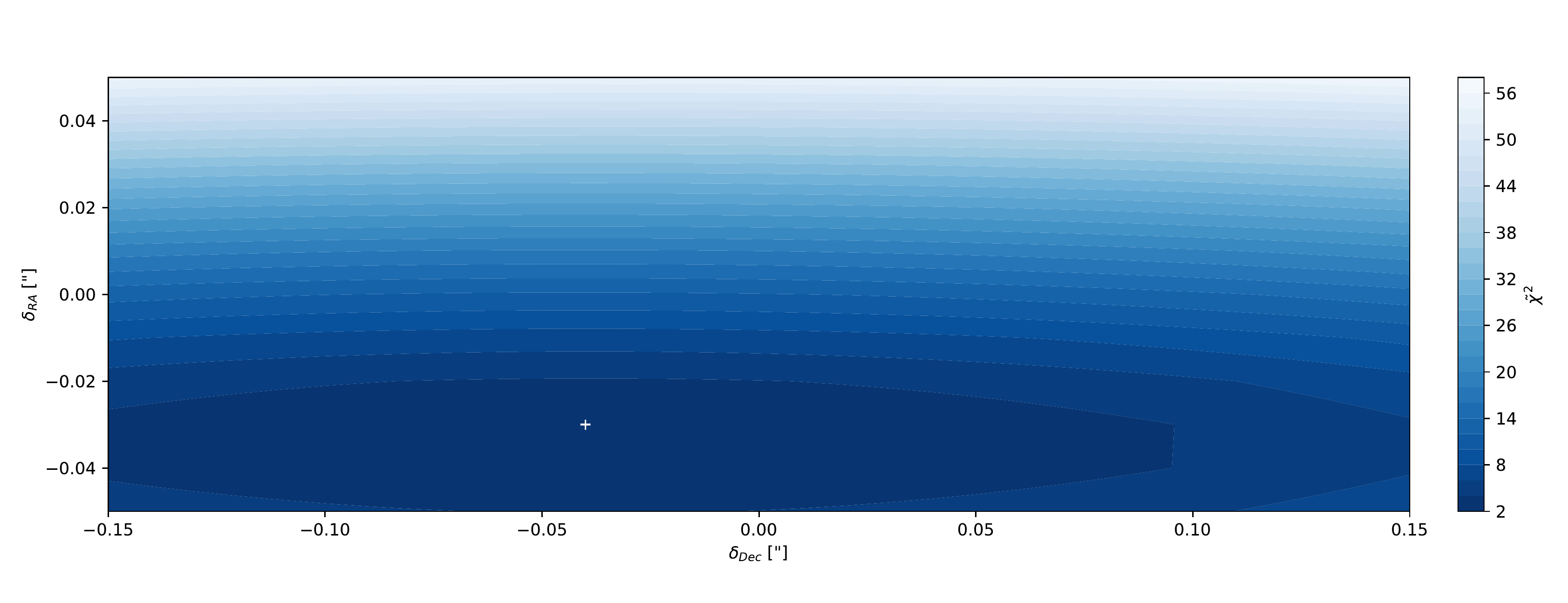}
    \caption{
        The $\Tilde{\chi}^{2}$ distribution from iterating $\delta_{\text{RA}}$ and $\delta_{\text{DEC}}$. The location with the minimum $\Tilde{\chi}^{2}$ is marked by a white cross. 
    }
    \label{fig:chisq2d_cen}
\end{figure}

\section{VLA Band Ka Image and Proper Motion} \label{sec:vla_bandka}

To account for proper motion, we utilize the Very Large Array (VLA) Ka-band observed in 2015 (PI: John Tobin; project code: 15A-381). Since the VLA Ka-band image has not been published elsewhere, we briefly describe the calibration procedure and resulting images below.

The VLA Ka-band data were observed in both B and A configuration, which have maximum baseline lengths of $\sim$11~km and $\sim$36~km, respectively. The B-configuration data were taken on 2015 Feb 16 with a 3 hour execution, and the A-configuration data were taken on 2015 Aug 16 and 2015 Sept 8 with $\sim$1.5 hour executions. The observations all used 3C84 as the bandpass calibrator, 3C147 as the flux density calibrator, and J0440+2728 as the complex gain calibrator. During the observations, pointing was updated approximately every hour using the source J0403+2600. The correlator was configured for 3-bit continuum mode with 4 GHz basebands centered at 28.97~GHz and 36.796~GHz. The bandwith was broken up into 64 spectral windows, each with 64 channels and 128MHz in width.

The data were processed using the scripted VLA calibration pipeline (version 1.3.1) in CASA 4.2.2. We ran the pipeline twice, we used the first run to identify data that required flagging. We then applied the necessary flags to the data and re-ran the pipeline on the edited dataset. Then to prepare for imaging the data, we combined the three measurement sets into a single measurement set using the CASA task \texttt{concat}.

We created the image with the robust set to 0.5 and \texttt{uvtaper} set to 3000k$\lambda$. The noise level is $\sigma=$6.3~$\mu$Jy beam$^{-1}$. The resolution is $0.13\arcsec \times 0.13\arcsec$ with a beam position angle of $-72^{\circ}$. The representative frequency is 33~GHz (9.1~mm). Following Section~\ref{sec:observations}, we assume a $10\%$ absolute flux calibration uncertainty, but only consider the statistical uncertainty for the rest of this section.

Fig.~\ref{fig:vla_continuum} (left) shows a continuum image that is centrally peaked and largely elongated along the north and south. The southern part of the disk appears slightly broader and brighter than the northern part. Using \texttt{imfit} from CASA, we fit a 2D Gaussian and obtain a center of %(04:33:16.4972, +022:53:20.33) in J2000 or 
(04:33:16.4952, +22:53:20.34) in ICRS. The integrated flux from the fitted 2D Gaussian is $490\pm30$~$\mu$Jy, while the integrated flux above $3\sigma$ is $380$~$\mu$Jy. The FWHM of the deconvolved major axis is $750 \pm 50$~mas and that of the minor axis is $150 \pm 20$~mas. The position angle is $175^{\circ} \pm 1^{\circ}$ which is consistent with the fitted 2D Gaussian for Band~6. The consistent position angle is evidence that the Band~Ka image still detects the edge-on dust disk, while the smaller major axis FWHM is expected as the disk is optically thinner at the longer wavelength \citep[e.g.][]{Lin2021}.

Fig.~\ref{fig:vla_continuum} (right) also shows the cuts along the major and minor axes using the position angle derived from Section~\ref{sec:continuum}. The major axis does not appear symmetric from the origin. For example, the secondary peak at $-0.25\arcsec$ corresponds to a trough at $+0.25\arcsec$ with values of $51$~$\mu$Jy beam$^{-1}$ and 21~$\mu$Jy beam$^{-1}$ respectively and the difference is $\sim 5\sigma$. Similarly, the secondary peak at $+0.35\arcsec$ corresponds to a trough at $-0.35\arcsec$. At face value, the secondary peaks could suggest the existence of substructure and rule out symmetric rings given the non-symmetric locations of the secondary peaks. However, as demonstrated in the case of L1527 IRS, secondary peaks may not correspond to physical substructure when observed with better integration times \citep{Nakatani2020ApJ...895L...2N, Sheehan2022ApJ...934...95S}.

The peak of the continuum is 0.115~mJy beam$^{-1}$ and, equivalently, the brightness temperature is $8.7$~K using the full Planck function. The brightness temperature is much lower than the $\sim 14$~K from Band~6 in Section~\ref{sec:continuum}, but slightly higher than $\sim 6.7$~K at Band~4 \citep{Villenave2020}. Although one may expect that, for an edge-on disk, the longer wavelength should trace the inner regions with higher temperature and lead to higher brightness temperature \citep{Lin2021}, the low value is likely because the vertical extent of the disk is unresolved. 

\begin{figure*}
    \centering
    \includegraphics[width=\textwidth]{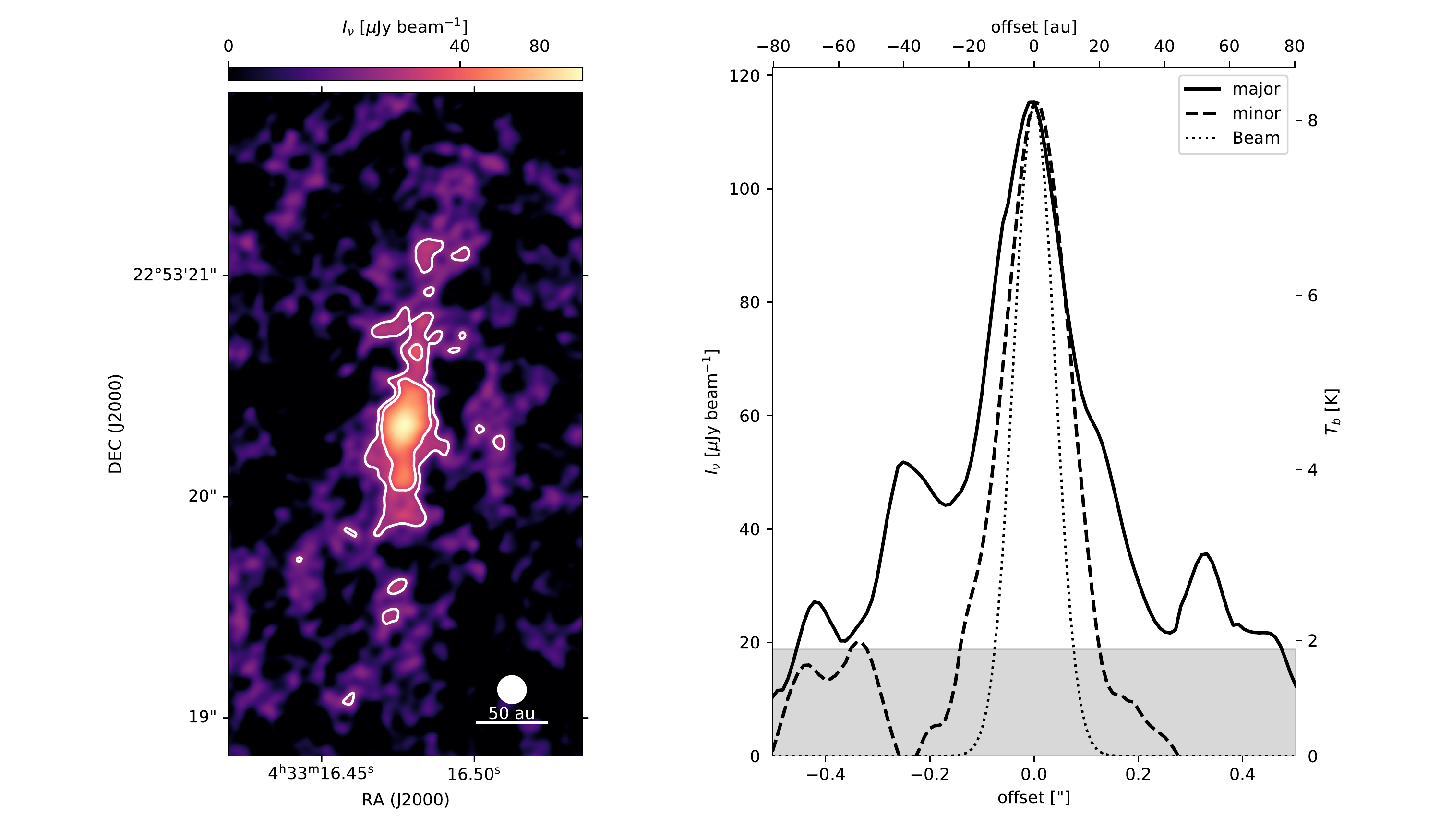}
    \caption{
        (left) The VLA Band~Ka continuum image. The contours mark the $3\sigma$ and $5\sigma$ levels (see Appendix~\ref{sec:vla_bandka} for $\sigma$). The white ellipse in the lower right represents the beam. The length scale is 50~au using the adopted distance of 160~pc. (right) The cuts along the major (solid line) and minor (dashed line) axes. The dotted line represents the Gaussian beam with the peak set to the peak of the cuts. The origin is at the center of the fitted 2D Gaussian. The bottom and top axes mark the offset from the origin along the cut in arcsec and au. The positive location for the major axis is along the northern part of the disk, while the positive location for the minor axis is along the eastern part of the disk. The left and right axes mark the intensity in $\mu$Jy beam$^{-1}$ and brightness temperature (using the full Planck function) in Kelvin respectively. The shaded region is the intensity below $3\sigma$. 
    }
    \label{fig:vla_continuum}
\end{figure*}

%The Band 4 image was observed on 2017 Sep 27 06:53 (UTC) with a fitted center of (04:33:16.50, +22:53:20.3) in ICRS. The Band 7 image was observed on 2017 Aug 18 11:30 (UTC) with a fitted center of (04:33:16.5, +22:53:20.3). In addition to Fitting across all four different centers at different observation times, 
By comparing with the Band 6 continuum from this work, we find a proper motion of $\sim (7, -17)$ mas yr$^{-1}$ in RA and Dec, respectively. The proper motion is similar to that of L1527 IRS which is in the same Taurus region \citep{Loinard2002ApJ...581L.109L}. 

% ==== bibliography ====
\bibliography{main}{}
\bibliographystyle{aasjournal}

\end{document}